# INFLUENCE OF SPEED LIMIT ON ROADWAY SAFETY IN INDIANA

A Thesis

Submitted to the Faculty

of

Purdue University

by

Nataliya V. Malyshkina

In Partial Fulfillment of the

Requirements for the Degree

of

Master of Science in Civil Engineering

To my mother, father and husband

# ACKNOWLEDGMENTS

I would first like to thank my advisor, Professor Fred Mannering. Without his expert advice and his support none of this research would be possible. Not only has he always helped and guided me, but he also carefully listened to my opinions and suggestions on the research. I am lucky to be his student.

I would like to thank Professor Kristofer Jennings and especially Professor Andrew Tarko for their very helpful comments and for carefully reading the thesis. I would also like to thank Dorothy Miller, Maeve Drummond and Marcie Duffin for their help in completing all administrative procedures and requirements for the M.S. thesis defense.

I am deeply indebted to my colleagues at the Ural State University of Railroad Transportation in Russia, where I obtained my first research and teaching experience. This thesis and my graduate studies at Purdue University would never be possible without their support and encouragement years ago.

Finally, I feel endless gratitude and love to my wonderful family – my mother, Nadezhda, my father, Vladimir, and my husband, Leonid. I owe everything I have to them and to their love and support.

# TABLE OF CONTENTS

# LIST OF TABLES

# LIST OF FIGURES

# ABSTRACT

Malyshkina, Nataliya V. M.S., Purdue University, December 2006.
Influence of Speed Limit on Roadway Safety in Indiana. Major Professor: Fred Mannering.

The influence of speed limits on roadway safety is an extremely important social issue and is subject to an extensive debate in the State of Indiana and nationwide. With around 800-900 fatalities and thousands of injuries annually in Indiana, traffic accidents place an incredible social and economic burden on the state. Still, speed limits posted on highways and other roads are routinely exceeded as individual drivers try to balance safety and mobility (speed). This research explores the relationship between speed limits and roadway safety. Namely, the research focuses on the influence of the posted speed limit on the causation and severity of accidents. Data on individual accidents from the Indiana Electronic Vehicle Crash Record System is used in the research, and appropriate statistical models are estimated for causation and severity of different types of accidents on all road classes. The results of the modeling show that speed limits do not have a statistically significant adverse effect on unsafe-speed-related causation of accidents on all roads, but generally increase the severity of accidents on the majority of roads other than highways (the accident severity on highways is unaffected by speed limits). Our findings can perhaps save both lives and travel time by helping the Indiana Department of Transportation determine optimal speed limit policies in the state.

# CHAPTER 1. INTRODUCTION

A new law, which took effect on July 1, 2005, made Indiana the 30th U.S. state to raise interstate speed limits up to 70 mph. The top speed limit value was increased on some portions of the state's interstate highway system from 65 mph to 70 mph. This increase intensified an important debate in the engineering community on the tradeoff between highway mobility (speed) and safety. On one hand, as speed increases, travel times decrease, which reduces transportation costs and leads to an increased productivity and a noticeable positive effect for the national economy. On the other hand, higher speed can possibly have a negative effect on roadway safety.

The relationship between speed limits and roadway safety is not as obvious as it seems. The reason is that there are several important issues in this relationship. On one hand, as speed increases, vehicles have higher kinetic energy, travel larger distances during human reaction times, and vehicles are exposed to stronger aerodynamic and centrifugal. This tends to increase accident frequency and severity. On the other hand, as speed increases, the variance of vehicle velocities may decrease, resulting in easier and safer driving conditions. As a result, the overall effect of a speed limit increase on road safety is complicated, and requires a thorough study. Such a study and a detailed analysis of the relationship between speed limits and safety on Indiana roads will be undertaken in this thesis.

In general, there are two measures of road safety that are commonly considered:

1. The first measure evaluates accident frequencies on roadway sections. The accident frequency on a road section is obtained by counting the number of accidents occurring on this section during a specified period of time. Then count-data statistical models (e.g. Poisson, negative binomial models and their zero-inflated counterparts) are estimated for accident frequencies on different road sections. The explanatory variables used in these models are the road section characteristics (e.g. road section length, curvature, slope, type, etc).
2. The second measure evaluates accident severity outcomes as determined by the injury level sustained by the most severely injured individual (if any) involved into the accident. This evaluation is done by using data on *individual* accidents and estimating discrete outcome statistical models (e.g. ordered probit and multinomial logit models) for the accident severity outcomes. The explanatory variables used in these models are the individual accident characteristics (e.g. time and location of an accident, weather conditions and road characteristics at the accident location, characteristics of the vehicles and drivers involved, etc).

These two measures of read safety are complementary. On one hand, an accident frequency study gives a statistical model of the probability of an accident occurring on a road section. On the other hand, an accident severity study gives a statistical model of the conditional probability of a severity outcome of an accident, given an accident occurred. The unconditional probability of the accident severity outcome is the product of its conditional probability and the accident probability.

The number of road safety studies that consider one or both of the two road safety measures described above is enormous. Some of the key studies include the following:

- Shankar et al. (1996) used a nested logit model for statistical analysis of accident severity outcomes on rural highways in Washington State. They found that environment conditions, highway design, accident type, driver and vehicle characteristics significantly influence accident severity. They found that overturn accidents, rear-end accidents on wet pavement, fixed-object accidents, and failures to use the restraint belt system lead to higher probabilities of injury or/and fatality accident outcomes, while icy pavement and single-vehicle collisions lead to higher probability of property damage only outcomes.
- Shankar et al. (1997) studied the distinction between safe and unsafe road sections by estimating zero-inflated Poisson and zero-inflated negative binomial models for accident frequencies in Washington State (for these models the zero state corresponds to near zero accident likelihood on safe road sections).
- Duncan et al. (1998) applied an ordered probit model to injury severity outcomes in truck-passenger car rear-end collisions in North Carolina. They found that injury severity is increased by darkness, high speed differentials, high speed limits, wet grades, drunk driving, and being female.
- Karlaftis and Tarko (1998) considered heterogeneous panel data for frequencies of accidents occurred in Indiana over a 6-year period. They developed an improved method of accident frequency modeling in panel data, which is based on a two-step approach: first, heterogeneous data is divided into homogeneous groups by determining (dis)similarities and using cluster analysis; second, negative binomial models are estimated separately for each homogeneous data group. The results obtained by Karlaftis and Tarko clearly indicate that there are significant differences between the accident frequency models estimated for urban, suburban and rural counties.

- Chang and Mannering (1999) focused on the effects of trucks and vehicle occupancies on accident severities. They estimated nested logit models for severity outcomes of truck-involved and non-truck-involved accidents in Washington State and found that accident injury severity is noticeably worsened if the accident has a truck involved, and that the effects of trucks are more significant for multi-occupant vehicles than for single-occupant vehicles.
- Carson and Mannering (2001) studied the effect of ice warning signs on ice-accident frequencies and severities in Washington State. They modeled accident frequencies and severities by using zero-inflated negative binomial and logit models respectively. They found that the presence of ice warning signs was not a significant factor in reducing ice-accident frequencies and severities.
- Khattak (2001) estimated ordered probit models for severity outcomes of multi-vehicle rear-end accidents in North Carolina. In particular, the results of his research indicate that in two-vehicle collisions the leading driver is more likely to be severely injured, in three-vehicle collisions the driver in the middle is more likely to be severely injured, and being in a newer vehicle protects the driver in rear-end collisions.
- Ulfarsson (2001) and Ulfarsson and Mannering (2004) focused on male and female differences in analysis of accident severity. They used multinomial logit models and accident data from Washington State. They found significant behavioral and physiological differences between genders, and also found that probability of fatal and disabling injuries is higher for females as compared to males.
- Kockelman and Kweon (2002) applied ordered probit models to modeling of driver injury severity outcomes. They used a nationwide accident data sample and found that pickups and sport utility vehicles are less (more) safe than passenger cars in single-vehicle (two-vehicle) collisions.

- Lee and Mannering (2002) estimated zero-inflated count-data models and nested logit models for frequencies and severities of run-off-roadway accidents in Washington State. They found that run-off-roadway accident frequencies can be reduced by avoiding cut side slopes, decreasing (increasing) the distance from outside shoulder edge to guardrail (light poles), and decreasing the number of isolated trees along roadway. The results of their research also show that run-off-roadway accident severity is increased by alcohol impaired driving, high speeds, and the presence of a guardrail.
- Abdel-Aty (2003) used ordered probit models for analysis of driver injury severity outcomes at different road locations (roadway sections, signalized intersections, toll plazas) in Central Florida. He found higher probabilities of severe accident outcomes for older drivers, male drivers, those not wearing seat belt, drivers who speed, those who drive vehicles struck at the driver's side, those who drive in rural areas, and drivers using electronic toll collection device (E-Pass) at toll plazas.
- Kweon and Kockelman (2003) studied probabilities of accidents and accident severity outcomes for a given fixed driver exposure (which is defined as the total miles driven). They used Poisson and ordered probit models, and considered a nationwide accident data sample. After normalizing accident rates by driver exposure, the results of their study indicate that young drivers are far more crash prone than other drivers, and that sport utility vehicles and pickups are more likely to be involved in rollover accidents.
- Yamamoto and Shankar (2004) applied bivariate ordered probit models to an analysis of driver's and passenger's injury severities in collisions with fixed objects. They considered a 4-year accident data sample from Washington State and found that collisions with leading ends of guardrail and trees tend to cause more severe injuries, while collisions with sign posts, faces of guardrail, concrete barrier or bridge and fences tend to cause less severe injuries. They also found that proper use of vehicle restraint system strongly decreases the probability of severe injuries and fatalities.

- Khorashadi et al. (2005) explored the differences of driver injury severities in rural and urban accidents involving large trucks. Using 4-years of California accident data and multinomial logit model approach, they found considerable differences between rural and urban accident injury severities. In particular, they found that the probability of severe/fatal injury increases by 26% in rural areas and by 700% in urban areas when a tractor-trailer combination is involved, as opposed to a single-unit truck being involved. They also found that in accidents where alcohol or drug use is identified, the probability of severe/fatal injury is increased by 250% and 800% in rural and urban areas respectively.
- Islam and Mannering (2006) studied driver aging and its effect on male and female single-vehicle accident injuries in Indiana. They employed multinomial logit models and found significant differences between different genders and age groups. Specifically, they found an increase in probabilities of fatality for young and middle-aged male drivers when they have passengers, an increase in probabilities of injury for middle-aged female drivers in vehicles 6 years old or older, and an increase in fatality probabilities for males older than 65 years old.
- Savolainen (2006), Savolainen and Mannering (2006a) and Savolainen and Mannering (2006b) focused on an important topic of motorcycle safety on Indiana roads. He used multinomial and nested logit models and found that poor visibility, unsafe speed, alcohol use, not wearing a helmet, right-angle and head-on collisions, and collisions with fixed objects cause more severe motorcycle-involved accidents.

As far, as the relationship between speed and road safety is concerned, it has been studied in the past by considering the two measures of road safety described above. Previous empirical studies of this relationship have generally found the following two results. First, on all road classes (urban streets, highways, etc) vehicle operating speeds exceed the posted speed limit (Renski et al., 1999; Khan, 2002). Second, there are no sure indications that a reasonable increase in speed limit has a considerable negative impact on traffic safety. For example, O'Donnell and Connor (1996) estimated logit and probit models for injury severity outcomes of accidents in Australia and determined that effects of an increase in vehicle speed from 42 to 100 kilometers per hour (from 26.1 mph to 62.1 mph) are surprisingly small. Shankar et al. (1997) used zero-inflated Poisson and zero-inflated negative binomial models for a study of accident frequencies. They found that a speed limit increase reduced accident frequencies on road sections in the Western part of Washington State, and had no statistically significant effect on accident frequencies in the Eastern part. Very similar results were obtained by Milton and Mannering (1998), who estimated negative binomial models for frequencies of accidents on sections of principal arterials in Washington State in 1992 and 1993 and found a reduction of the frequencies with a speed limit increase. Renski et al. (1999) specifically addressed the effect of speed limit on injury severity outcomes in single-vehicle accidents on interstate highways in North Carolina. They used a paired-comparison analysis and ordered probit modeling. They found that while increasing speed limits from 55 to 60 mph and from 55 to 65 mph increased the probability of sustaining minor and non-incapacitating injuries, increasing speed limits from 65 to 70 mph did not have a significant effect on accident severity. A thorough analysis of speed limit policies for Indiana was recently carried out by Khan (2002). He found that while previous upward changes in speed limits in Indiana during the past two decades did increase speeds observed on roads, there was no statistically significant evidence to indicate that such increases had a negative impact on safety.

In the present study we focus on the relationship between speed and road safety. We consider data on individual accidents and use the methodologies of statistical modeling within the framework of the accident discrete outcome analysis (refer to the second measure of road safety discussed on page 2 above). However, our study differs from the previous studies in that we analyze both the severity and causation of accidents. We will compile and use data from Indiana for different types of accidents (single-vehicle accidents, car or SUV versus truck accidents, etc) on all classes of roads (interstate highways, urban streets, US routes, etc). To analyze and understand the effect of speed limit on roadway safety, in our study we will use the following two statistical modeling approaches:

1. In the first approach we will focus on causation of accidents. The idea is to study a relationship between the posted speed limit and the probability of unsafe and/or excessive speed being the primary cause of the accident. This is done by estimation of appropriate statistical models for the unsafe-speed-related accident causation.
2. In the second approach we will undertake a traditional accident severity study. We will estimate statistical models for the level of accident severity (determined by the injury level sustained by the most critically injured individual in the accident). Then we will test whether the posted speed limit has any effect on accident severity.

To reveal the effect of speed limits on safety, while modeling accident causation and severity, we will control for other possible confounding effects, such as road characteristics, weather conditions, driver characteristics, and so on. To increase the predictive power of our models, we will consider accidents separately for each combination of accident type and road class (e.g. single-vehicle accidents on urban streets will be considered separately from car-truck accidents on interstate highways). The use of the above two accident modeling approaches will provide important new insights and sufficient statistical evidence on the effect of the posted speed limit on roadway safety.

The thesis is organized as follows. In the next chapter we will briefly describe the methodology of statistical modeling used in our study. Detailed descriptions and simple descriptive statistics of the accident data used are given in CHAPTER 3. In CHAPTER 4 we consider influence of speed limit on accident causation related to unsafe and/or excessive speed. In CHAPTER 5 we consider influence of speed limit on accident severity level. Finally, in CHAPTER 6 we summarize and discuss the main results of our study, and consider implications for optimal speed limit policies in Indiana State. All details on the study results, including the estimated statistical models for accident causation and severity, are given in the appendices.

# CHAPTER 2. METHODOLOGY OF STATISTICAL MODELING

Our study deals with accident causation and accident severity, both are non-quantitative discrete outcomes of traffic accidents. The most widely used statistical models for non-count data that is composed of discrete outcomes are the multinomial logit model and the ordered probit model. However, there are two potential problems with applying ordered probability models to accident severity outcomes (Savolainen and Mannering 2006b). The first is related to the fact that non-injury accidents are likely to be under-reported in accident data because they are less likely to be reported to authorities. The presence of under-reporting in an ordered probability model can result in biased and inconsistent model coefficient estimates. In contrast, the coefficient estimates of an unordered multinomial logit probability model are consistent except for the constant terms (Washington et. al. 2003, page 279). The second problem is related to undesirable restrictions that ordered probability models place on influences of the explanatory variables (Washington et. al. 2003, page 294). As a result, in our research study we use and estimate binary and multinomial logit models for accident causation and severity.

The multinomial logit model can be introduced as follows. Let there be $N$ available data observations and $I$ possible discrete outcomes in each observation. Then in the multinomial logit model the probability $P_n^{(i)}$ of the $i$ th outcome in the $n$ th observation is specified by equation (Washington et al., 2003, page 263)

$$P_n^{(i)} = \frac{\exp(\boldsymbol{\beta}_\mathbf{i}'\mathbf{X}_\mathbf{in})}{\sum_{j=1}^{I}\exp(\boldsymbol{\beta}_\mathbf{j}'\mathbf{X}_\mathbf{jn})}, \qquad i = 1,2,3,...,I\ , \quad n = 1,2,3,...,N\ . \qquad \text{Eq. 2.1}$$

Here $\mathbf{X_{in}}$ is the vector of explanatory variables for the $n^{\text{th}}$ observation and $\boldsymbol{\beta}_\mathbf{i}$ is the vector of model coefficients to be estimated ($\boldsymbol{\beta}'_\mathbf{i}$ is the transpose of $\boldsymbol{\beta}_\mathbf{i}$). We use a conventional assumption that the first component of vector $\mathbf{X_{in}}$ is equal to unity, and therefore, the first component of vector $\boldsymbol{\beta}_\mathbf{i}$ is the intercept in linear product $\boldsymbol{\beta}'_\mathbf{i}\mathbf{X_{in}}$. Note that $P_n^{(i)}$, given by Equation (2.1), is a valid probability set for $I$ discrete outcomes because the necessary and sufficient conditions $P_n^{(i)} \geq 0$ and $\sum_{i=1}^{I} P_n^{(i)} = 1$ are obviously satisfied[1].

We can multiply the numerator and denominator of the fraction in Equation (2.1) by an arbitrary number without any change of the probabilities. As a result, without any loss of generality we can set one of the intercepts to zero. We choose the first component of vector $\boldsymbol{\beta}_I$ to be zero in this case. Moreover, if the vector of explanatory variables does not depend on discrete outcomes, i.e. if $\mathbf{X_{in}} \equiv \mathbf{X_n}$, then without any loss of generality we can set one of vectors of model coefficients to zero. We choose vector $\boldsymbol{\beta}_I$ to be zero in this case.

Because accidents are independent events, the likelihood function $L$ and the log-likelihood function $LL$ for the set of probabilities given in Equation (2.1) are obviously equal to

$$L = \prod_{n=1}^{N} \prod_{i=1}^{I} [P_n^{(i)}]^{\delta_{in}}, \quad LL = \sum_{n=1}^{N} \sum_{i=1}^{I} \delta_{in} P_n^{(i)}, \qquad \text{Eq. 2.2}$$

where $\delta_{in}$ is defined to be equal to unity if the $i^{\text{th}}$ discrete outcome is observed in the $n^{\text{th}}$ observation and to zero otherwise.

---

[1] Equation (2.1) can formally be derived by using a linear specification $U_{in} \equiv \boldsymbol{\beta}'_\mathbf{i}\mathbf{X_{in}} + \tilde{\varepsilon}_{in}$, by defining $P_n^{(i)} = \text{Prob}\left\{ U_{in} \geq \max_{\forall j \neq i}(U_{jn}) \right\}$ and by choosing the Gumbel (Type I) extreme value distribution for the i.i.d. random error terms $\tilde{\varepsilon}_{in}$. For details see Washington et al., 2003.

Now we assume that the explanatory variables vector is independent of the discrete outcomes, $\mathbf{X_{in}} \equiv \mathbf{X_n}$, and consider two simple special cases of the multinomial logit model. First, if there are just two possible discrete outcomes, $I = 2$ and $i = 1,2$, then in this case the model becomes a binary logit model, and Equation (2.1) simplifies to

$$P_n^{(1)} = \frac{\exp(\boldsymbol{\beta}_1'\mathbf{X_n})}{\exp(\boldsymbol{\beta}_1'\mathbf{X_n}) + 1}, \qquad P_n^{(2)} = \frac{1}{\exp(\boldsymbol{\beta}_1'\mathbf{X_n}) + 1}, \qquad \text{Eq. 2.3}$$

where there is only one coefficient vector $\boldsymbol{\beta}_1$ to be estimated. Second, if there are three possible discrete outcomes, $I = 3$ and $i = 1,2,3$, then in this case Equation (2.1) simplifies to

$$\begin{aligned} P_n^{(1)} &= \frac{\exp(\boldsymbol{\beta}_1'\mathbf{X_n})}{\exp(\boldsymbol{\beta}_1'\mathbf{X_n}) + \exp(\boldsymbol{\beta}_2'\mathbf{X_n}) + 1}, \\ P_n^{(2)} &= \frac{\exp(\boldsymbol{\beta}_2'\mathbf{X_n})}{\exp(\boldsymbol{\beta}_1'\mathbf{X_n}) + \exp(\boldsymbol{\beta}_2'\mathbf{X_n}) + 1}, \\ P_n^{(3)} &= \frac{1}{\exp(\boldsymbol{\beta}_1'\mathbf{X_n}) + \exp(\boldsymbol{\beta}_2'\mathbf{X_n}) + 1}, \end{aligned} \qquad \text{Eq. 2.4}$$

where there are two coefficient vectors $\boldsymbol{\beta}_1$ and $\boldsymbol{\beta}_2$ to be estimated. We will use these special-case logit models in the next two chapters.

It is customary to use the maximum likelihood method to estimate unknown vectors of coefficients $\boldsymbol{\beta}_\mathbf{i}$ in the logit models given by Equations (2.1), (2.3) and (2.4). Namely, one finds such values of the unknown coefficients that the likelihood function (and correspondingly the log-likelihood function) given by Equation (2.2) reaches its global maximum. In the present study we use econometric software package LIMDEP for all model estimations by means of the maximum likelihood method[2]. We also use MATLAB software package for initial processing of data.

---

[2] LIMDEP can be found at http://www.limdep.com, we use Version 7.0 in our study.

In the next chapters we will need to compare several estimated models in order to infer if there are statistically significant differences among these models. As a result, here we would like to demonstrate how model comparisons are done by using a likelihood ratio test. Assume that we have divided a data sample into different data bins. The likelihood ratio test uses the model estimated for the whole data sample and the models separately estimated for each data bin. The test statistic is (Washington et al., 2003, page 244)

$$-2\left[LL(\boldsymbol{\beta})-\sum_{m=1}^{M}LL(\boldsymbol{\beta}_{\mathbf{m}})\right] \sim \chi^2_{\mathrm{df}=(M-1)\times K}\ , \qquad \text{Eq. 2.5}$$

where $LL(\boldsymbol{\beta})$ is the log-likelihood of the model estimated for the whole data sample and $\boldsymbol{\beta}$ is the vector of coefficients estimated for this model; $LL(\boldsymbol{\beta}_{\mathbf{m}})$ is the log-likelihood of the model estimated for observations in the $m^{\mathrm{th}}$ data bin and $\boldsymbol{\beta}_{\mathbf{m}}$ is the vector of coefficients estimated for this model ($m=1,2,3,...,M$); $M$ is the number of the data bins; $K$ is the number of coefficients estimated for each model (i.e. $K$ is the length of vectors $\boldsymbol{\beta}$ and $\boldsymbol{\beta}_{\mathbf{m}}$)[3]; $\chi^2_{\mathrm{df}=(M-1)\times K}$ is the chi-squared distribution with $(M-1)\times K$ degrees of freedom (df). The zero-hypothesis for the test statistic given by Equation (2.5) is that the model estimated for the whole data sample and the combination of the $M$ models separately estimated for the data bins, are statistically the same. In other words, for a chosen confidence level $\pi$ if the left-hand-side of Equation (2.5) is between zero and the (1-$\pi$)$^{\mathrm{th}}$ percentile of the chi-squared distribution given on the right-hand-side, then we conclude that the division of the data into different bins makes no statistically significant difference for the model estimation. We conclude that there is a difference otherwise.

---

[3] Note that the left-hand-side of Equation (2.5) is always non-negative because a combination of models separately estimated for data bins always provides a fit which is at least as good as the fit for the whole data sample.

At the end of this chapter we describe how the magnitude of the influence of specific explanatory variables on the discrete outcome probabilities can be measured. This is done by elasticity computations (Washington et al., 2003, page 271). Elasticities $E_{X_{jn,k}}^{P_n^{(i)}}$ are computed from the partial derivatives of the outcome probabilities for the $n^{\text{th}}$ observation as

$$E_{X_{jn,k}}^{P_n^{(i)}} = \frac{\partial P_n^{(i)}}{\partial X_{jn,k}} \cdot \frac{X_{jn,k}}{P_n^{(i)}}, \quad i, j = 1,...,I, \quad n = 1,...,N, \quad k = 1,...,K. \qquad \text{Eq. 2.6}$$

Here $P_n^{(i)}$ is the probability of outcome $i$ given by Equation (2.1), $X_{jn,k}$ is the $k^{\text{th}}$ component of the vector of explanatory variables $\mathbf{X_{jn}}$ that enters the formula for the probability of outcome $j$, and $K$ is the length of this vector. If $j = i$, then the elasticity given by Equation (2.6) is called *direct* elasticity, otherwise, if $j \neq i$, then the elasticity is called *cross* elasticity. The direct elasticity of the outcome probability $P_n^{(i)}$ with respect to variable $X_{in,k}$ measures the percent change in $P_n^{(i)}$ that results from an infinitesimal percentage change in $X_{in,k}$. Note that $X_{in,k}$ directly enters the numerator of the formula for $P_n^{(i)}$, as given by Equation (2.1). The cross elasticity of $P_n^{(i)}$ with respect to variable $X_{jn,k}$ measures the percent change in $P_n^{(i)}$ that results from an infinitesimal percentage change in $X_{jn,k}$. Note that $X_{jn,k}$ enters the numerator of the formula for the probability $P_n^{(j)}$ of the outcome $j$, which is different from outcome $i$. Thus, cross elasticities measure indirect effects that arise from the fact that the outcome probabilities must sum to unity, $\sum_{i=1}^{I} P_n^{(i)} = 1$. If the absolute value of the computed elasticity $E_{X_{jn,k}}^{P_n^{(i)}}$ of explanatory variable $X_{jn,k}$ is less than unity, then this variable is said to be inelastic, and the resulting percentage change in the outcome probability $P_n^{(i)}$ will be less (in its absolute value) than a percentage change in the variable. Otherwise, the variable is said to be elastic.

Using Equation (2.1) and calculating the derivatives in Equation (2.6), we obtain the formulas for the direct and cross elasticities of explanatory variables in the multinomial logit model:

$$E^{P_n^{(i)}}_{X_{in,k}} = \left[1 - P_n^{(i)}\right] \cdot \beta_{i,k} X_{in,k} \qquad \text{for direct elasticities;}$$
$$E^{P_n^{(i)}}_{X_{jn,k}} = -P_n^{(j)} \cdot \beta_{j,k} X_{jn,k} \qquad \text{for cross elasticities, } j \neq i \text{ .} \qquad \text{Eq. 2.7}$$

Here $\beta_{i,k}$ is the $k$ th component of the vector of the model estimable coefficients in the formula for the probability $P_n^{(i)}$ of outcome $i$ [refer to Equation (2.1)]. If the explanatory variables vector is independent of the discrete outcomes, $\mathbf{X_{in}} \equiv \mathbf{X_n}$, then Equations (2.7) stay valid with $X_{in,k} \equiv X_{jn,k} \equiv X_{n,k}$ .

It is customary to report averaged elasticities, which are the elasticities averaged over all observations (i.e. averaged over $n = 1,2,3,...,N$). Let us consider the cases of two and three possible discrete outcomes, given by Equations (2.3) and (2.4) respectively, and let us average the elasticities given by Equations 2.7) over all observations. Then we find the averaged direct and cross elasticities. In the case of two discrete outcomes ($i = 1,2$) we obtain

$$\overline{E}^{(1)}_{1;X_k} = \left\langle E^{P_n^{(1)}}_{X_{1n,k}} \right\rangle_n = \left\langle \left[1 - P_n^{(1)}\right] \cdot \beta_{1,k} X_{n,k} \right\rangle_n \qquad \text{averaged direct elasticity;}$$
$$\overline{E}^{(2)}_{1;X_k} = \left\langle E^{P_n^{(2)}}_{X_{1n,k}} \right\rangle_n = -\left\langle P_n^{(1)} \cdot \beta_{1,k} X_{n,k} \right\rangle_n \qquad \text{averaged cross elasticity.} \qquad \text{Eq. 2.8}$$

In the case of three discrete outcomes ($i = 1,2,3$) we obtain

$$\left.\begin{aligned} \overline{E}^{(1)}_{1;X_k} &= \left\langle E^{P_n^{(1)}}_{X_{1n,k}} \right\rangle_n = \left\langle \left[1 - P_n^{(1)}\right] \cdot \beta_{1,k} X_{n,k} \right\rangle_n \\ \overline{E}^{(2)}_{2;X_k} &= \left\langle E^{P_n^{(2)}}_{X_{2n,k}} \right\rangle_n = \left\langle \left[1 - P_n^{(2)}\right] \cdot \beta_{2,k} X_{n,k} \right\rangle_n \end{aligned}\right\} \text{averaged direct elasticities;}$$
$$\left.\begin{aligned} \overline{E}^{(2)}_{1;X_k} = \overline{E}^{(3)}_{1;X_k} &= \left\langle E^{P_n^{(2,3)}}_{X_{1n,k}} \right\rangle_n = -\left\langle P_n^{(1)} \cdot \beta_{1,k} X_{n,k} \right\rangle_n \\ \overline{E}^{(1)}_{2;X_k} = \overline{E}^{(3)}_{2;X_k} &= \left\langle E^{P_n^{(1,3)}}_{X_{2n,k}} \right\rangle_n = -\left\langle P_n^{(2)} \cdot \beta_{2,k} X_{n,k} \right\rangle_n \end{aligned}\right\} \text{averaged cross elasticities.} \qquad \text{Eq. 2.9}$$

Here brackets $\langle ... \rangle_n$ means averaging over all observations $n = 1,2,3,...,N$ .

All elasticity formulas given above are applicable only when explanatory variable $X_{jn,k}$ used in the outcome probability model is continuous. In the case when $X_{jn,k}$ takes on discrete values, the elasticities given by Equation (2.6) can not be calculated, and they are replaced by *pseudo-elasticities* (for example, see Washington et al., 2003, page 272). The later are given by the following equation, which is an obvious discrete counterpart of Equation (2.6),

$$E_{X_{jn,k}}^{P_n^{(i)}} = \frac{\Delta P_n^{(i)}}{\Delta X_{jn,k}} \cdot \frac{X_{jn,k}}{P_n^{(i)}}, \quad i, j = 1,...,I, \quad n = 1,...,N, \quad k = 1,...,K. \qquad \text{Eq. 2.10}$$

Here $\Delta P_n^{(i)}$ denotes the resulting discrete change in the probability of outcome $i$ due to discrete change $\Delta X_{jn,k}$ in variable $X_{jn,k}$. We will neither calculate nor use pseudo-elasticities in the present research study.

# CHAPTER 3. DATA DESCRIPTION

The accident data used in the present study is from the Indiana Electronic Vehicle Crash Record System (EVCRS). The EVCRS was launched in 2004 and includes available information on all accidents investigated by Indiana police starting from January 1, 2003.

The information on accidents included into the EVCRS can be divided into three major categories[4]:

1. An Environmental Record – it includes information on circumstances related to an accident. For example, weather, roadway and traffic conditions, number of dead and injured people involved, etc.
2. A Vehicle and Driver Record – it includes information on all vehicles involved into an accident and on all drivers of these vehicles. For example, accident contributing factors by each vehicle, type and model of each vehicle, posted speed limit for each vehicle, driver's injury status, driver's age and gender, driver's name and address, etc.
3. Non-driver Individual Record – it includes information on all people who are involved into an accident but are not drivers. This record includes only the name and address of those people, but it does not include any information on their injuries (if any).

---

[4] Note that accident data is subject to missing observations and typos. In addition, there can be misidentification errors on police crash reports due police officers' mistakes and prejudices. We eliminate obvious typos during initial data processing and exclude missing observations, but we do not correct for concealed typos and unobserved misidentification errors. Such correction can be done under the Bayesian statistics and Markov Chain Monte Carlo (MCMC) simulations framework, in which one introduces and estimates auxiliary unobserved state variables that indicate unobserved errors (Tsay, 2002, page 413). This is beyond the scope of our study. We assume that police misidentification errors are sufficiently small not to affect our final results.

In our study we use only information from the first two categories above. These two categories include 127 variables for each accident, which is an abundance of data. However, we do not need to consider all these variables. Indeed, because our study focuses on accident causation and severity, we choose all information and all data variables that can reasonably be related to the subject of our study, and we consider only these variables. For example, we do not consider the name of the road where an accident took place and the license plate numbers of the vehicles involved because we can reasonably expect that these variables do not contribute to the accident cause and severity. The list of all variables that we consider and their explanation is given in Appendix A.

In the present study we use data on 204,382 accidents that occurred in 2004 and 182,922 accidents that occurred in 2006. We do not consider 2005 accidents because in 2005 the top speed limit value was raised on some portions of Indiana interstate highways from 65 to 70 mph, and we would like to separate our research results and conclusions from the effects of drivers' adjustment to new speed limit values.

## 3.1. Accident data for year 2004

The percentage distributions of 2004 accidents by road class and by accident type are given in Figure 3.1 and Figure 3.2 respectively[5].

[5] For convenience, from each of the percentage distribution plot we exclude accidents for which the considered descriptive variable (e.g. road class or accident type) is unknown.

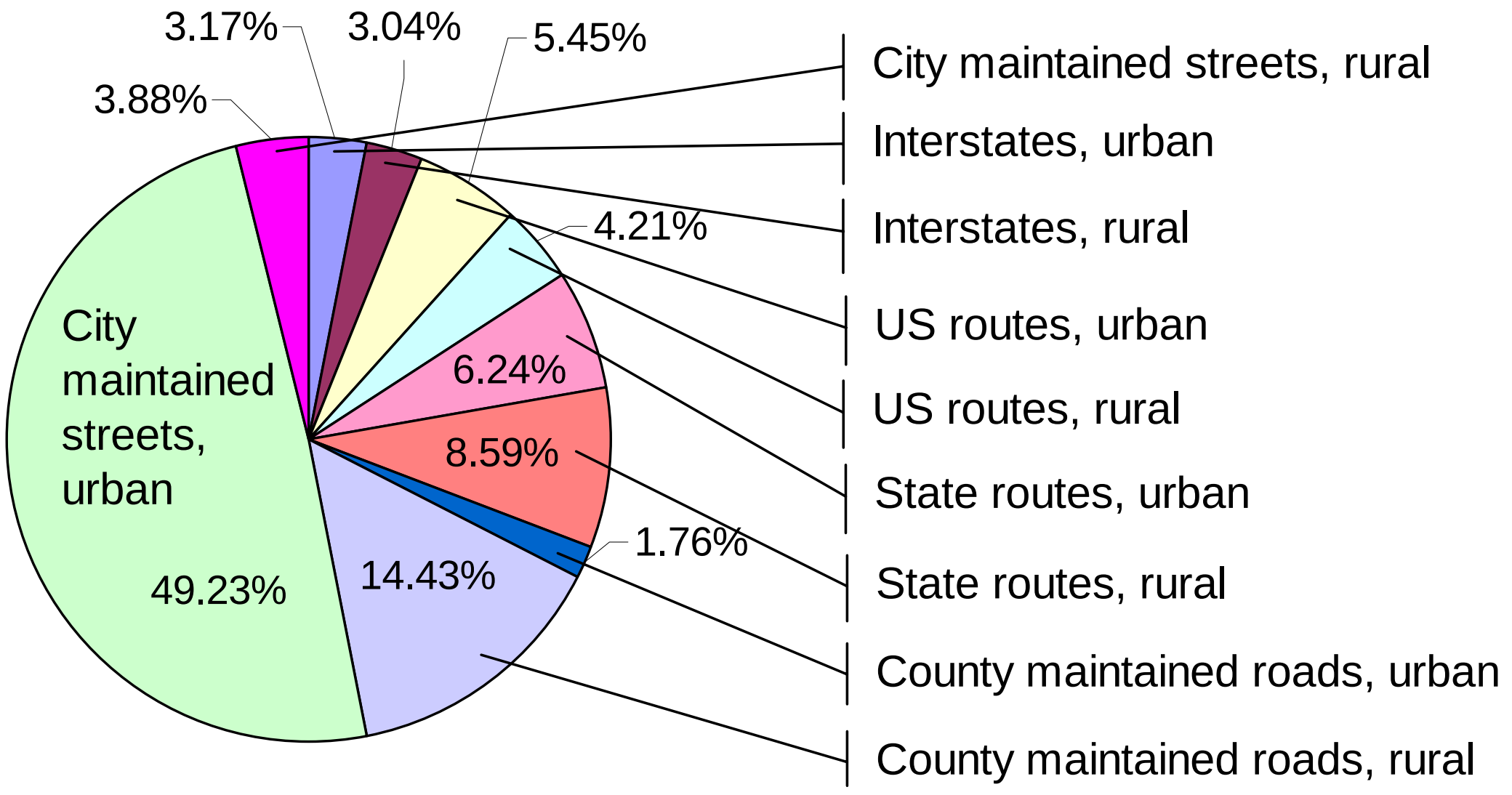

Figure 3.1 Percentage distribution of 2004 accidents by road class

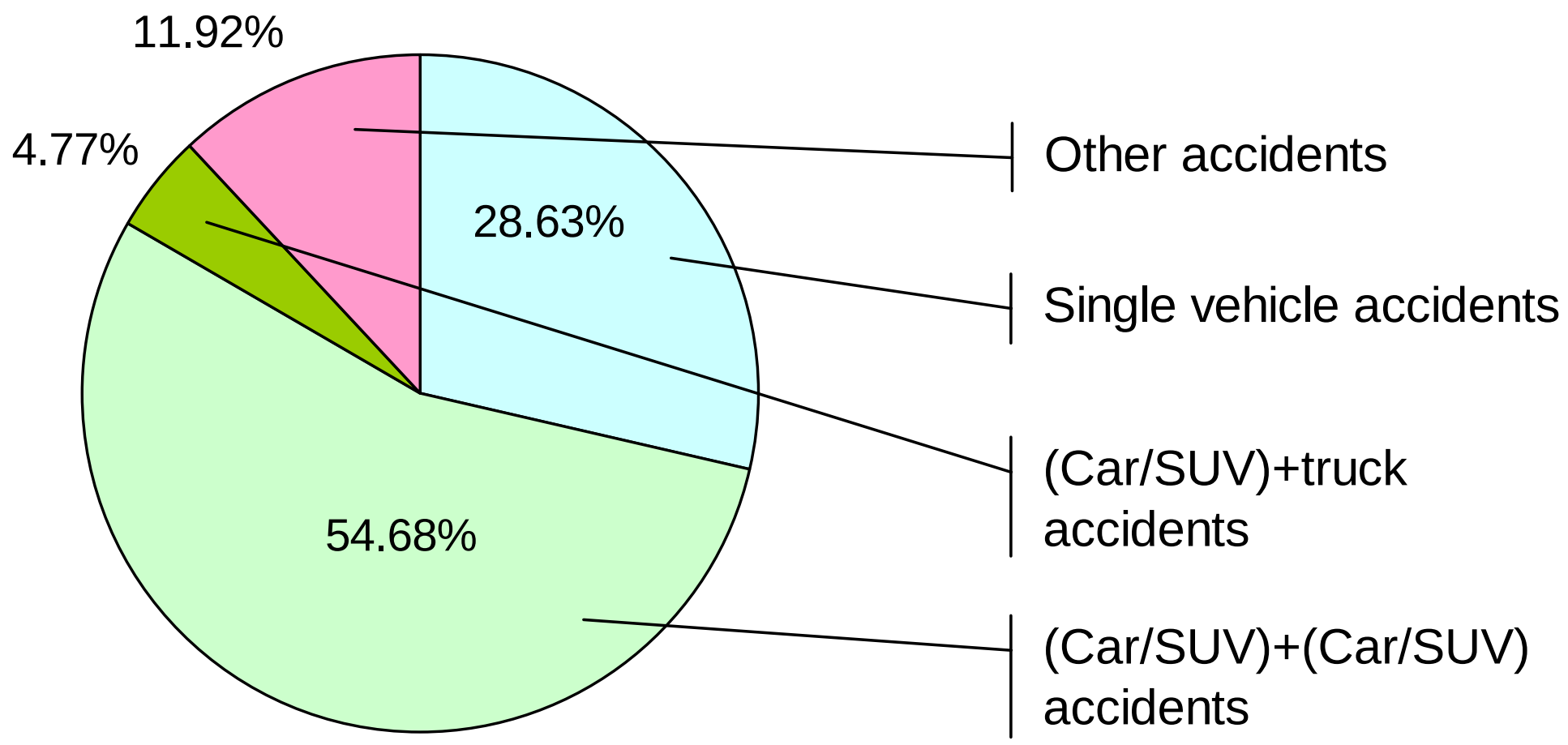

Figure 3.2 Percentage distribution of 2004 accidents by their type

As stated above, the goal of our study is to analyze the effect of speed limit on unsafe-speed-related causation and severity of accidents. As a result, first, we plot the percentage distributions of all 2004 accidents by their causation and severity level in Figure 3.3 and Figure 3.4 respectively. Second, we divide 2004 accidents into four different speed limit data bins, which respectively include accidents that occurred on roads with low ($\leq 30$ mph), medium-low ($> 30$ mph but $\leq 50$ mph), medium-high ($> 50$ mph but $\leq 60$ mph) and high ($> 60$ mph) speed limits. Finally, we plot the percentage distributions by accident causation and severity level separately for accidents in each of these chosen speed limit bins. The plots are given in Figure 3.5 and Figure 3.6.

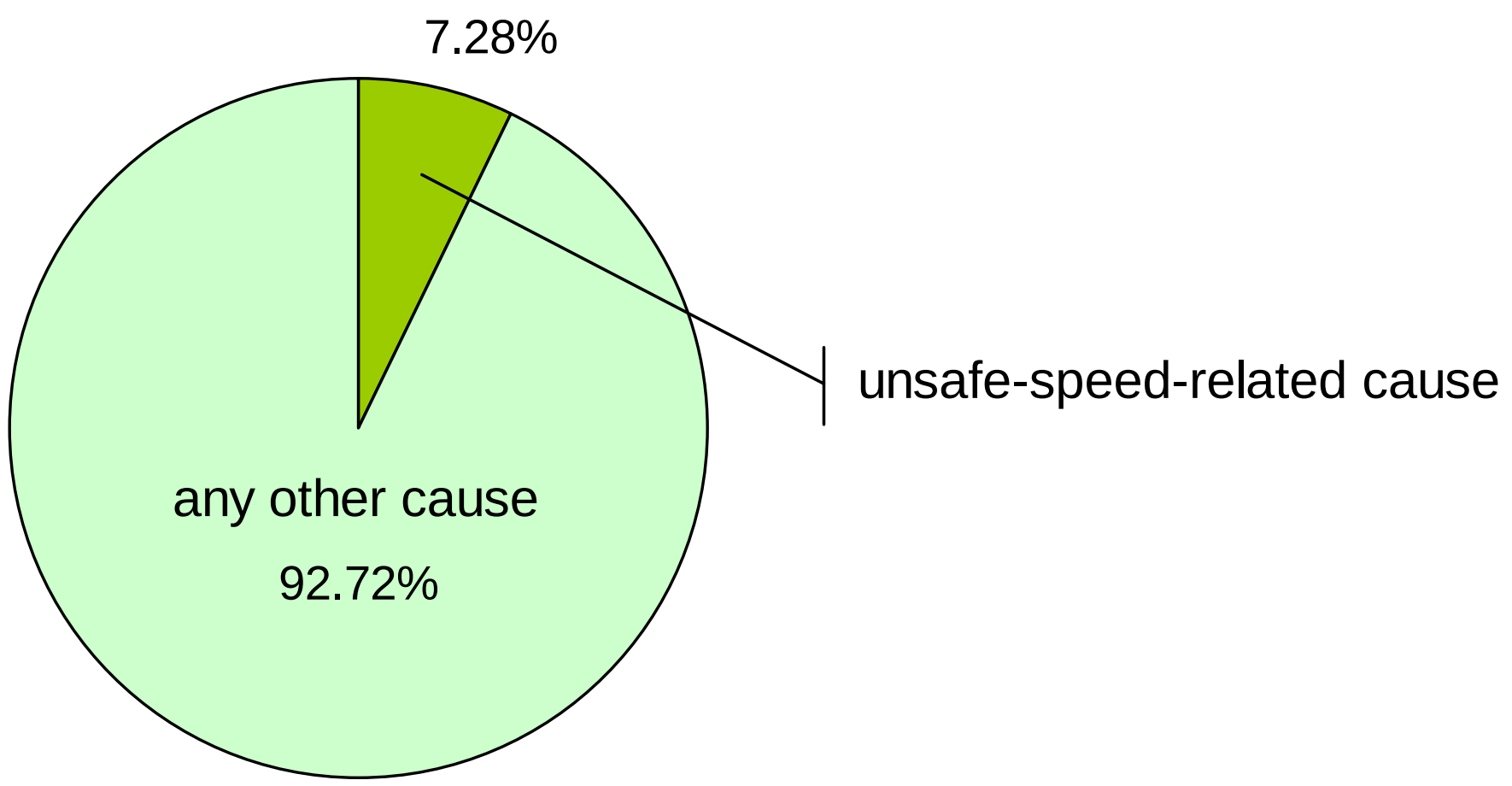

Figure 3.3 Percentage distribution of 2004 accidents by their causation

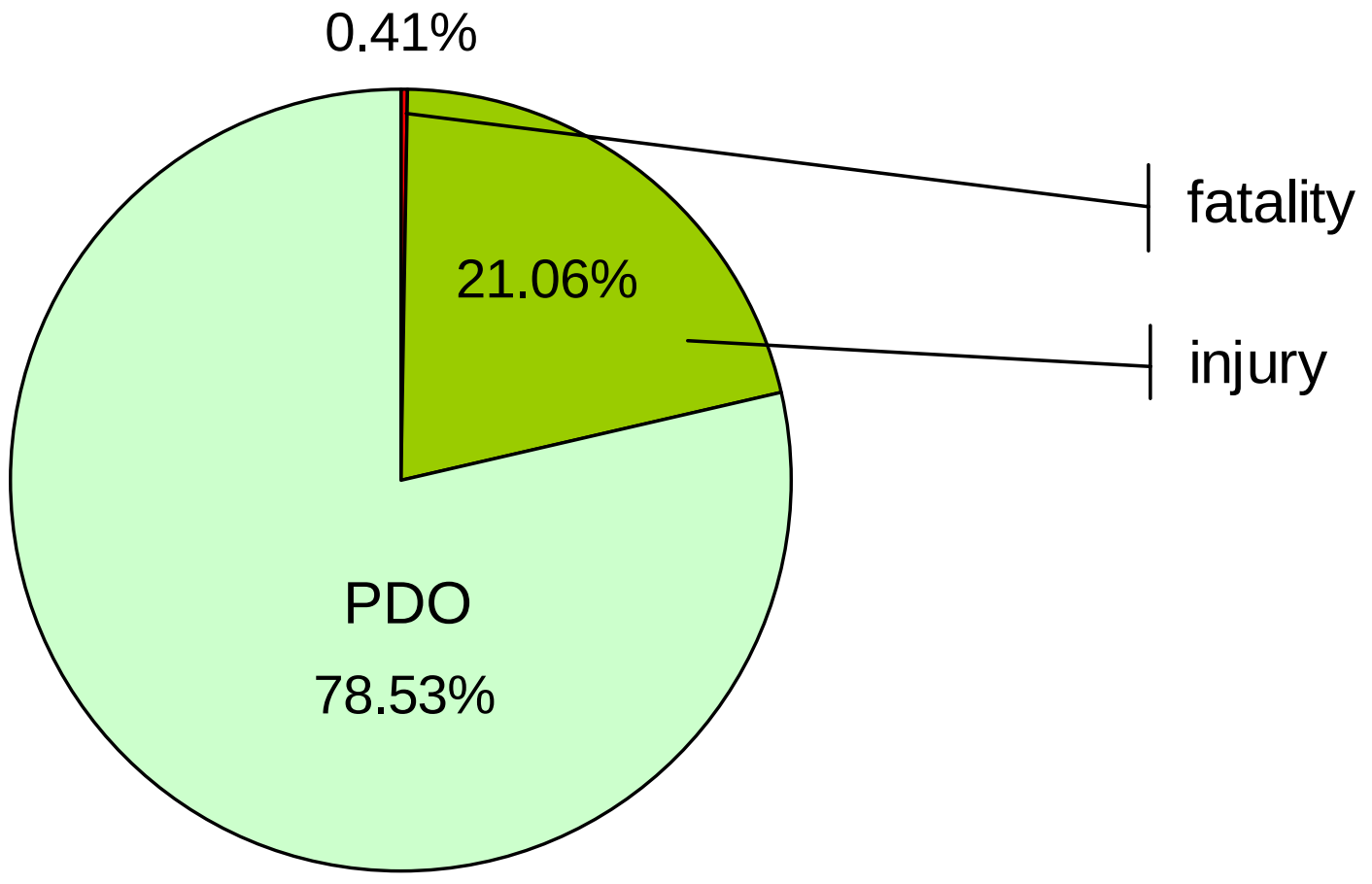

"PDO" means property damage only (no injury)

Figure 3.4 Percentage distribution of 2004 accidents by their severity level

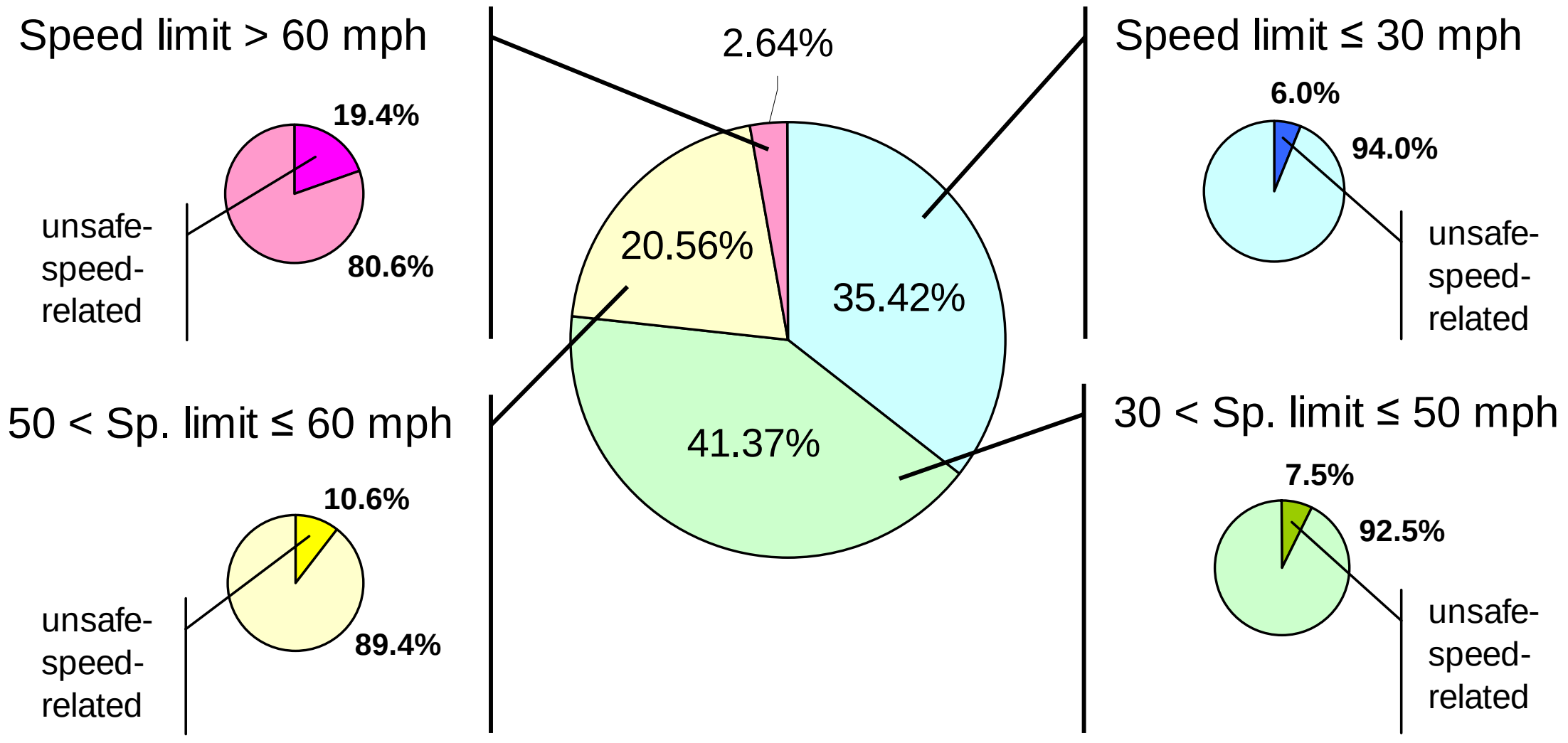

Figure 3.5 Percentage distributions of 2004 accidents by their causation in four different speed limit data bins

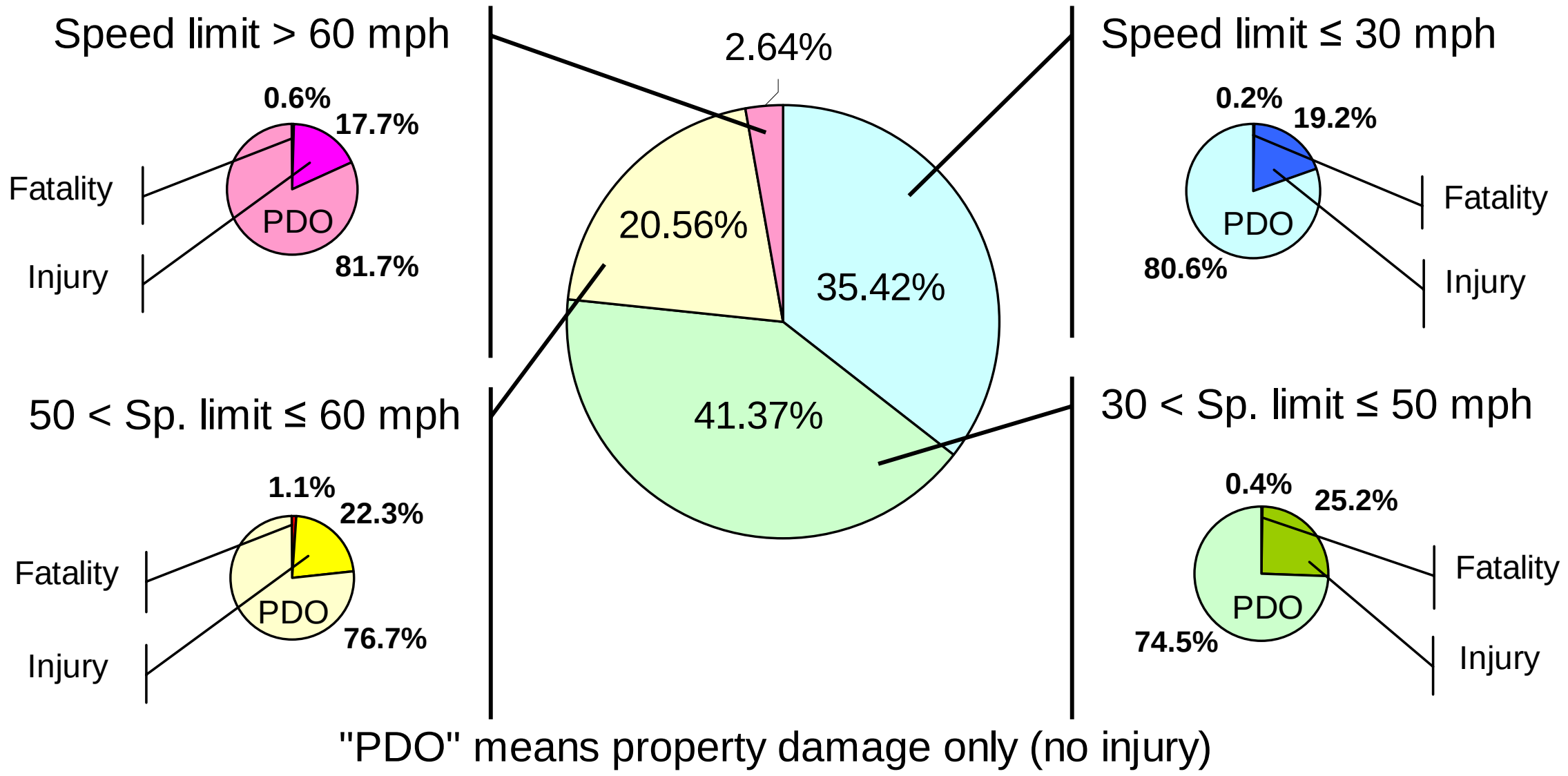

Figure 3.6 Percentage distributions of 2004 accidents by their severity level in four different speed limit data bins

We can make some interesting observations by using the plots in Figure 3.5 and Figure 3.6. First, from Figure 3.5 it seems that the probability of unsafe and/or excessive speed being the primary cause of an accident grows with speed limit. Second, from Figure 3.6 it seems that the posted speed limit does not have a clearly pronounced and easily understandable effect on the severity level of an accident. Indeed, the probabilities of fatality and injury appear to decrease for very high speed limit values ($>60$ mph). However, we must keep in mind that mathematical relations (or absence of them) inferred from simple descriptive statistics can be spurious. The main reason is that different explanatory variables can be (and usually are) mutually dependent, which greatly complicates the inference problem. Thus, it can well be the case that some other variables impact accident causation and severity, while speed limit simply happens to be correlated with these other variables. As a result, to truly understand the effect of speed limit on accident causation and severity, one has to control for all other relevant variables in making an inference about the effect of speed limit. This is done by building appropriate statistical models, and this is the main subject of our research, which is presented in the next two Chapters.

### 3.2. Accident data for year 2006

Now let us describe 2006 accident data that we use. The percentage distributions of 2006 accidents by road class and by accident type are given in Figure 3.7 and Figure 3.8 respectively. The percentage distributions of 2006 accidents by their causation and severity level are plotted in Figure 3.9 and Figure 3.10 respectively. We divide 2006 accidents into four different speed limit data bins the same way as we divided 2004 accidents. The percentage distributions by accident causation and severity level are calculated for 2006 accidents that are inside each of these four speed limit bins and are plotted in Figure 3.11 and Figure 3.12.

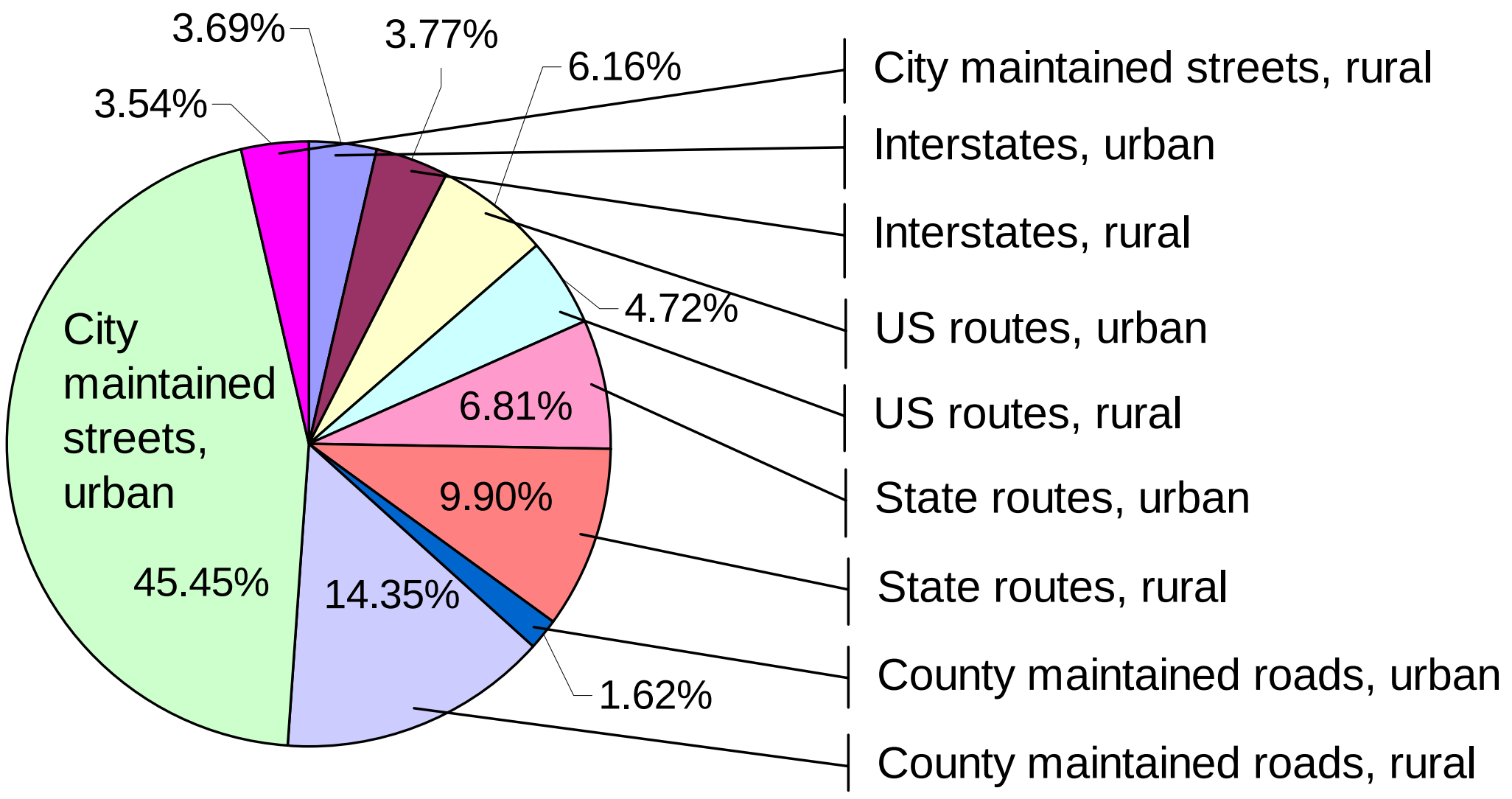

Figure 3.7 Percentage distribution of 2006 accidents by road class

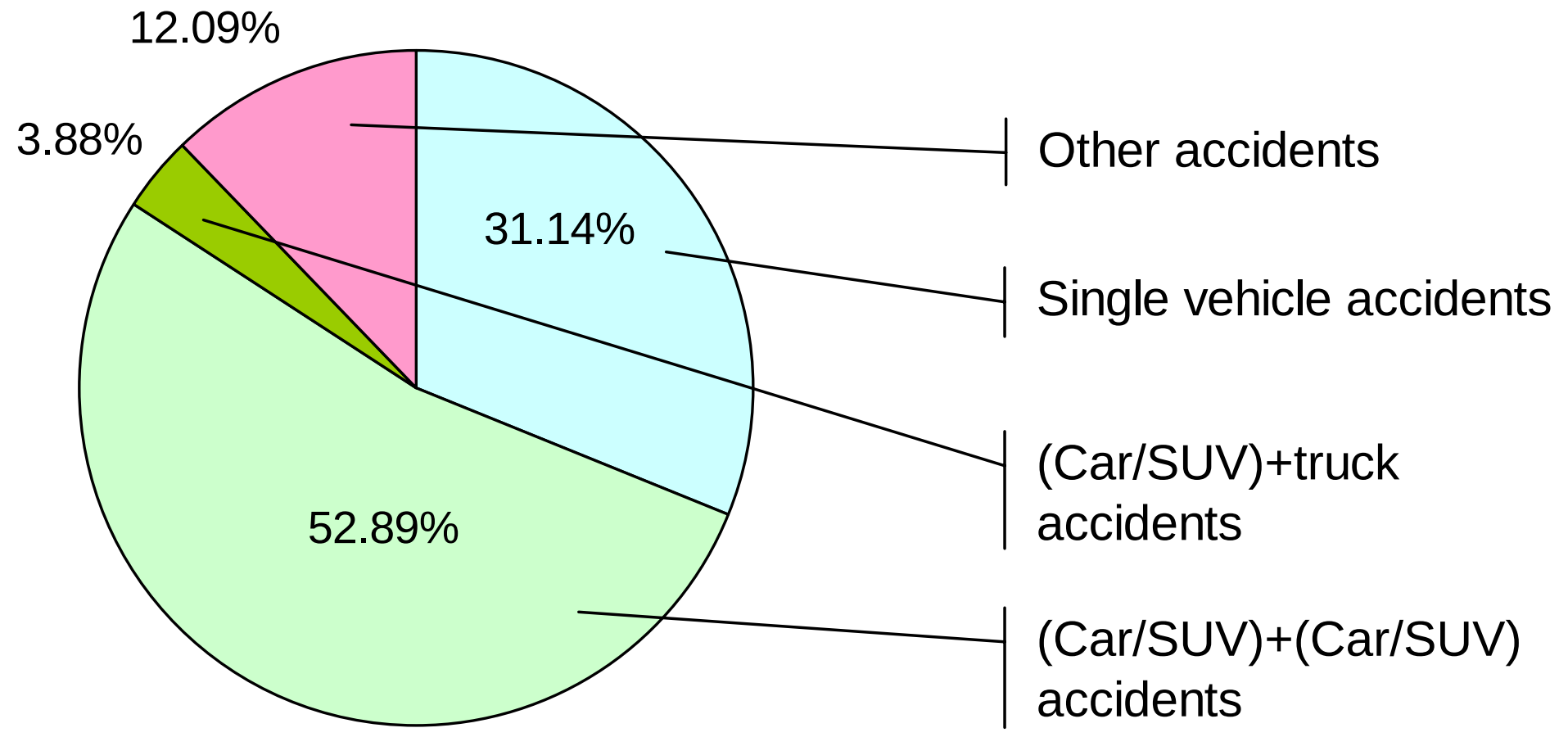

Figure 3.8 Percentage distribution of 2006 accidents by their type

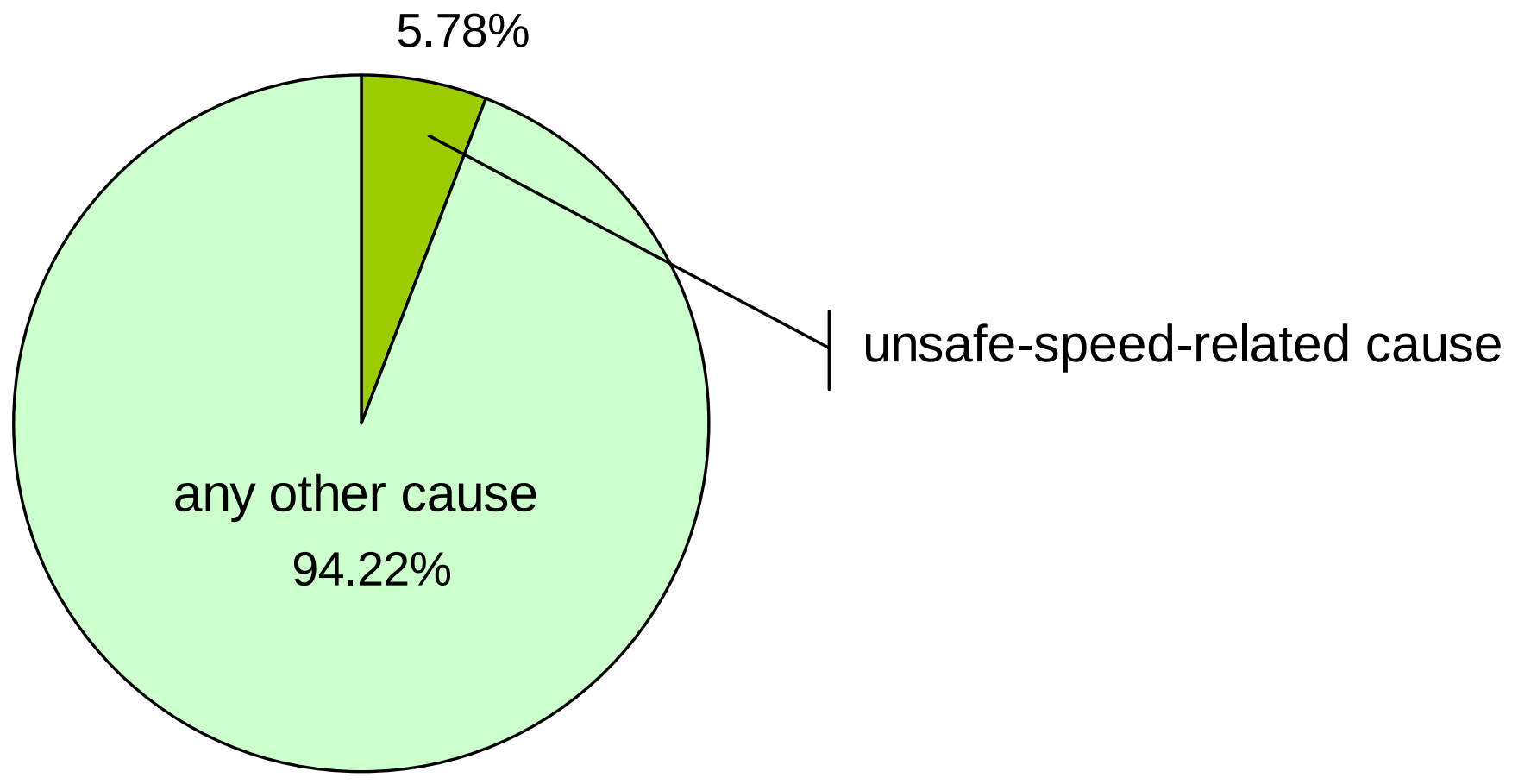

Figure 3.9 Percentage distribution of 2006 accidents by their causation

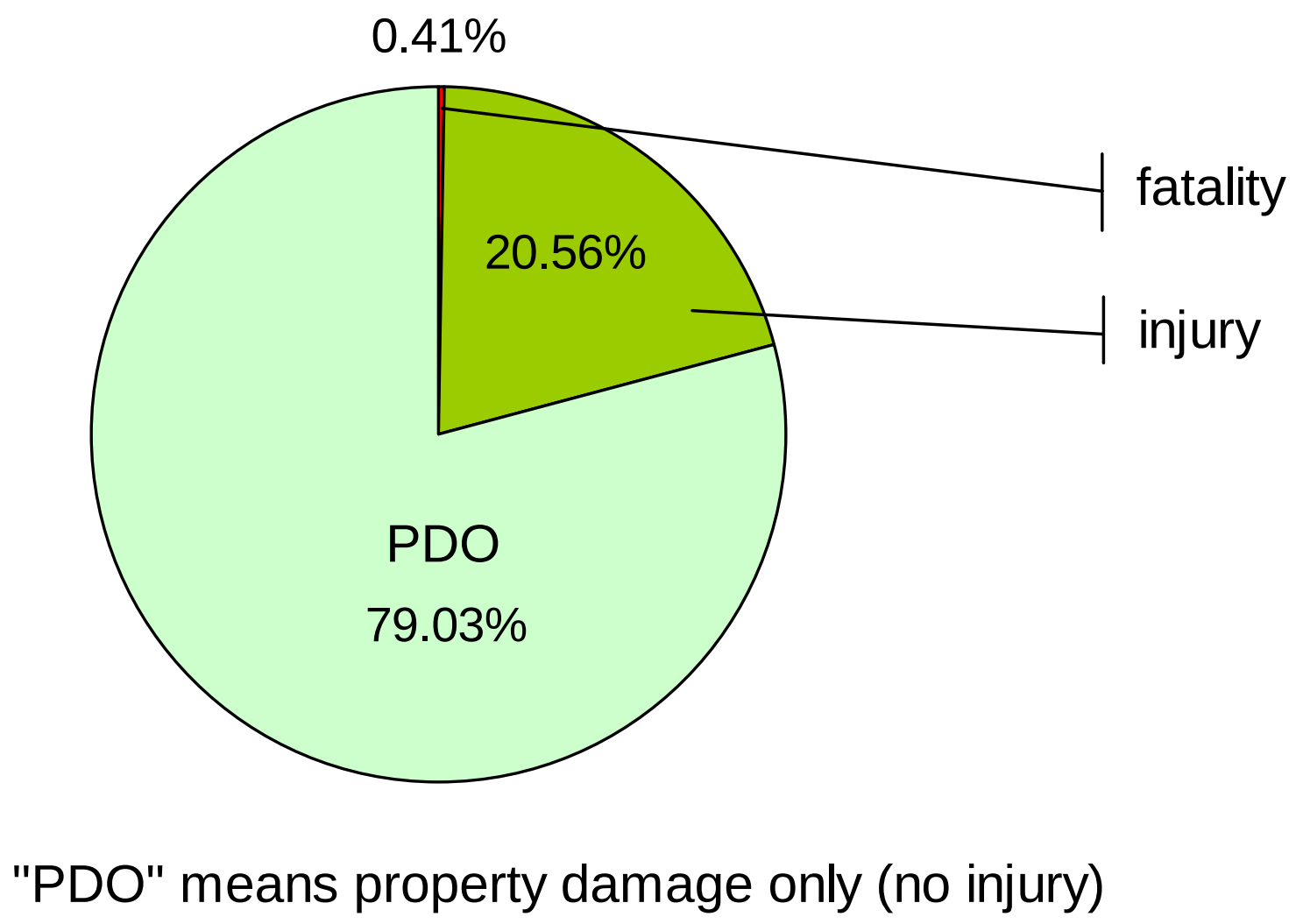

Figure 3.10 Percentage distribution of 2006 accidents by their severity level

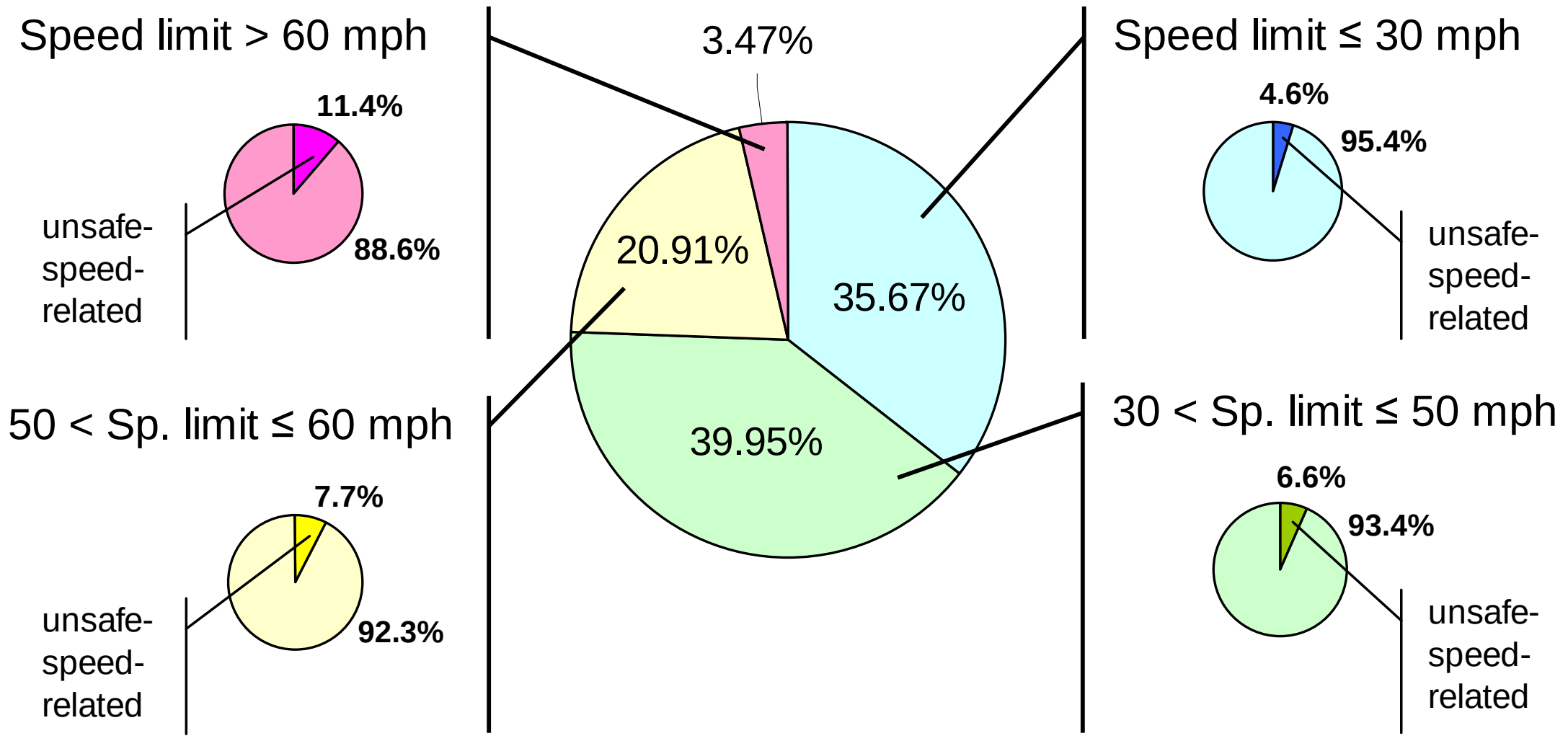

Figure 3.11 Percentage distributions of 2006 accidents by their causation in four different speed limit data bins

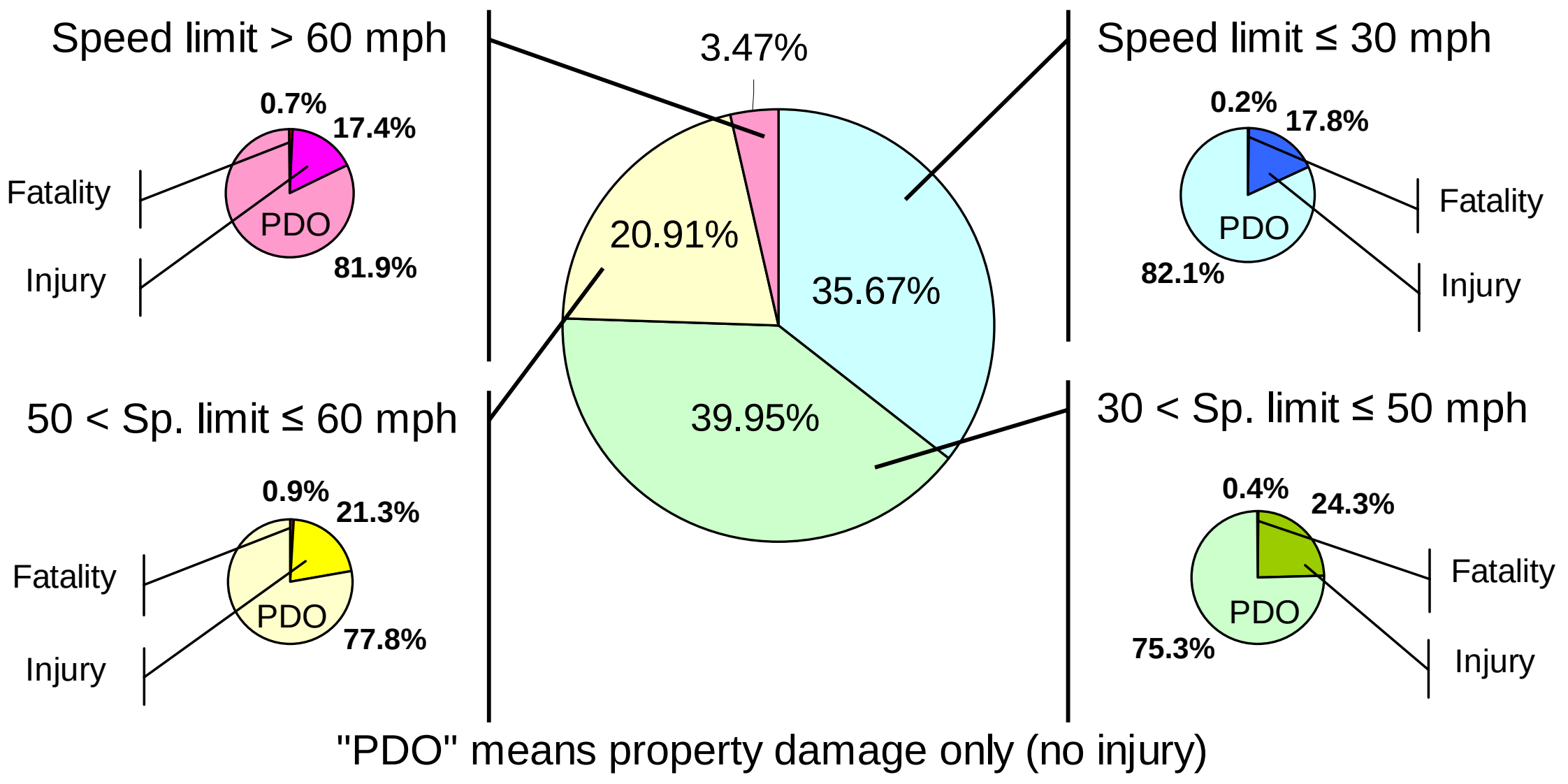

Figure 3.12 Percentage distributions of 2006 accidents by their severity level in four different speed limit data bins

Using the plots in Figure 3.11 and Figure 3.12, we make the same observations for 2006 accidents as those made for 2004 accidents. First, it again seems that the probability of unsafe and/or excessive speed being the primary cause of an accident grows with speed limit (refer to Figure 3.11). Second, from Figure 3.12 it seems that the posted speed limit does not have a clearly pronounced effect on the severity level of an accident because the probabilities of fatality and injury appear to decrease for very high speed limit values ($>60$ mph). However, we again can not make definite inference on the effect of the speed limit from these observations without building appropriate statistical models for accident causation and severity.

# CHAPTER 4. ACCIDENT CAUSATION STUDY

In this chapter we study the unsafe-speed-related causation of accidents and its dependence on the posted speed limit and other explanatory variables that characterize accidents. Below, we first explain how we use the available accident data and estimate statistical models for unsafe-speed-related causation. Then, we present the results obtained from the estimation of these models for accidents that happened in Indiana in 2004 and 2006.

## 4.1. Modeling Procedures: accident causation

There exists one primary cause of each accident, as identified by a police officer in his report on this accident[6]. All possible accident primary causes are classified into three categories:

1. Driver-related contributing circumstances (e.g. unsafe speed, speed too fast for weather conditions, driver illness, improper passing, etc.).
2. Vehicle-related contributing circumstances (e.g. tire failure or defective, brake failure or defective, etc.).
3. Environment-related contributing circumstances (e.g. animal on roadway, roadway surface condition, glare, etc.).

Here we are interested in an unsafe and/or excessive speed being the primary cause of an accident and its dependence on the posted speed limit. As a result, we introduce an indicator (dummy) variable that is equal to unity if the primary cause of an accident is either "unsafe speed" or "speed too fast for weather conditions" and is equal to zero for any other primary cause. We then estimate binary logit models with two possible outcomes that are determined by this indicator variable, refer to equation (2.3).

---

[6] For potential problems with primary cause identification see footnote 4 on page 17.

To uncover the direct influence of the posted speed limit on the accident primary cause, we need to control for other explanatory variables that might also affect accident causation. Examples of these other variables are weather conditions, accident time and date, vehicle and driver characteristics, and so on. All explanatory variables can be divided into two distinct types. First, there are indicator (dummy) variables that are equal to unity if some particular conditions are satisfied, and are equal to zero otherwise. Examples of indicator variables are driver's gender indicator, weekend indicator and precipitation indicator. Second, there are quantitative variables that take on meaningful quantitative values, such as driver's age, speed limit and number of fatalities. In addition, one can easily define derivative indicator variables that are obtained from quantitative variables. For example, one can define a "young driver" indicator as being equal to unity if the driver's age is below 25. When estimating models, we frequently define and use the most useful (as judged by the model likelihood function) new derivative indicator variables that are based on quantitative variables.

Because results of safety analysis vary significantly across different road classes and accident types (Karlaftis and Tarko, 1998; Chang and Mannering, 1999; Khan, 2002; Kweon and Kockelman, 2003; Ulfarsson and Mannering, 2004; Khorashadi et al., 2005), we divide accident data by road class and accident type as shown in Figure 4.1, and we estimate the accident causation models separately for each road-class-accident-type combination. Note that we do not consider accidents with two trucks involved and with more than two vehicles involved (there are less than 12.1% of such accidents, see Figure 3.2 and Figure 3.8). For all two-vehicle accident types other than two-truck accidents, we test whether cars and SUVs can be considered together or must be considered separately (refer to the additional division shown inside the dotted box in Figure 4.1). This test is done by using the likelihood ratio test, which is explained in CHAPTER 2. The complete list of combinations of different road classes and accident types that we consider in our causation study of 2004 and 2006 accidents can be found in Table B.1 and Table B.2 in Appendix B.

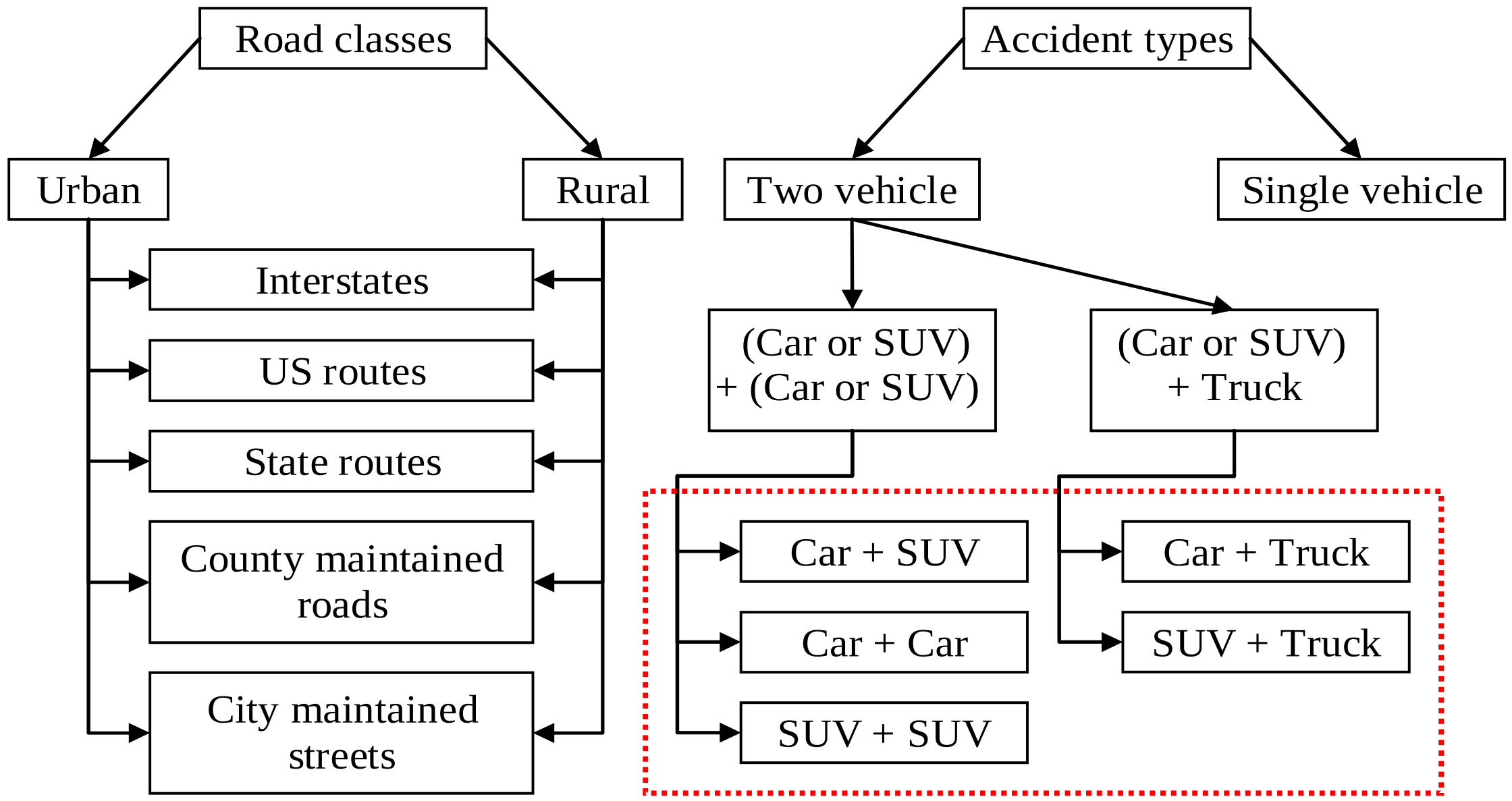

"SUV" means sport utility vehicles, pickups and vans. "Truck" means any possible kind of a truck or a tractor. SUVs and cars are considered together unless their additional division, as shown inside the dotted box, is required by the likelihood ratio test.

Figure 4.1 Data division by road class and by accident type[7]

We check statistical significance of the explanatory variables in all logit models by using 5% significance level for the two-tailed t-test of a large data sample. In other words, coefficients with t-ratios between -1.96 and +1.96 are considered to be statistically insignificant. Note that the explanatory variables can be mutually dependent (e.g. a quantitative variable and its derivative indicator variable are strongly mutually dependent).

[7] We consider US routes and State routes separately even though they have similar design and other properties. The reason is that our final logit models for unsafe-speed-related accident causation on US and State routes turn out to be statistically different from each other. We use the likelihood ratio test to check this difference (but we do not report the test results in this thesis).

Statistical models are (usually) estimated by maximizing the model's log-likelihood function. However, one can not rely on the log-likelihood maximization alone in order to choose the optimal number of explanatory variables to be included into a statistical model. The reason is that the log-likelihood (LL) function is always maximized when all available explanatory variables are included into the model. This is because a removal of any explanatory variable is equivalent to restricting its value to zero, which always either decreases the maximum of LL or leaves it the same. As a result, in the present study we use the Akaike Information Criterion (AIC), minimization of which ensures an optimal choice of explanatory variables in a model (Tsay, 2002, page 37; Washington et al., 2003, page 212; Wikipedia). The main idea behind the AIC is to examine the complexity of a model together with goodness of its fit to the data sample, and to find a balance between the two. A model with too few explanatory variables will provide a poor fit to the data sample. A model with too many variables will provide a very good fit, but will lack necessary robustness and will perform poorly in out-of-the-sample data. The preferred model with the optimal number of explanatory variables is the model with the lowest AIC value, which is given by equation

$$AIC = -2LL + 2K \text{ ,} \qquad \text{Eq. 4.1}$$

where LL is the log-likelihood value of a model, and K is the number of estimable coefficients in the model (one coefficient for each explanatory variable, including the intercepts).

In our research we estimate all logit models by using one of the two procedures A and B shown in Figure 4.2. Procedure A is as follows:

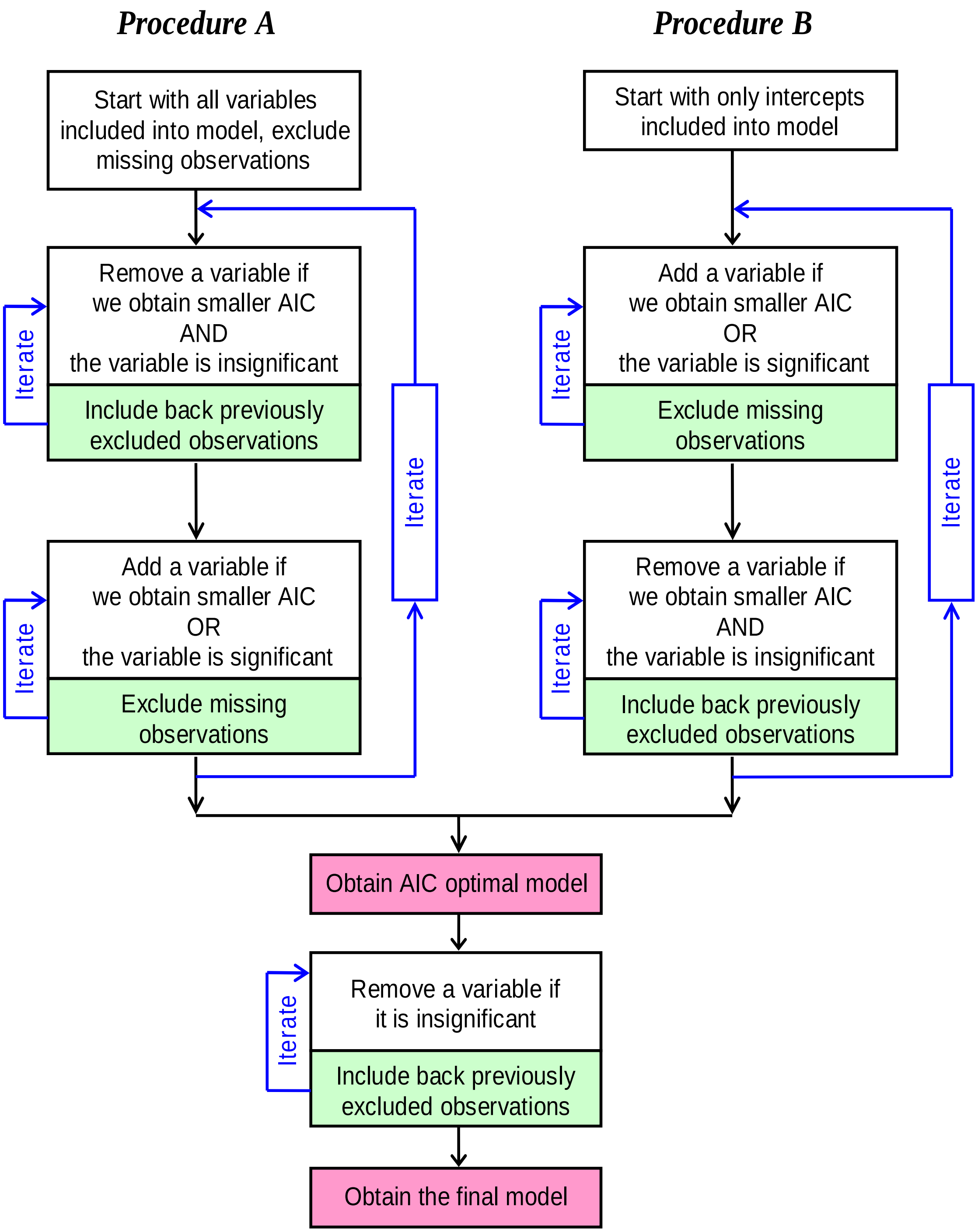

Figure 4.2 Model estimation procedures

A. We start with all explanatory variables initially included into a logit model. Note that, when estimating a model, we have to exclude observations that are missing for any of the included variables. Next, we obtain the final model by using three steps of model estimation. The first step is

1. We remove the least statistically significant explanatory variables (as judged by their t-ratios) one by one if *both* of the following two conditions are satisfied: the removal of a variable decreases the AIC value *and* the removed variable is statistically insignificant (under the 5% confidence level)[8]. Note that while using the Akaike information criterion, we always keep the number of data sample observations constant in order to calculate the changes of the AIC value correctly. Each time when we have removed several (usually four) least significant explanatory variables from a model, we include some of the previously excluded observations back into the data sample because now the model includes fewer variables with missing observations. We keep removing insignificant explanatory variables one by one, periodically including previously excluded observations back into the data sample, until we can not remove any additional variable under the two conditions listed above.

After we removed all variables that we could, we need to check if any of the removed variables can be added back into the model. This is because variables are mutually dependent and "interact" in the model. Therefore, we proceed to the second step of model estimation:

[8] If the asymptotic normality of maximum likelihood estimates holds, then the AIC value does not change with removal (addition) of a variable whose coefficient has 15.73% p-value for the two-tailed test (15.73% p-value corresponds to $\pm\sqrt{2}$ t-ratio for a normal variate). In this case the 5% confidence level test of the variable is redundant, and the AIC test alone can be used for removal and addition of variables in model estimation steps 1 and 2. Nevertheless, we use both tests to make our estimation procedures more robust in case the normality of maximum likelihood estimates does not hold.

2. We add explanatory variables one by one if *at least one* of the following two conditions is satisfied: *either* the addition of a variable decreases the AIC value *or* the added variable is significant[9]. As usual, the AIC values are compared under the condition that the number of observations is kept constant. As the number of the explanatory variables included into the model grows, the data sample size shrinks because of a larger number of missing observations associated with the included variables. We add explanatory variables one by one until no any additional variable can be added to the model.

Next we return back to the first estimation step given above and remove variables that can be removed. We iterate between steps 1 and 2 until we can neither remove nor add any more variables. At this point we arrive at the model that we call the "AIC optimal model" (refer to Figure 4.2). Next, we proceed to the third and final step of model estimation:

3. To make our final results more robust, we drop from the AIC optimal model all remaining statistically insignificant variables (judged by the 5% significance level for the two-tailed t-test). As a result, we obtain the final model, which is our best model (according to the estimation procedures chosen by us).

Now we describe procedure B:

B. In this procedure we start with only intercepts (constant terms) initially included into a logit model (refer to Figure 4.2). Next, we proceed in a way very similar to that used in procedure A. We run step 2 of model estimation and add explanatory variables into the model. Then, we iterate between steps 1 and 2 until we can neither remove nor add any more variables, at which point we arrive at the AIC optimal model. Finally, we run step 3 of model estimation and obtain the best final model.

---

[9] We first search for and add AIC decreasing variables, and afterwards we add significant variables if there are any.

By default we always use procedure A for model estimation, and only if we can not use it (usually when the available data sample is too small for the initial model estimation with all explanatory variables included), then we resort to procedure B.

## 4.2. Results: accident causation models

For each of the road-class-accident-type combinations listed in Table B.1 and Table B.2 in Appendix B, we find and estimate the best binary logit model by using either procedure A or procedure B described above. The binary logit models are given in Equation (2.3), where outcome “1” corresponds to the case when the primary cause of an accident is either “unsafe speed” or “speed too fast for weather conditions”, and outcome “2” corresponds to any other primary cause of the accident. The results of the estimation of the best models are given in Table B.3 and Table B.4 for 2004 and 2006 accidents respectively (see Appendix B).

In Table B.5 and Table B.6 in Appendix B we give the results of testing whether, in 2004 and 2006 two-vehicle accidents, cars and SUVs can be considered together or must be considered separately. This testing is done for the best models by using the likelihood ratio test given in Equation (2.5). According to the results shown in Table B.5 and Table B.6, we find that in our unsafe-speed-related accident causation study cars and SUVs can be considered together in all 2004 two-vehicle accidents on all road classes, but they must be considered separately in the case of several road-class-accident-type combinations for 2006 two-vehicle accidents.

Let us now examine the model estimation results, which are given in Table B.3 and Table B.4 for 2004 and 2006 accidents respectively. We will consider the effects of the posted speed limit and other explanatory variables on the probability of an unsafe and/or excessive speed being the primary cause of an accident. Since our primary interest is the effect of the speed limit, we focus on it first.

### 4.2.1. Effect of Speed Limit

We assume that the speed limit posted at the location of an accident is known only if it is indicated as known and the same for all vehicles involved into the accident. The speed limit is variable $X_{29}$ in Appendix A. Its coefficients and averaged elasticities in the best final binary logit models for 2004 and 2006 unsafe-speed-related accident causation are given in Table 4.1 and Table 4.2 below. In order to understand the results reported in these tables, please refer to Equations (2.3) and (2.8). These equations give the binary logit model and the corresponding elasticities that we calculate. The outcomes “1” and “2” in the binary models correspond to the “unsafe-speed-related cause” and “any other cause” of an accident. In Table 4.1 and Table 4.2 we report all *statistically significant* coefficients of the speed limit variable (these coefficients are copied from Table B.3 on page 67 and Table B.4 on page 78) and the corresponding elasticities. In addition, in these tables we report all statistically *insignificant* coefficients of the speed limit variable (without elasticities). These insignificant coefficients are shown in the square brackets and are obtained by test-adding the speed limit variable into the AIC optimal logit models (note that this is done only as a test; in Table 4.1, Table 4.2, Table B.3 and Table B.4 all significant coefficients and the corresponding elasticities are reported for the final models, which themselves do not contain any insignificant variables).

We find the following results for the effects of speed limit on accident causation:

Table 4.1 2004 accident causation models: results for speed limit[10]

| # | Model name | | | Speed limit coefficient (t-ratio) | Averaged elasticities of speed limit (SL) | |
|---|---|---|---|---|---|---|
| | | | | | $\overline{E}^{(1)}_{1;SL}$ | $\overline{E}^{(2)}_{1;SL}$ |
| 1 | County road | rural | (car/SUV)+(car/SUV) | [.00943 (1.67)] | | |
| 2 | County road | rural | (car/SUV)+(truck) | [-.00555 (-.223)] | | |
| 3 | County road | rural | one vehicle | .00859 (2.83) | .337 | -.061 |
| 4 | County road | urban | (car/SUV)+(car/SUV) | .0368 (2.24) | 1.23 | -.094 |
| 5 | County road | urban | (car/SUV)+(truck) | [.177 (1.43)] | | |
| 6 | County road | urban | one vehicle | [.00986 (.827)] | | |
| 7 | Interstate | rural | (car/SUV)+(car/SUV) | [.0172 (1.27)] | | |
| 8 | Interstate | rural | (car/SUV)+(truck) | [-.0127 (-.317)] | | |
| 9 | Interstate | rural | one vehicle | [-.0294 (-1.95)] | | |
| 10 | Interstate | urban | (car/SUV)+(car/SUV) | [.0182 (1.19)] | | |
| 11 | Interstate | urban | (car/SUV)+(truck) | [.0539 (1.42)] | | |
| 12 | Interstate | urban | one vehicle | [-.00985 (-.873)] | | |
| 13 | State route | rural | (car/SUV)+(car/SUV) | [-.00230 (-.175)] | | |
| 14 | State route | rural | (car/SUV)+(truck) | [.00806 (.254)] | | |
| 15 | State route | rural | one vehicle | [-.0108 (-1.30)] | | |
| 16 | State route | urban | (car/SUV)+(car/SUV) | .0212 (2.51) | .769 | -.034 |
| 17 | State route | urban | (car/SUV)+(truck) | [.0337 (1.05)] | | |
| 18 | State route | urban | one vehicle | [.000799 (.066)] | | |
| 19 | City street | rural | (car/SUV)+(car/SUV) | .0225 (2.50) | .773 | -.050 |
| 20 | City street | rural | (car/SUV)+(truck) | [.102 (1.25)] | | |
| 21 | City street | rural | one vehicle | [-.00158 (-.248)] | | |
| 22 | City street | urban | (car/SUV)+(car/SUV) | [-.00504 (-1.14)] | | |
| 23 | City street | urban | (car/SUV)+(truck) | [.0113 (.543)] | | |
| 24 | City street | urban | one vehicle | -.0117 (-3.82) | -.323 | .053 |
| 25 | US route | rural | (car/SUV)+(car/SUV) | [.00721 (.509)] | | |
| 26 | US route | rural | (car/SUV)+(truck) | [-.0356 (-1.23)] | | |
| 27 | US route | rural | one vehicle | -.0422 (-3.17) | -2.05 | .204 |
| 28 | US route | urban | (car/SUV)+(car/SUV) | .0181(2.33) | .679 | -.040 |
| 29 | US route | urban | (car/SUV)+(truck) | [.0517 (1.50)] | | |
| 30 | US route | urban | one vehicle | [.00636 (.394)] | | |

[10] Refer to Equations (2.3) and (2.8), where outcomes "1" and "2" correspond to the "unsafe-speed-related" and "any other" accident causes. We report statistically significant coefficients of the speed limit variable and the corresponding elasticities. In addition, in the square brackets we report statistically insignificant coefficients (obtained by test-adding the speed limit variable into the AIC optimal models). All coefficients are the components of vector $\boldsymbol{\beta}_1$ that are multiplied by the speed limit variable in Equation (2.3).

Table 4.2 2006 accident causation models: results for speed limit

| # | Model name | | | Speed limit coefficient (t-ratio) | Averaged elasticities of speed limit (SL) | |
|---|---|---|---|---|---|---|
| | | | | | $\overline{E}^{(1)}_{1;SL}$ | $\overline{E}^{(2)}_{1;SL}$ |
| 1 | County road | rural | (car/SUV)+(car/SUV) | [-.0222 (-1.57)] | | |
| 2a | | | (car)+(truck) | [-.115 (-1.36)] | | |
| 2b | | | (SUV)+(truck) | [.246 (.917)] | | |
| 3 | | | one vehicle | [.000990 (.155)] | | |
| 4 | | urban | (car/SUV)+(car/SUV) | [-.0101 (-.309)] | | |
| 5 | | | (car/SUV)+(truck) | [0.00792 (.113)] | | |
| 6 | | | one vehicle | [-.00975 (-.755)] | | |
| 7a | Interstate | rural | (car)+(car) | [-.00840 (-.346)] | | |
| 7b | | | (car)+(SUV) | [.0110 (.569)] | | |
| 7c | | | (SUV)+(SUV) | [.0296 (.958)] | | |
| 8 | | | (car/SUV)+(truck) | [-.0176 (-.816)] | | |
| 9 | | | one vehicle | -.0439 (-5.72) | -2.51 | .370 |
| 10 | | urban | (car/SUV)+(car/SUV) | [.0232 (1.85)] | | |
| 11a | | | (car)+(truck) | [-.0479 (-1.49)] | | |
| 11b | | | (SUV)+(truck) | [.113 (1.27)] | | |
| 12 | | | one vehicle | [-.0158 (-.811)] | | |
| 13 | State route | rural | (car/SUV)+(car/SUV) | [.000526 (.0230)] | | |
| 14 | | | (car/SUV)+(truck) | [.00874 (.284)] | | |
| 15 | | | one vehicle | -.0373 (-5.34) | -1.87 | .109 |
| 16 | | urban | (car/SUV)+(car/SUV) | .0277 (3.47) | .999 | -.0390 |
| 17 | | | (car/SUV)+(truck) | [.0250 (.907)] | | |
| 18 | | | one vehicle | [-.0102 (-1.000)] | | |
| 19 | City street | rural | (car/SUV)+(car/SUV) | [-.00521 (-.241)] | | |
| 20a | | | (car)+(truck) | [.0239 (.381)] | | |
| 20b | | | (SUV)+(truck) | [-.142 (-.634)] | | |
| 21 | | | one vehicle | [.00475 (.612)] | | |
| 22 | | urban | (car/SUV)+(car/SUV) | [.00811 (1.010)] | | |
| 23a | | | (car)+(truck) | [.0325 (.889)] | | |
| 23b | | | (SUV)+(truck) | [.118 (1.910)] | | |
| 24 | | | one vehicle | [.00226 (.301)] | | |
| 25 | US route | rural | (car/SUV)+(car/SUV) | [.00969 (.321)] | | |
| 26 | | | (car/SUV)+(truck) | [.00144 (.042)] | | |
| 27 | | | one vehicle | [-.00537 (-.464)] | | |
| 28 | | urban | (car/SUV)+(car/SUV) | [.0122 (.646)] | | |
| 29a | | | (car)+(truck) | [-.0127 (-.294)] | | |
| 29b | | | (car)+(truck) | [.362 (1.73)] | | |
| 30 | | | one vehicle | [-.0101 (-.814)] | | |

- Speed limit does not have any statistically significant effect on unsafe-speed-related causation of accidents of any types[11] on interstate highways (urban and rural), except for the case of 2006 one-vehicle accidents on rural interstates. In this single case the probability of unsafe speed being the primary cause of an accident actually decreases with an increase in the posted speed limit.
- Speed limit does not also have a statistically significant effect on unsafe-speed-related accident causation for the majority of other accident types on the majority of road classes other than the interstate highways.
- There are only ten combinations of different accident types and road classes for which speed limit turns out to have a statistically significant effect on unsafe-speed-related causation of 2004 and 2006 accidents. For convenience, in
- Figure 4.3 and Figure 4.4 we present in graphical form the t-ratios of the speed limit coefficients for these ten combinations. We see that there are mixed effects of the speed limit on unsafe-speed-related accident causation. On one hand, the probability of unsafe speed being the primary cause of an accident rises with an increase in the posted speed limit for 2004 single-vehicle accidents on rural county maintained roads, for 2004 car/SUV-car/SUV accidents on rural city maintained streets, urban US routes and urban county maintained roads, as well as for 2004 & 2006 car/SUV-car/SUV accidents on urban state routes. On the other hand, the probability decreases with an increase in the speed limit for 2004 single-vehicle accidents on rural US routes and urban city maintained streets, and for 2006 single-vehicle accidents on rural state routes and rural interstates.
- The speed limit variable is elastic for only one of the six road-class-accident-type combinations that display a statistically significant increase of the unsafe speed accident causation probability with an increase of the posted speed limit (this single combination is 2004 car/SUV-car/SUV accidents on urban county maintained roads).

[11] Note that we consider only single-vehicle accidents and all two-vehicle accidents except those that involve two trucks.

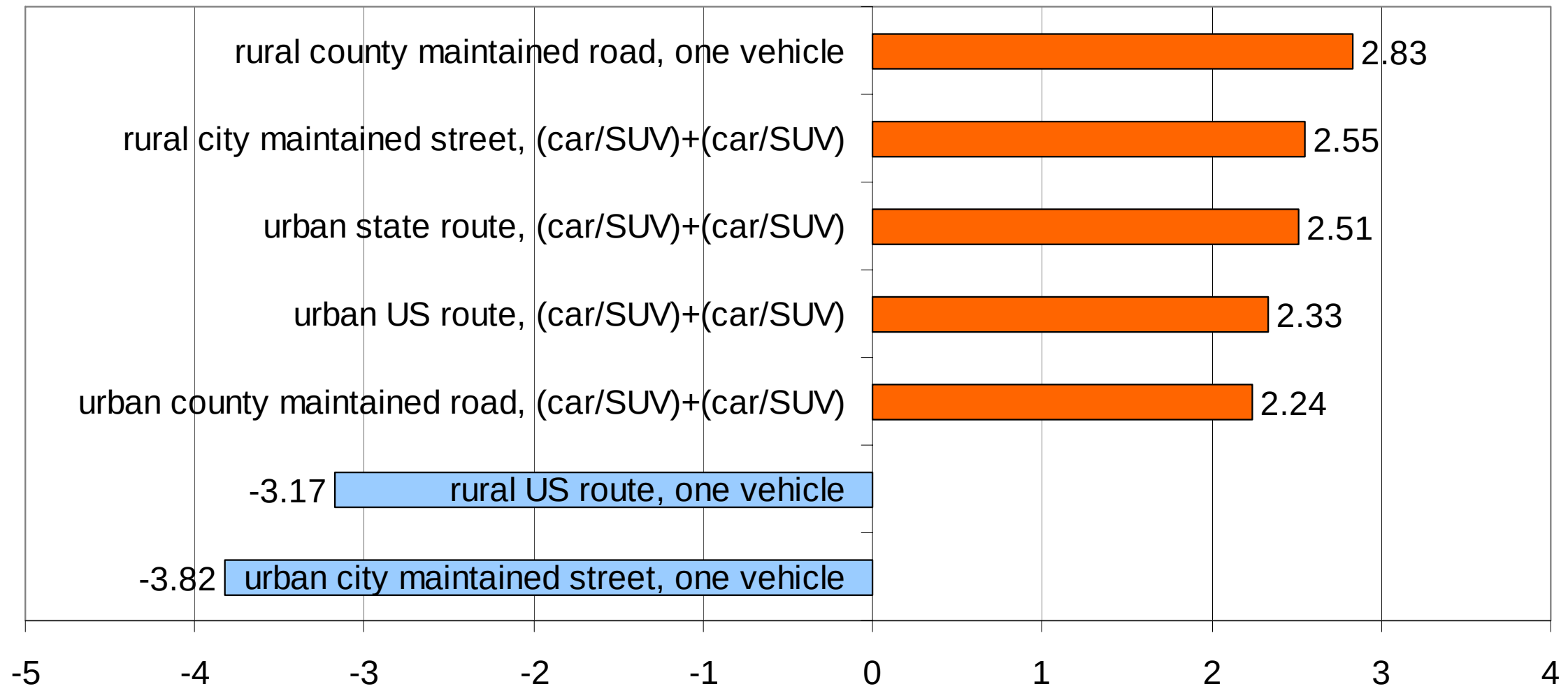

Figure 4.3 T-ratios of statistically significant speed limit coefficients in 2004 accident causation models

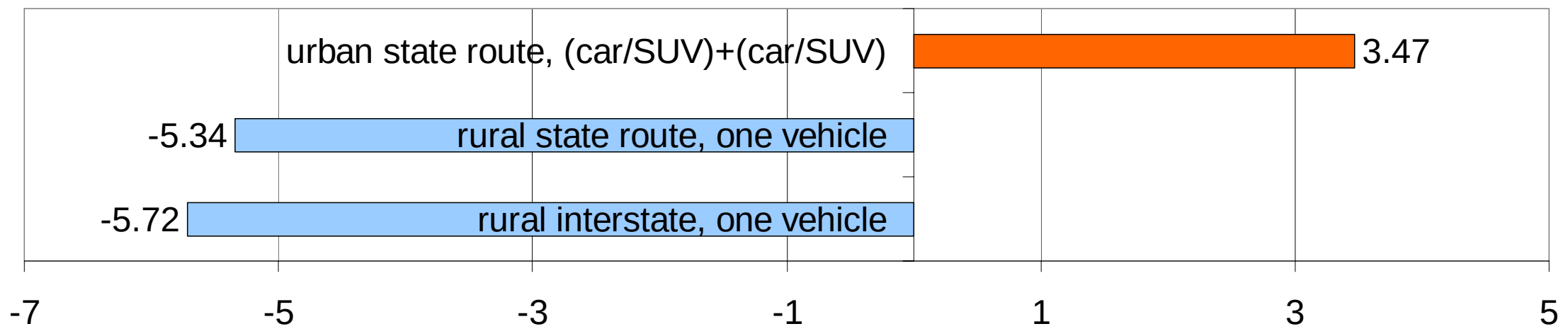

Figure 4.4 T-ratios of statistically significant speed limit coefficients in 2006 accident causation models

We postpone a discussion of the above findings until the last chapter, which discusses our results for both accident causation and accident severity.

### 4.2.2. Effect of Other Explanatory Variables

Now we use model estimation results given in Table B.3 and Table B.4 in Appendix B to study the influence of explanatory variables other than the posted speed limit on unsafe-speed-related causation of 2004 and 2006 accidents. We limit our consideration to several of the most important variables, which are statistically significant not just in a few but in many models for different road classes and accident types[12].

- **Variable “wint” (see pages 68 and 79):** We find that the probability of unsafe-speed-related cause of an accident increases during winter season. This seasonal effect is very strong. It is reasonable because driving conditions in Indiana worsen during winters, while some drivers apparently fail to adjust their driving speeds accordingly.
- **Variables "precip", “snow” and "dry" (see pages 70 and 81):** We find that the probability of unsafe-speed-related cause of an accident increases when precipitation and/or snow is observed at the accident location, and it decreases when the road surface is dry. This is a very strong effect on all road classes, and it can clearly be explained by drivers not being adjusting their speed appropriately when the weather conditions are less safe for driving.
- **Variable "nojun" (see pages 71 and 82):** The absence or presence of a road junction at the accident location has a mixed effect on the probability of unsafe-speed-related cause of an accident. The probability of an unsafe-speed-related accident with only one vehicle involved decreases when no junction is present on the road, but the probability increases when two vehicles are involved. This can be explained by two concurrent effects. On one hand, the road is safer in the absence of junctions. On the other hand, drivers might have a habit of slowing down a bit when they approach a junction and see another vehicle(s) on or near the junction[13].

---

[12] One of the reasons for this limitation is that the number of available accident observations in our data sample is relatively small for several road-class-accident-type combinations, refer to Table B.1 and Table B.2.

- **Variables "curve" (see pages 72 and 83):** We find an expected result that the probability of unsafe-speed-related cause of an accident decreases when the road is straight and increases when the road is at curve. Clearly curved roads are less safe because of centrifugal forces acting onto moving vehicles in the rotating coordinate frame of reference.
- **Variable "heavy" (see pages 73 and 85):** We find that heavy trucks and tractors are less likely to cause unsafe-speed-related accidents than other vehicles are. The reason can be that drivers of trucks and tractors are more professional and are less likely to speed.
- **Variables "stopsig" (see pages 76 and 87):** We find that the presence of a stop sign generally reduces the probability of unsafe-speed-related accidents, which can be due to effectiveness of stop signs in controlling traffic flows on streets and minor roads.
- **Variable "$X_{34}$" (see pages 77 and 88):** We find that the probability of unsafe-speed-related cause of an accident decreases with the age of the driver at fault. This is a very strong effect on all road classes. Apparently older drivers are more experienced, more careful and much less likely to exceed safe driving speed.

[13] Note that police officer's misidentification errors can be stronger for a single-vehicle accident because in this case the officer has to rely on testimonies of occupants of the single vehicle involved into the accident. See footnote 4 on page 17 for discussion of misidentification errors.

# CHAPTER 5. ACCIDENT SEVERITY STUDY

In this chapter our goal is to reveal and study the impact of the posted speed limit on the severity level of an accident. Similar to the previous chapter, we first describe the procedures of accident severity statistical modeling that we use, and second, we present and discuss the results that we obtain for 2004 and 2006 accidents.

## 5.1. Modeling Procedures: accident severity

For each accident, the severity level is determined by the injury level sustained by the most injured individual (if any) involved into the accident. By using the available individual accident data, we are able to distinguish between three levels of accident severity. Listed in increasing order, these are

1. no-injury or property damage only (PDO),
2. injury,
3. fatality,

refer to Figure 3.4 and Figure 3.10. As a result, for the statistical modeling of accident severity we use a multinomial logit model with three possible outcomes that correspond to these three levels of accident severity. This multinomial logit model is given by Equation (2.4), where the outcomes “1”, “2” and “3” correspond to “fatality”, “injury” and “PDO” levels of accident severity respectively.

We estimate multinomial logit models for accident severity in a way similar to the estimation of binary logit models for accident causation considered in the previous chapter. Namely, we again consider different road classes and accident types separately, as shown in Figure 4.1. The list of all combinations of different road classes and accident types that we consider in our accident severity study can be found in Table C.1 and Table C.2 for 2004 and 2006 accidents respectively (see Appendix C). We again use 5% significance level for the two-tailed t-test of a large data sample in order to make inference on statistical significance of all indicator and quantitative explanatory variables in the accident severity logit models. We also use the same AIC-based procedures (A and B) for all severity model estimations, as described in CHAPTER 4 on pages 32 and 33 and in Figure 4.2.

Thus, to study the impact of speed limit on the resulting accident severity, for each road-class-accident-type combination we proceed as follows:

1. First, using the data on accidents that constitute the considered road-class-accident-type combination and the procedures described above, we find the best multinomial logit model with three possible accident severity outcomes (fatality, injury, PDO). From this model we can immediately see whether there is any statistically significant effect of the posted speed limit on the resulting accident severity level.
2. Second, we divide the accident data into separate speed limit data bins, according to the posted speed limit at the place of an accident[14]. The speed limit bins chosen by us for 2004 and 2006 accidents are given in Table C.3 and Table C.4 in Appendix C.
3. Third, we take the best logit model obtained in the first step[15] and re-estimate it separately for each of the speed-limit data bins chosen in the second step. Then we test to see if there are statistically significant differences among the models estimated for the different speed-limit bins. This is done by using the likelihood ratio test, which is given by Equation (2.5) and explained in the end of CHAPTER 2. We use 5% confidence level for the likelihood ratio test statistic in Equation (2.5) to make inference on whether the collection of models estimated separately for the speed-limit data bins is statistically the same as the model estimated for the whole data sample which includes all speed limits together. In other words, if the left-hand-side of Equation (2.5) is between zero and the 95$^{th}$ percentile of the chi-squared distribution given on the right-hand-side, then we conclude that the posted speed limit makes no statistically significant difference for the structure of accident severity models in the case of the considered road-class-accident-type combination. We conclude that there is a difference otherwise.

---

[14] We disregard accidents with more than one posted speed limit at the accident location.

[15] In the third step we first remove the speed limit variable (if any) from the best final model because the speed limit is usually constant inside the speed limit data bins. In some cases of the 2004 accident severity models we have to remove additional explanatory variables that are constant inside some of the speed limit bins. For specific cases of the removal see Table 5.3.

## 5.2. Results: accident severity models

For each of the road-class-accident-type combinations listed in Table C.1 and Table C.2 in Appendix C, we find and estimate the best multinomial logit model, as given in Equation (2.3) with the outcomes “1”, “2” and “3” corresponding to “fatality”, “injury” and “PDO” accident severity levels respectively. Table C.5 and Table C.6 in Appendix C give the estimation results for the best models in the cases of 2004 and 2006 accidents.

In Table C.7 and Table C.8 we show the results of testing whether in two-vehicle accidents cars and SUVs can be considered together or must be considered separately in our accident severity study. This testing is done by using the likelihood ratio test in exactly the same way as done in the accident causation study. If the likelihood ratio test indicates that cars and SUVs should be considered separately, then we apply the additional division shown inside the dotted box in Figure 3.2 and find the best model separately for each of the sub-categories obtained by this additional division. For example, from the test results given in Table C.7 we see that the 2004 "car/SUV-truck accidents on rural interstates" category has to be divided into sub-categories 8a (car-truck accidents) and 8b (SUV-truck accidents).

### 5.2.1. Effect of Speed Limit

To judge whether speed limit makes any statistically significant difference for the resulting accident severity outcomes, we first study the severity model estimation results for the speed limit variable and its elasticities reported in Table 5.1 and Table 5.2 for 2004 and 2006 accidents respectively. In order to understand the results presented in these tables, refer to Equations (2.4) and (2.9), where outcomes "1", "2" and "3" correspond to "fatality", "injury" and "PDO" accident severity outcomes respectively. In Table 5.1 and Table 5.2 the elasticities are reported only for statistically significant coefficients of the speed limit variable (which are copied from Table C.5 on page 96 and Table C.6 on page 126). In Table 5.1 and Table 5.2 we also report all statistically insignificant coefficients of the speed limit variable, which are shown in the square brackets and are obtained by test-adding the speed limit variable into the AIC optimal models (note that this is done only as a test; in Table 5.1, Table 5.2, Table C.5 and Table C.6 all significant coefficients and the corresponding elasticities are reported for the final models, which themselves do not contain any insignificant variables).

We find that

- Speed limit does not have any statistically significant effect on severity of 2004 and 2006 accidents of any type[16] on interstate highways (both urban and rural).

[16] Remember that we consider only single- and two-vehicle accidents except two-truck accidents.

Table 5.1 2004 accident severity models: results for speed limit[17]

| # | Model name* | | | Speed limit coefficient (t-ratio) | | Averaged elasticities of speed limit (SL) | | | |
|---|---|---|---|---|---|---|---|---|---|
| | | | | fatality [$\boldsymbol{\beta}_1$] | injury [$\boldsymbol{\beta}_2$] | $\overline{E}^{(1)}_{1;SL}$ | $\overline{E}^{(2)}_{1;SL} = \overline{E}^{(3)}_{1;SL}$ | $\overline{E}^{(2)}_{2;SL}$ | $\overline{E}^{(1)}_{2;SL} = \overline{E}^{(3)}_{2;SL}$ |
| 1 | County road | rural | (C/S)+(C/S) | .108 (3.61) | .0255 (5.15) | 4.49 | -.042 | .776 | -.297 |
| 2 | County road | rural | (C/S)+(T) | [.337 (1.34)] | .0414 (3.42) | | | 1.35 | -.405 |
| 3 | County road | rural | one vehicle | .0382 (3.47) | [-.00315 (-1.03)] | 1.75 | -.022 | | |
| 4 | County road | urban | (C/S)+(C/S) | .0323 (3.75) | .0323 (3.75) | 1.19 | -.001 | .927 | -.259 |
| 5 | County road | urban | (C/S)+(T) | [-.00511 (.000)] | [.0536 (1.08)] | | | | |
| 6 | County road | urban | one vehicle | [-.0974 (-.853)] | [.00469 (.354)] | | | | |
| 7 | Interstate | rural | (C/S)+(C/S) | [.000351 (.000)] | [.0116 (.646)] | | | | |
| 8a | Interstate | rural | (C)+(T) | [.00133 (.000)] | [.00870 (.358)] | | | | |
| 8b | Interstate | rural | (S)+(T) | [.0272 (.000)] | [.0374 (1.33)] | | | | |
| 9 | Interstate | rural | one vehicle | [-.0186 (-1.49)] | [-.0186 (-1.49)] | | | | |
| 10 | Interstate | urban | (C/S)+(C/S) | [.0171 (1.41)] | [.0171 (1.41)] | | | | |
| 11 | Interstate | urban | (C/S)+(T) | [.0798 (.518)] | [.000717 (.026)] | | | | |
| 12 | Interstate | urban | one vehicle | [.0366 (.938)] | [-.00262 (-.240)] | | | | |
| 13 | State route | rural | (C/S)+(C/S) | [.0628 (1.43)] | .0306 (3.90) | | | 1.03 | -.505 |
| 14 | State route | rural | (C/S)+(T) | [.168 (1.88)] | [.0239 (1.36)] | | | | |
| 15 | State route | rural | one vehicle | [.0313 (1.09)] | [-.00313 (-.422)] | | | | |
| 16a | State route | urban | (C)+(C) | .0340 (5.14) | .0340 (5.14) | 1.27 | -.001 | .949 | -.319 |
| 16b | State route | urban | (C)+(S) | .0225 (4.36) | .0225 (4.36) | .853 | -.001 | .652 | -.201 |
| 16c | State route | urban | (S)+(S) | .0315 (3.39) | .0315 (3.39) | 1.21 | -.005 | .951 | -.265 |
| 17 | State route | urban | (C/S)+(T) | .0418 (2.89) | .0418 (2.89) | 1.64 | -.010 | 1.29 | -.363 |
| 18 | State route | urban | one vehicle | [.0448 (1.19)] | [.00242 (.306)] | | | | |
| 19 | City street | rural | (C/S)+(C/S) | .114 (2.53) | .0273 (5.40) | 4.23 | -.009 | .775 | -.240 |
| 20 | City street | rural | (C/S)+(T) | [-.0733 (-.667)] | .0676 (3.17) | | | 2.11 | -.597 |
| 21 | City street | rural | one vehicle | [-.0218 (-.800)] | [-.00128 (-.244)] | | | | |
| 22 | City street | urban | (C/S)+(C/S) | .0938 (2.76) | .0304 (11.8) | 3.14 | -.003 | .785 | -.232 |
| 23a | City street | urban | (C)+(T) | .0469 (3.99) | .0469 (3.99) | 1.58 | -.003 | 1.30 | -.290 |
| 23b | City street | urban | (S)+(T) | .0640 (4.41) | .0640 (4.41) | 2.15 | -.005 | 1.82 | -.329 |
| 24 | City street | urban | one vehicle | [.0132 (.789)] | [-.000526(-.136)] | | | | |
| 25 | US route | rural | (C/S)+(C/S) | .340 (2.48) | .0409 (4.54) | 17.0 | -.212 | 1.39 | -.685 |
| 26 | US route | rural | (C/S)+(T) | [-.00980(-.271)] | .0720 (3.02) | | | 2.37 | -1.38 |
| 27 | US route | rural | one vehicle | [-.00249(-.075)] | [.00304 (.275)] | | | | |
| 28 | US route | urban | (C/S)+(C/S) | .0263 (5.76) | .0263 (5.76) | 1.04 | -.001 | .779 | -.265 |
| 29 | US route | urban | (C/S)+(T) | .0307 (2.52) | .0307 (2.52) | 1.26 | -.002 | .986 | -.278 |
| 30 | US route | urban | one vehicle | [.00230 (.055)] | [-.0139 (-1.37)] | | | | |

* "C", "S" and "T" mean car, SUV and truck respectively.

---

[17] Refer to Equations (2.4) and (2.9), where outcomes "1", "2", "3" correspond to "fatality", "injury", "PDO". We report statistically significant coefficients of the speed limit variable and the corresponding elasticities. In the square brackets we report insignificant coefficients (obtained by test-adding the speed limit variable into the AIC optimal models). All coefficients are the components of vectors $\boldsymbol{\beta}_1$ and $\boldsymbol{\beta}_2$, multiplied by the speed limit variable in Equation (2.4).

Table 5.2 2006 accident severity models: results for speed limit

| # | Model name* | | | Speed limit coefficient (t-ratio) | | Averaged elasticities of speed limit (SL) | | | |
|---|---|---|---|---|---|---|---|---|---|
| | | | | fatality [$\boldsymbol{\beta}_1$] | injury [$\boldsymbol{\beta}_2$] | $\bar{E}^{(1)}_{1;SL}$ | $\bar{E}^{(2)}_{1;SL} = \bar{E}^{(3)}_{1;SL}$ | $\bar{E}^{(2)}_{2;SL}$ | $\bar{E}^{(1)}_{2;SL} = \bar{E}^{(3)}_{2;SL}$ |
| 1 | County road | rural | (C/S)+(C/S) | .0396(5.48) | .0396(5.48) | 1.61 | -.016 | 1.20 | -.429 |
| 2 | County road | rural | (C/S)+(T) | .0648(3.06) | .0648(3.06) | 2.77 | -.032 | 2.35 | -.453 |
| 3 | County road | rural | one vehicle | .00506(2.04) | .00506(2.04) | .235 | -.002 | .185 | -.052 |
| 4a | County road | urban | (C)+(C) | [.00689(.000)] | [.00507(.321)] | | | | |
| 4b | County road | urban | (C)+(S) | [.0231(.000)] | .0613(2.43) | | | 1.80 | -.405 |
| 4c | County road | urban | (S)+(S) | [.0110(.000)] | [.0269(.843)] | | | | |
| 5 | County road | urban | (C/S)+(T) | [-.5454(-.518)] | [-.0288(-.301)] | | | | |
| 6 | County road | urban | one vehicle | [-.0852(-1.23)] | [.000725(.069)] | | | | |
| 7 | Interstate | rural | (C/S)+(C/S) | [.103(1.28)] | [.00872(.884)] | | | | |
| 8 | Interstate | rural | (C/S)+(T) | [.150(.908)] | [.00133(.063)] | | | | |
| 9 | Interstate | rural | one vehicle | [-.0237(-1.55)] | [-.0237(-1.55)] | | | | |
| 10 | Interstate | urban | (C/S)+(C/S) | [11.04(.000)] | [-.00108(-.135)] | | | | |
| 11 | Interstate | urban | (C/S)+(T) | [-.00188(-.011)] | [.0120(.519)] | | | | |
| 12 | Interstate | urban | one vehicle | [.00776(.197)] | [.00384(.476)] | | | | |
| 13 | State route | rural | (C/S)+(C/S) | .248(3.48) | .0416(3.25) | 11.9 | -.495 | 1.32 | -.759 |
| 14 | State route | rural | (C/S)+(T) | .127(2.50) | .127(2.50) | 5.79 | -.541 | 5.36 | -.970 |
| 15 | State route | rural | one vehicle | 0.0636(2.34) | [.0127(2.25)] | 3.34 | -.029 | | |
| 16 | State route | urban | (C/S)+(C/S) | .251(3.35) | .0290(7.95) | 9.40 | -.015 | .835 | -.253 |
| 17 | State route | urban | (C/S)+(T) | [5.60(.000)] | [.452(1.73)] | | | | |
| 18 | State route | urban | one vehicle | [.0268(.418)] | [-.0115(-.827)] | | | | |
| 19 | City street | rural | (C/S)+(C/S) | .0414(6.13) | .0414(6.13) | 1.46 | -.001 | 1.12 | -.349 |
| 20 | City street | rural | (C/S)+(T) | [.0185(.000)] | [.540(1.43)] | | | | |
| 21 | City street | rural | one vehicle | [-.0800(-1.46)] | [-.00409(-.381)] | | | | |
| 22a | City street | urban | (C)+(C) | .0251(6.33) | .0251(6.33) | .810 | .000 | .626 | -.184 |
| 22b | City street | urban | (C)+(S) | .0218(4.65) | .0218(4.65) | .727 | -.001 | .560 | -.167 |
| 22c | City street | urban | (S)+(S) | .0343(4.20) | .0343(4.20) | 1.14 | -.001 | .865 | -.279 |
| 23 | City street | urban | (C/S)+(T) | .0284(2.34) | .0284(2.34) | .937 | -.001 | .831 | -.107 |
| 24 | City street | urban | one vehicle | [.00968(.382)] | [-.00128(-.240)] | | | | |
| 25 | US route | rural | (C/S)+(C/S) | [.0644(1.32)] | [.0272(1.84)] | | | | |
| 26 | US route | rural | (C/S)+(T) | .0608(3.07) | .0608(3.07) | 3.12 | -.078 | 2.28 | -.912 |
| 27 | US route | rural | one vehicle | [.0137(1.56)] | [.0137(1.56)] | | | | |
| 28 | US route | urban | (C/S)+(C/S) | .0154(2.14) | .0154(2.14) | .613 | -.001 | .443 | -.171 |
| 29 | US route | urban | (C/S)+(T) | .0586(3.60) | .0586(3.60) | 2.33 | -.027 | 2.02 | -.337 |
| 30 | US route | urban | one vehicle | [.0327(.450)] | [.0134(.878)] | | | | |

* “C”, “S” and “T” mean car, SUV and truck respectively.

Table 5.3 Speed limit effect on structure of 2004 accident severity models[18]

| # | Model name | | | $M$ | $K$ | $LL(\boldsymbol{\beta}_m)$ | $\sum LL(\boldsymbol{\beta}_m)$ | df | p-value | conclusion* |
|---|---|---|---|---|---|---|---|---|---|---|
| 1 | County road | rural | (car/SUV)+(car/SUV) | 6 | 15 | -1656.7 | -1588.2 | 75 | 1.7e-5 | SL effect |
| 2 | | | (car/SUV)+(truck) | 5 | 8 | -256.74 | -235.02 | 32 | 0.085 | |
| 3 | | | one vehicle | 7 | 20 | -5066.7 | -4990.7 | 120 | 0.026 | SL effect |
| 4 | | urban | (car/SUV)+(car/SUV) | 7 | 9 | -691.21 | -657.69 | 54 | 0.11 | |
| 5 | | | (car/SUV)+(truck) | 3 | 5 | -90.86 | -86.40 | 10 | 0.54 | |
| 6 | | | one vehicle | 4 | 8 | -332.07 | -320.00 | 24 | 0.45 | |
| 7 | Interstate | rural | (car/SUV)+(car/SUV) | 4 | 5 | -414.78 | -404.68 | 15 | 0.16 | |
| 8a** | | | (car)+(truck) | 3 | 5 | -84.02 | -79.32 | 10 | 0.49 | |
| 8b | | | (SUV)+(truck) | 2 | 8 | -49.65 | -45.94 | 8 | 0.49 | |
| 9 | | | one vehicle | 4 | 11 | -1346.2 | -1324.2 | 33 | 0.17 | |
| 10 | | urban | (car/SUV)+(car/SUV) | 3 | 13 | -684.32 | -666.77 | 26 | 0.11 | |
| 11 | | | (car/SUV)+(truck) | 2 | 6 | -299.29 | -296.98 | 6 | 0.60 | |
| 12 | | | one vehicle | 5 | 12 | -761.47 | -738.41 | 48 | 0.55 | |
| 13 | Sate route | rural | (car/SUV)+(car/SUV) | 2 | 11 | -1310.1 | -1293.1 | 11 | 3.8e-4 | SL effect |
| 14*** | | | (car/SUV)+(truck) | 4 | 10 | -381.87 | -368.11 | 30 | 0.60 | |
| 15 | | | one vehicle | 6 | 13 | -2153.5 | -2117.0 | 65 | 0.23 | |
| 16a | | urban | (car)+(car) | 8 | 7 | -1129.5 | -1094.6 | 49 | 0.026 | SL effect |
| 16b | | | (car)+(SUV) | 7 | 10 | -1557.2 | -1512.4 | 60 | 7.8e-3 | SL effect |
| 16c | | | (SUV)+(SUV) | 3 | 8 | -497.85 | -485.13 | 16 | 0.063 | |
| 17 | | | (car/SUV)+(truck) | 6 | 4 | -177.67 | -162.33 | 20 | 0.059 | |
| 18 | | | one vehicle | 7 | 10 | -618.56 | -569.55 | 60 | 1.4e-3 | SL effect |
| 19 | City street | rural | (car/SUV)+(car/SUV) | 10 | 10 | -1376.0 | -1313.4 | 90 | 8.3e-3 | SL effect |
| 20 | | | (car/SUV)+(truck) | 4 | 8 | -73.63 | -63.33 | 24 | 0.66 | |
| 21 | | | one vehicle | 6 | 13 | -946.00 | -905.80 | 65 | 0.095 | |
| 22 | | urban | (car/SUV)+(car/SUV) | 10 | 20 | -14620 | -14435 | 180 | 3.3e-14 | SL effect |
| 23a | | | (car)+(truck) | 4 | 12 | -534.10 | -509.25 | 36 | 0.064 | |
| 23b | | | (SUV)+(truck) | 6 | 6 | -354.56 | -323.86 | 30 | 6.2e-4 | SL effect |
| 24 | | | one vehicle | 5 | 22 | -3627.5 | -3560.3 | 88 | 1.0e-3 | SL effect |
| 25 | US route | rural | (car/SUV)+(car/SUV) | 6 | 8 | -996.60 | -969.57 | 40 | 0.068 | |
| 26 | | | (car/SUV)+(truck) | 2 | 11 | -275.63 | -262.02 | 11 | 4.3e-3 | SL effect |
| 27 | | | one vehicle | 5 | 10 | -1236.7 | -1219.5 | 40 | 0.72 | |
| 28 | | urban | (car/SUV)+(car/SUV) | 5 | 11 | -2361.9 | -2326.9 | 44 | 7.6e-3 | SL effect |
| 29 | | | (car/SUV)+(truck) | 7 | 8 | -314.86 | -287.88 | 48 | 0.26 | |
| 30 | | | one vehicle | 7 | 10 | -493.83 | -458.36 | 60 | 0.16 | |

* – For models with "SL effect" conclusion speed limit is statistically significant for the structure of accident severity models, it is not significant otherwise.

** – Variables X33f and X15 are constant inside speed limit bins and have been removed from the best final logit model before carrying out the test.

*** – Variable X33f is constant and has been removed from the best final logit model.

[18] The tests of the effect are done by using the likelihood ratio, refer to paragraph 3 on page 43.

Table 5.4 Speed limit effect on structure of 2006 accident severity models

| # | Model name | | | $M$ | $K$ | $LL(\boldsymbol{\beta}_\mathbf{m})$ | $\sum LL(\boldsymbol{\beta}_\mathbf{m})$ | df | p-value | conclusion* |
|---|---|---|---|---|---|---|---|---|---|---|
| 1 | County road | rural | (car/SUV)+(car/SUV) | 4 | 9 | -847.91 | -813.01 | 27 | 1.2e-4 | SL effect |
| 2 | | | (car/SUV)+(truck) | 5 | 7 | -119.31 | -100.96 | 28 | 0.13 | |
| 3 | | | one vehicle | 5 | 19 | -7154.2 | -7106.0 | 76 | 0.058 | |
| 4a | | urban | (car)+(car) | 7 | 4 | -303.42 | -291.38 | 24 | 0.46 | |
| 4b | | | (car)+(SUV) | 4 | 3 | -235.35 | -226.24 | 9 | 0.033 | SL effect |
| 4c | | | (SUV)+(SUV) | 4 | 5 | -161.81 | -151.13 | 15 | 0.13 | |
| 5 | | | (car/SUV)+(truck) | 3 | 4 | -15.658 | -98.097 | 8 | 0.17 | |
| 6 | | | one vehicle | 5 | 8 | -342.22 | -330.11 | 32 | 0.84 | |
| 7 | Interstate | rural | (car/SUV)+(car/SUV) | 4 | 9 | -451.87 | -438.45 | 27 | 0.47 | |
| 8 | | | (car/SUV)+(truck) | 2 | 6 | -98.695 | -97.460 | 6 | 0.87 | |
| 9 | | | one vehicle | 5 | 12 | -1474.1 | -1465.6 | 48 | 0.53 | |
| 10 | | urban | (car/SUV)+(car/SUV) | 8 | 9 | -836.55 | -811.72 | 63 | 0.89 | |
| 11 | | | (car/SUV)+(truck) | 2 | 9 | -221.17 | -215.30 | 9 | 0.23 | |
| 12 | | | one vehicle | 5 | 11 | -882.75 | -862.84 | 44 | 0.65 | |
| 13 | Sate route | rural | (car/SUV)+(car/SUV) | 2 | 9 | -575.69 | -559.57 | 9 | 1.8e-4 | SL effect |
| 14 | | | (car/SUV)+(truck) | 2 | 4 | -76.769 | -72.761 | 4 | 0.091 | |
| 15 | | | one vehicle | 5 | 16 | -3707.8 | -3657.2 | 64 | 0.00204 | SL effect |
| 16 | | urban | (car/SUV)+(car/SUV) | 8 | 10 | -3268.3 | -3203.0 | 70 | 1.5e-4 | SL effect |
| 17 | | | (car/SUV)+(truck) | 5 | 5 | -153.32 | -146.63 | 20 | 0.86 | |
| 18 | | | one vehicle | 6 | 8 | -747.88 | -715.34 | 40 | 7.4e-3 | SL effect |
| 19 | City street | rural | (car/SUV)+(car/SUV) | 7 | 7 | -1088.1 | -1046.1 | 42 | 1.3e-3 | SL effect |
| 20 | | | (car/SUV)+(truck) | 3 | 2 | -2851.8 | -2834.5 | 4 | 5.3e-6 | SL effect |
| 21 | | | one vehicle | 2 | 11 | -288.38 | -283.71 | 11 | 0.59 | |
| 22a | | urban | (car)+(car) | 6 | 15 | -6848.4 | -6782.1 | 75 | 4.8e-4 | SL effect |
| 22b | | | (car)+(SUV) | 6 | 14 | -4300.4 | -4247.2 | 70 | 3.3e-2 | SL effect |
| 22c | | | (SUV)+(SUV) | 2 | 10 | -1350.6 | -1338.5 | 10 | 7.4e-2 | SL effect |
| 23 | | | (car/SUV)+(truck) | 6 | 11 | -523.35 | -498.99 | 55 | 0.71 | |
| 24 | | | one vehicle | 4 | 20 | -2432.0 | -2393.4 | 60 | 0.065 | |
| 25 | US route | rural | (car/SUV)+(car/SUV) | 4 | 9 | -329.22 | -311.62 | 27 | 0.13 | |
| 26 | | | (car/SUV)+(truck) | 3 | 7 | -243.30 | -233.82 | 14 | 0.17 | |
| 27 | | | one vehicle | 2 | 14 | -1484.2 | 1472.3 | 14 | 0.049 | SL effect |
| 28 | | urban | (car/SUV)+(car/SUV) | 2 | 7 | -969.51 | -961.10 | 7 | 0.019 | SL effect |
| 29 | | | (car/SUV)+(truck) | 2 | 6 | -192.74 | -184.75 | 6 | 0.014 | SL effect |
| 30 | | | one vehicle | 3 | 9 | -173.27 | -166.32 | 18 | 0.74 | |

* – For models with “SL effect” conclusion speed limit is statistically significant for the structure of accident severity models, it is not significant otherwise.

- Higher speed limit values do generally lead to higher probabilities of more severe accidents (fatality and injury) on road classes other than interstate highways. This effect is especially strong for 2004 and 2006 accidents on rural country roads, urban city maintained streets and urban U.S. routes.
- The speed limit variable seems to be inelastic for some of the road-class-accident-type combinations that display a statistically significant relationship between speed limit and accident severity. For others, especially in the case of fatal accident outcomes, elasticities of the speed limit can be quite high.

Next let us refer to Table 5.3 and Table 5.4, which are related to 2004 and 2006 accident data respectively. These tables give the results for the log-likelihood ratio tests of statistically significant differences among the models estimated for different speed-limit bins for each of the considered road-class-accident-type combination (the bins themselves are given in Table C.3 and Table C.4 in Appendix C). We find the following results:

- Speed limit does not have any statistically significant effect on the structure of severity models for 2004 and 2006 accidents on interstate highways (both urban and rural).
- Speed limit has a statistically significant effect on the structure of severity models for accidents on all other road classes (with the exception of 2004 accidents on urban county maintained roads, for which speed limit has no any statistically significant effect on the severity model structure).

### 5.2.2. Effect of Other Explanatory Variables

Now we use model estimation results given in Table C.5 and Table C.6 in Appendix C and consider the influence of explanatory variables other than the posted speed limit on severity of 2004 and 2006 accidents. Similar to the accident causation study, we again limit our consideration to several of the most important and most statistically significant variables.

- **Variables “wint” and “sum” (see pages 97 and 127):** We find that the probability of higher severity of an accident generally decreases during winters and increases during summers (this effect was stronger is year 2004). This result appears to be an unexpected result. However, it can be explained. First, drivers do take some extra precautions during bad winter weather. Although these precautions are not sufficient to keep the drivers safe (refer to accident causation study results in Section 4.2.2), they can reduce probabilities of very severe accident outcomes. Second, it is very likely that the number of minor (PDO) accidents sharply increases during winters due to less safe weather and roadway conditions. This increase shifts the outcome probabilities (conditioned on the fact that an accident occurred and observed) toward less severe accident outcomes. In other words, the number of serious accidents (e.g. fatalities) might increase during winters, but the number of minor accidents is likely to increase much more. The summer seasonal effect is the opposite of the winter seasonal effect.
- **Variables “dark” and “darklamp” (see pages 102, 103, 132 and 133):** We find that the probability of higher severity of an accident generally increases when the road is dark, even if there are street lights. The explanation can be that during night time drivers have harder time controlling their vehicles and holding the road. Thus, drivers become more likely to be involved into serious accidents (such as head-on collisions, collisions with stationary objects, and rollovers).
- **Variables “nojun” and “way4” (see pages 107, 137 and 138):** We find that accidents are generally more likely to be more severe if they occur at road junctions, and especially at 4-way intersections. This effect mainly concerns two-vehicle accidents and it can be explained as follows. On one hand, two-vehicle collisions in which one vehicle hits a side of the other vehicle are most likely to occur at road junctions. On the other hand, side impacts are highly dangerous due to high driver and passenger vulnerability during such impacts.

- **Variables "driver" and "env" (see pages 111, 112 and 139):** We find that the probability of higher severity of an accident generally increases when the primary cause of the accident is driver-related, while the probability decreases when the primary cause is environment-related. This is a relatively strong effect. It is due to human mistakes being especially dangerous because they are totally unpredictable, while environment (e.g. weather-related) factors are observable and can be accounted for by taking additional precautions.
- **Variables "hl5", "hl10" and "hl20" (see pages 112, 113, 140 & 141):** Depending on the road class and accident type, we find that an accident has higher probability of a severe outcome if help arrives more quickly. This effect can be due to a data selection bias because help is not needed at all in the case of minor accidents.
- **Variable "moto" (see pages 114 and 141):** Accidents caused by a motorcycle are typically more severe. This is a strong effect. It is explained by very high vulnerability of motorcycle riders.
- **Variable "vage" and "voldo" (see pages 116, 124, 142 and 149):** We find that the probability of higher severity of an accident increases with the age of a vehicle involved into the accident. This effect exists because obviously older vehicles are less safe than newer vehicles are.
- **Variable "$X_{27}$" and "maxpass" (see pages 117, 124, 143 and 150):** We find that the probability of higher severity of an accident typically increases with the number of occupants in vehicles involved into the accident. This effect is expected for three reasons. First, the likelihoods of at least one death and at least one injury increase when there are more occupants in a colliding vehicle. Second, occupants' bodies hit each other during a collision. Third, a more heavily occupied vehicle has higher mass and higher kinetic energy to dissipate during a collision.

- **Variable "priv" (see pages 119 and 144):** We find that accidents occurred in private drives are typically less severe. Such accidents are minor because vehicles generally travel at low speeds in private drives.
- **Variable "$X_{33}$" (see pages 121 and 147):** We find that the probability of higher severity of an accident considerably increases when at least one of the vehicles involved into the accident is on fire. This is a very strong effect. Obviously, fire is very dangerous for drivers and passengers involved into an accident.
- **Variables "$X_{35}$", "ff" and "mm" (see pages 123, 125, 149 and 152):** We find that generally the probability of higher severity of an accident increases when the driver at fault is female. In addition, when two female drivers are involved into a two-vehicle accident, then this accident is more likely to be reported as severe, as compared to the case when two male drivers are involved. We attribute this to a possibility that females are more likely than males to report non-evident injuries. Females might also be less likely to survive in very severe accidents.

# CHAPTER 6. DISCUSSION

Let us summarize and discuss the results of our speed-safety relationship study, consider implementations for the optimal speed limit polices in Indiana State, and suggest possible directions for future speed-safety research.

First, we find that speed limits have no statistically significant effect on either the unsafe-speed-related causation or the severity of accidents that occurred on interstate highways, except for a single case of 2006 one-vehicle accidents on rural interstates. In this single case the probability of unsafe speed being the primary cause of an accident *decreases* with an increase in the posted speed limit. This is a very interesting and significant result because interstates are the roads with the highest posted speed limits and are of great importance to national trade and commerce. Let us also note that the top speed limit on some portions of Indiana highways was 70 mph in 2006 as opposed to only 65 mph in 2004. Our results for the speed-safety relationship on interstate highways can possibly be understood by considering the following two counteractive effects:

1. As the speed limit posted on a highway increases, the average speed of the traveling vehicles obviously also increases (Khan, 2002). As a result, vehicles travel larger distances during human reaction times. Thus, drivers have less time to react to changing conditions on the road (such as deer or an object on the roadway surface), resulting in an increase in the frequency of unsafe-speed-related accidents. In addition, a vehicle generally looses stability and roadway traction with an increase in speed because of increased aerodynamic and centrifugal forces acting onto the vehicle. This effect also leads to an increase in frequency of unsafe-speed-related accidents. Finally, the average kinetic energy of vehicles increases with their speeds. Since this energy must be dissipated during an accident collision with a stationary object or during a head-on collision, such accident collisions become more severe as speeds increase.

2. As the speed limit posted on a highway increases, the variance of the speed of the traveling vehicles may decrease (Renski et al., 1999). Below we will refer to this effect as to the “speed variance reduction effect”. This effect can be explained by the fact that a majority of sensible drivers have in their minds some psychological approximate upper value of speed that they do not want to exceed. For example, let us consider a case when a speed limit is increased by 10 mph from 60 mph to 70 mph. In this case slow drivers, who usually obey the speed limit law, will increase their speeds by about 10 mph. At the same time, the fastest drivers, who usually drive in significant excess of the posted speed limits, will probably increase their speeds by smaller increments or even not increase their speeds at all. As a result, the speed variance may decreases when the posted speed limit and average speed increase[19]. Now we note that as the speed variance decreases, the spread between velocities of different vehicles decreases as well. In other words, the vehicle velocities decrease in the co-moving coordinate frame of reference (the later can be defined as the coordinate system moving with the average velocity of the vehicles, or as the center of mass reference frame of the vehicles). This decrease leads to an increase in time available for human reaction response and to a decrease of the dissipated kinetic energy in all mutual collisions of vehicles traveling in the same direction[20]. Therefore, the frequency and severity of such accident collisions also decrease.

[19] An increase of the averaged speed might also decrease speed probability distribution moments that are higher than the second moment. In this case, the frequency of accidents caused by extremely fast drivers, whose speeds are in the right tail of the distribution, might also decrease.

Thus, when speed limits increase, the average vehicle speed increases while the speed variance possibly decreases. For interstate highway accidents these two effects may roughly balance each other. This is a possible explanation of the result of our study that an increase in the posted speed limit has no any statistically significant effect on safety in all cases except one, in which the unsafe-speed-related accident causation probability actually decreases with speed limit increase.

Another possible explanation of the absence of adverse effects of speed limit on safety on interstate highways is that interstates are of limited access and are specially designed for high speed traffic flows. In other words, interstate highways have "error-forgiving" design, which can explain why higher speed limit values do not affect safety significantly.

The second important result of our study concerns accidents that occur on roads other than interstate highways. We find that for these accidents, the speed limit typically has no statistically significant adverse effect on their unsafe-speed-related causation. At the same time we find that the speed limit does generally increase the severity of accidents on roads other than interstate highways. Thus, the speed limit seems to affect safety on non-highway roads in an indirect way. On one hand, a reasonable speed increment does not generally increase the likelihood of unsafe-speed-related accidents. Perhaps, this is because many drivers might not pay much attention to the posted speed limit and, instead, might choose the driving speeds at which they feel themselves comfortable. If most drivers are rational, then overall they make reasonable speed choices according to the road and driving conditions. On the other hand, the average speed of vehicles rises with an increase in the speed limit posted on a road. As a result, accidents tend to be more severe, even if these accidents happen for reasons other than unsafe speed. It is interesting that while there is no statistically significant relationship between speed and accident severity on interstate highways, such relationship exists on other roads. This difference can be due to the speed variance reduction effect being weak on roads other than interstates. In addition, interstate highways are very different from the other roads. As mentioned above, interstates are better designed for high speeds and more “error-forgiving” than other roads.

---

[20] Note that the kinetic energy, which must be dissipated during a mutual collision between two or more vehicles, is determined by the vehicle masses and the squares of the vehicle velocities in the center of mass reference frame of the colliding vehicles.

Our findings have the following implications for speed limit polices in Indiana:

- A reasonable increase in speed limits on interstate highways may increase mobility and productivity without a considerable adverse effect on road safety.
- As far as the speed limit policies on roads other than interstate highways are concerned, we suggest caution be exercised and any speed limit changes be done on case-to-case basis.

There are two clear possible extensions of the present research. First, we only use data on individual accidents that occurred in Indiana in 2004 and 2006. A study with larger statistical data sample that includes additional years will have a greater statistical power and can be beneficial. Second, in addition to statistical modeling of the probabilities of accident causes and severity levels, in the future we might want to consider accident frequencies as well. The reason is that the logit model probabilities, which we use and which are given by Equations (2.1), (2.3) and (2.4), are the conditioned on an accident occurring. As explained on page 2 in the introductory chapter, the unconditional probability of an accident outcome is equal to the product of the corresponding conditional probability of this outcome and the probability of the accident to occur. As a result, a study of conditional probabilities of accident outcomes can be enhanced by considering accident frequencies[21].

---

[21] For example, let us assume that the number of serious accidents on a road does not change, while the number of minor accidents increases. In this case, although the conditional probability of a severe accident outcome falls, the accident probability increases and the road obviously becomes less safe.

# LIST OF REFERENCES

Appendix A.

**List of explanatory variables:**

$X_3$ – Collision date

$X_4$ – Day of the week

$X_5$ – Collision time

$X_{13}$ – Construction
(*no; yes; buck-up of traffic outside of but due to construction zone*)

$X_{14}$ – Light condition
(*daylight; dawn / dusk; dark with street lights on; dark with no lights*)

$X_{15}$ – Weather condition
(*clear; cloudy; sleet/hail / freezing rain; fog / smoke / smog; rain; snow; severe cross wind*)

$X_{16}$ – Surface condition
(*dry; wet; muddy; snow / slush; ice; loose material on roadway; water*)

$X_{17}$ – Type of median
(*drivable; curbed; barrier wall; none*)

$X_{18}$ – Type of roadway junction
(*no junction involved; four-way intersection; ramp T-intersection; Y-intersection; traffic circle / roundabout; five point or more; interchange*)

$X_{19}$ – Road character
(*straight / level; straight / grade; straight / hillcrest; curve / level; curve / grade; curve / hillcrest; non roadway crash*)

$X_{20}$ – Primary contributing circumstance
(*alcoholic beverages; illegal drugs; driver asleep or fatigue; prescription drugs; driver illness;* ***unsafe speed****; failure to yield right of way; disregard signal / red signal; left of center; improper passing; improper turning; improper lane usage; following too closely; unsafe backing; overcorrecting / oversteering; ran off road right; ran off road left; wrong way on one way; pedestrian action; passenger distraction; violation of license restriction; jackknifing; cell phone usage; other telematics in use; other (explain in narrative); driver distracted [explain in narrative];* ***speed too fast for weather conditions****; engine failure or defective; accelerator failure or defective; brake failure or defective; tire failure or defective;*

*headlight defective or not on; other lights defective; steering failure; window / windshield defective; oversize / overweight load; insecure / leaky load; tow hitch failure; other explained in narrative; glare; roadway surface condition; holes / ruts in surface; shoulder defective; road under construction; severe crosswinds; obstruction not marked; lane marking obscured; view obstructed; animal on roadway; traffic control problem; other [explained in narrative]; utility work*)

$X_{22}$ – Time when help arrived

$X_{25}$ – Vehicle type, considered for the vehicle at fault, i.e. for the vehicle that contributed to the primary cause of an accident
(*passenger car / station wagon; pickup; van; sport utility vehicle; truck [single 2 axle, 6 tires]; truck [single 3 or more axles]; truck / trailer [not semi]; tractor / one semi trailer; tractor / double trailer; tractor / triple trailer tractor [cab only, no trailer]; motor home / recreational vehicle; motorcycle; bus/seats 9-15 persons with driver; bus / seats 15+ persons with driver; school bus; unknown type; farm vehicle; combination vehicle; pedestrian; bicycle*)

$X_{26}$ – Vehicle year, considered for all vehicles involved

$X_{27}$ – Number of occupants, considered for all vehicles involved

$X_{28}$ – Vehicle license state, considered for the vehicle at fault, i.e. for the vehicle that contributed to the primary cause of an accident
(*Indiana; Indiana's neighboring states [IL, KY, OH, MI]; other US states; Canada / Mexico / U.S. Territories; other foreign countries*)

$X_{29}$ – Speed limit, considered only if known and the same speed limit value for all vehicles involved

$X_{30}$ – Road type, considered for the vehicle at fault, i.e. for the vehicle contributed to the primary cause of an accident
(*one lane [one way]; two lanes [one way]; multi-lanes [one way]; two lanes [two way]; multi-lane undivided [two way]; multi-lane undivided 2-way left [two way]; multi-lane divided 3 or more lanes [two way]; alley; private drive*)

$X_{31}$ – Traffic control, considered for the vehicle at fault, i.e. for the vehicle contributed to the primary cause of an accident
(*officer / crossing guard / flagman; RR crossing gate / flagman; RR crossing flashing signal; RR crossing sign; traffic control signal; flashing signal; stop sign; yield sign; lane control; no passing zone; other regulatory sign / marking; none*)

$X_{33}$ – Fire, considered for all vehicles involved
(*no; yes*)

$X_{34}$ – Driver age, considered for all drivers involved

$X_{35}$ – Driver gender, considered for all drivers involved

# Appendix B.

Table B.1 Road classes & accident types in 2004 accident causation study

| # | Road-class-accident-type combination | | | Number of observations: all | available for the models*: total | unsafe-speed-related | other causes |
|---|---|---|---|---|---|---|---|
| 1 | County road | rural | (car/SUV**)+(car/SUV) | 7249 | 5198 | 518 | 4680 |
| 2 | | | (car/SUV)+(truck***) | 647 | 617 | 28 | 589 |
| 3 | | | one vehicle | 18045 | 11998 | 1877 | 10121 |
| 4 | | urban | (car/SUV)+(car/SUV) | 1854 | 1490 | 97 | 1393 |
| 5 | | | (car/SUV)+(truck) | 143 | 121 | 7 | 114 |
| 6 | | | one vehicle | 972 | 689 | 142 | 547 |
| 7 | Interstate | rural | (car/SUV)+(car/SUV) | 1041 | 995 | 168 | 827 |
| 8 | | | (car/SUV)+(truck) | 811 | 338 | 77 | 311 |
| 9 | | | one vehicle | 3347 | 1617 | 430 | 1187 |
| 10 | | urban | (car/SUV)+(car/SUV) | 2227 | 1386 | 131 | 1255 |
| 11 | | | (car/SUV)+(truck) | 3306 | 922 | 84 | 838 |
| 12 | | | one vehicle | 1605 | 1442 | 463 | 979 |
| 13 | State route | rural | (car/SUV)+(car/SUV) | 4774 | 2311 | 170 | 2141 |
| 14 | | | (car/SUV)+(truck) | 682 | 665 | 41 | 624 |
| 15 | | | one vehicle | 9775 | 6432 | 540 | 5892 |
| 16 | | urban | (car/SUV)+(car/SUV) | 7999 | 4698 | 191 | 4507 |
| 17 | | | (car/SUV)+(truck) | 636 | 633 | 16 | 617 |
| 18 | | | one vehicle | 1488 | 1004 | 99 | 905 |
| 19 | City street | rural | (car/SUV)+(car/SUV) | 3778 | 2648 | 155 | 2493 |
| 20 | | | (car/SUV)+(truck) | 261 | 261 | 9 | 252 |
| 21 | | | one vehicle | 2387 | 2187 | 265 | 1922 |
| 22 | | urban | (car/SUV)+(car/SUV) | 62701 | 50180 | 1901 | 48279 |
| 23 | | | (car/SUV)+(truck) | 3574 | 3105 | 85 | 3200 |
| 24 | | | one vehicle | 12205 | 7988 | 1134 | 6854 |
| 25 | US route | rural | (car/SUV)+(car/SUV) | 2588 | 2005 | 152 | 1853 |
| 26 | | | (car/SUV)+(truck) | 566 | 563 | 52 | 511 |
| 27 | | | one vehicle | 4202 | 2667 | 247 | 2420 |
| 28 | | urban | (car/SUV)+(car/SUV) | 6895 | 6462 | 328 | 6134 |
| 29 | | | (car/SUV)+(truck) | 750 | 734 | 37 | 697 |
| 30 | | | one vehicle | 1061 | 988 | 97 | 891 |

* – observations available for the best estimated statistical models after exclusion of all missing observations

** – "SUV" includes sport utility vehicles, pickups and vans

*** – "truck" includes any possible kind of truck or tractor

Table B.2 Road classes & accident types in 2006 accident causation study

| # | Road-class-accident-type combination | | | Number of observations | | | |
|---|---|---|---|---|---|---|---|
| | | | | all | available for the models*: total | unsafe-speed-related | other causes |
| 1 | County road | rural | (car/SUV**)+(car/SUV) | 5956 | 1698 | 111 | 1587 |
| 2a | County road | rural | (car)+(truck***) | 194 | 194 | 6 | 188 |
| 2b | County road | rural | (SUV)+(truck***) | 150 | 126 | 2 | 124 |
| 3 | County road | rural | one vehicle | 16132 | 3518 | 500 | 3018 |
| 4 | County road | urban | (car/SUV)+(car/SUV) | 1485 | 1483 | 66 | 1417 |
| 5 | County road | urban | (car/SUV)+(truck) | 83 | 79 | 5 | 74 |
| 6 | County road | urban | one vehicle | 797 | 752 | 103 | 649 |
| 7a | Interstate | rural | (car)+(car) | 395 | 354 | 37 | 317 |
| 7b | Interstate | rural | (car)+(SUV) | 518 | 476 | 59 | 417 |
| 7c | Interstate | rural | (SUV)+(SUV) | 210 | 209 | 26 | 183 |
| 8 | Interstate | rural | (car/SUV)+(truck) | 757 | 742 | 97 | 645 |
| 9 | Interstate | rural | one vehicle | 3730 | 3637 | 489 | 3148 |
| 10 | Interstate | urban | (car/SUV)+(car/SUV) | 2392 | 2203 | 172 | 2031 |
| 11a | Interstate | urban | (car)+(truck) | 627 | 541 | 47 | 494 |
| 11b | Interstate | urban | (SUV)+(truck) | 220 | 209 | 20 | 189 |
| 12 | Interstate | urban | one vehicle | 1883 | 303 | 75 | 228 |
| 13 | State route | rural | (car/SUV)+(car/SUV) | 4577 | 4377 | 217 | 4160 |
| 14 | State route | rural | (car/SUV)+(truck) | 522 | 511 | 30 | 481 |
| 15 | State route | rural | one vehicle | 10155 | 9421 | 546 | 8875 |
| 16 | State route | urban | (car/SUV)+(car/SUV) | 7461 | 6241 | 220 | 6021 |
| 17 | State route | urban | (car/SUV)+(truck) | 509 | 508 | 23 | 485 |
| 18 | State route | urban | one vehicle | 1700 | 1603 | 126 | 1477 |
| 19 | City street | rural | (car/SUV)+(car/SUV) | 3778 | 1294 | 55 | 1239 |
| 20a | City street | rural | (car)+(truck) | 97 | 97 | 4 | 93 |
| 20b | City street | rural | (SUV)+(truck) | 57 | 57 | 3 | 54 |
| 21 | City street | rural | one vehicle | 2104 | 1929 | 225 | 1704 |
| 22 | City street | urban | (car/SUV)+(car/SUV) | 50034 | 20704 | 563 | 20141 |
| 23a | City street | urban | (car)+(truck) | 1526 | 945 | 25 | 920 |
| 23b | City street | urban | (SUV)+(truck) | 788 | 481 | 8 | 473 |
| 24 | City street | urban | one vehicle | 110682 | 9011 | 945 | 8066 |
| 25 | US route | rural | (car/SUV)+(car/SUV) | 2478 | 2353 | 129 | 2224 |
| 26 | US route | rural | (car/SUV)+(truck) | 474 | 457 | 27 | 430 |
| 27 | US route | rural | one vehicle | 4204 | 4151 | 262 | 3889 |
| 28 | US route | urban | (car/SUV)+(car/SUV) | 6925 | 6140 | 206 | 5934 |
| 29a | US route | urban | (car)+(truck) | 346 | 342 | 12 | 330 |
| 29b | US route | urban | (SUV)+(truck) | 226 | 226 | 2 | 224 |
| 30 | US route | urban | one vehicle | 1209 | 1148 | 90 | 1058 |

* – observations available for the best estimated statistical models after exclusion of all missing observations
** – "SUV" includes sport utility vehicles, pickups and vans
*** – "truck" includes any possible kind of truck or tractor

Table B.3 Binary logit models for 2004 accident causation[22]

| # | Model name | | | Log-likelihood | | $R^2$ | Coefficient (t-ratio) | |
|---|---|---|---|---|---|---|---|---|
| | | | | model | restricted* | | $X_{29}$ | constant |
| 1 | County road | rural | (car/SUV)+(car/SUV) | -1426.7 | -1685.9 | .154 | | -1.14(-6.71) |
| 2 | County road | rural | (car/SUV)+(truck) | -87.04 | -113.95 | .236 | | |
| 3 | County road | rural | one vehicle | -4468.1 | -5203.8 | .141 | .00859(2.83) | -.743(-3.30) |
| 4 | County road | urban | (car/SUV)+(car/SUV) | -246.94 | -307.15 | .196 | .0368(2.24) | -1.42(-2.12) |
| ***5*** | County road | urban | (car/SUV)+(truck) | -16.14 | -26.74 | .396 | | -2.57(-2.00) |
| 6 | County road | urban | one vehicle | -305.79 | -350.52 | .128 | | |
| 7 | Interstate | rural | (car/SUV)+(car/SUV) | -327.44 | -451.78 | .275 | | .985(-2.40) |
| ***8*** | Interstate | rural | (car/SUV)+(truck) | -128.79 | -193.32 | .334 | | |
| 9 | Interstate | rural | one vehicle | -564.03 | -936.51 | .398 | | |
| 10 | Interstate | urban | (car/SUV)+(car/SUV) | -339.99 | -436.39 | .221 | | -1.52(-4.83) |
| 11 | Interstate | urban | (car/SUV)+(truck) | -203.66 | -306.50 | .336 | | -.928(-2.92) |
| 12 | Interstate | urban | one vehicle | -700.30 | -942.58 | .257 | | |
| 13 | State route | rural | (car/SUV)+(car/SUV) | -537.67 | -607.23 | .115 | | -1.72(-6.68) |
| ***14*** | State route | rural | (car/SUV)+(truck) | -128.69 | -153.94 | .164 | | -2.08(-6.50) |
| 15 | State route | rural | one vehicle | -1452.8 | -1854.5 | .217 | | |
| 16 | State route | urban | (car/SUV)+(car/SUV) | -716.67 | -798.76 | .103 | .0212(2.51) | -2.77(-7.83) |
| ***17*** | State route | urban | (car/SUV)+(truck) | -57.96 | -74.64 | .224 | | -4.05(-8.74) |
| 18 | State route | urban | one vehicle | -274.39 | -323.30 | .103 | | -1.03(-2.08) |
| 19 | City street | rural | (car/SUV)+(car/SUV) | -525.56 | -586.00 | .094 | .0225(2.50) | -2.87(-5.48) |
| ***20*** | City street | rural | (car/SUV)+(truck) | -35.71 | -39.15 | .088 | | -3.69(-8.93) |
| 21 | City street | rural | one vehicle | -681.68 | -807.55 | .156 | | |
| 22 | City street | urban | (car/SUV)+(car/SUV) | -6259.1 | -7117.7 | .121 | | -2.43(-20.9) |
| 23 | City street | urban | (car/SUV)+(truck) | -323.90 | -389.67 | .169 | | -.983(-2.22) |
| 24 | City street | urban | one vehicle | -2776.2 | -3263.2 | .149 | -.0117(-3.82) | |
| 25 | US route | rural | (car/SUV)+(car/SUV) | -404.95 | -475.98 | .149 | | -2.14(-8.87) |
| 26 | US route | rural | (car/SUV)+(truck) | -147.16 | -173.39 | .151 | | -.822(-3.51) |
| 27 | US route | rural | one vehicle | -593.93 | -822.88 | .278 | -.0422(-3.17) | 3.48(4.66) |
| 28 | US route | urban | (car/SUV)+(car/SUV) | -1088.6 | -1200.4 | .093 | .0181(2.33) | -2.32(-6.06) |
| 29 | US route | urban | (car/SUV)+(truck) | -117.72 | -146.59 | .197 | | |
| 30 | US route | urban | one vehicle | -206.07 | -255.30 | .193 | | -.916(-2.15) |

.... – positive coefficient
.... – negative coefficient

* – restricted log-likelihood found by setting all coefficients except intercepts to zero
** – models are estimated by using procedure A on page 32, except the models marked by bold numbers and estimated by using procedure B on page 33

"$X_{29}$" – "posted speed limit (if the same for all vehicles involved)" quantitative variable
"constant" – "constant term (intercept)" quantitative variable

[22] Refer to Equation (2.3), where outcomes "1" and "2" correspond to "unsafe-speed-related cause" and "any other cause". Only statistically significant coefficients, which are components of vector $\boldsymbol{\beta}_1$ in Equation (2.3), are given in the table.

Table B.3 (Continued)

| # | Coefficient (t-ratio) | | | | | |
|---|---|---|---|---|---|---|
| | wint [$X_3$] | mon [$X_4$] | tues [$X_4$] | wed [$X_4$] | fr [$X_4$] | sat [$X_4$] |
| 1 | | .382(2.91) | | | | |
| 2 | | | | | | |
| 3 | .370(6.32) | | | | | |
| 4 | | | | | | |
| ***5*** | | | | | | |
| 6 | | | | | | |
| 7 | .916(4.40) | | | | | |
| ***8*** | | | | | | |
| 9 | 1.34(8.52) | | | | | |
| 10 | | | | | | |
| 11 | .630(4.44) | | | | | |
| 12 | .662(4.59) | | | | | |
| 13 | .613(3.57) | | | | | |
| ***14*** | | | | | | |
| 15 | .562(5.60) | | | | | -.291(-2.00) |
| 16 | | | | | | |
| ***17*** | 1.59(2.39) | | | | | |
| 18 | | | | | | |
| 19 | | | | | | |
| ***20*** | | | | | | |
| 21 | .426(2.94) | | | -.608(-2.23) | | |
| 22 | .425(7.71) | | | | | |
| 23 | | | | | | |
| 24 | .262(3.57) | | | | | |
| 25 | | | | | | |
| 26 | | | | | | |
| 27 | .514(3.13) | | | | -.639(-2.75) | |
| 28 | .431(3.52) | | | | | |
| 29 | | | | | | |
| 30 | | | | | | |

…. – positive coefficient
…. – negative coefficient

---

"wint" – "winter season" indicator variable
"mon" – "Monday" indicator variable
"tues" – "Tuesday" indicator variable
"wed" – "Wednesday" indicator variable
"fr" – "Friday" indicator variable
"sat" – "Saturday" indicator variable

Table B.3 (Continued)

| # | Coefficient (t-ratio) | | | | | | |
|---|---|---|---|---|---|---|---|
| | sund [$X_4$] | wday [$X_4$] | jobend [$X_5$] | peak [$X_5$] | nigh [$X_5$] | nocons [$X_{13}$] | light [$X_{14}$] |
| 1 | | | | | | | |
| 2 | | | | | | | |
| 3 | | | | | | | .416(7.64) |
| 4 | | | | | | | -.515(-2.28) |
| ***5*** | | | | | | | |
| 6 | | | | .535(2.27) | | | |
| 7 | | | | | | | -.509(-2.35) |
| ***8*** | 1.14(2.29) | | | | | | |
| 9 | | | | | | | |
| 10 | | -.660(-2.90) | | | | | |
| 11 | | | | | | | |
| 12 | | -.333(-2.51) | | | | .962(4.98) | |
| 13 | | | | | | | |
| ***14*** | | | | | | | |
| 15 | | | | | | | .693(7.26) |
| 16 | | | | | 1.64(3.67) | | |
| ***17*** | | | | | | | |
| 18 | | | | | | | .745(2.50) |
| 19 | .597(2.39) | | .382(2.07) | | | | |
| ***20*** | | | | | | | |
| 21 | | | | | | | |
| 22 | | | | | | | |
| 23 | | .677(-2.25) | | | | | |
| 24 | | | | | | | |
| 25 | | | | | | | |
| 26 | | | | | | | |
| 27 | | | | | | | |
| 28 | | -.280(-2.03) | | | | | |
| 29 | | -1.08(-3.42) | | | | | |
| 30 | | | | | | | |

…. – positive coefficient
…. – negative coefficient

"sund" – "Sunday" indicator variable
"wday" – "weekday (Monday through Friday)" indicator variable
"jobend" – "evening rush hours: from 16:00 to 19:00" indicator variable[23]
"peak" – "rush hours: 7:00 to 9:00 OR 17:00 to 19:00" indicator variable
"nigh" – "late night hours: 1:00 to 5:00" indicator variable
"nocons" – "no construction at the accident location" indicator variable
"light" – "daylight time OR street lights lit up during dark time" indicator variable

---

[23] We use military 24-hour time everywhere in our research.

Table B.3 (Continued)

| # | Coefficient (t-ratio) | | | | | |
|---|---|---|---|---|---|---|
| | dark [$X_{14}$] | day [$X_{14}$] | precip [$X_{15}$] | snow [$X_{15}$] | dry [$X_{16}$] | slush [$X_{16}$] |
| 1 | | | | | -1.90(-17.7) | |
| 2 | | | | | -2.32(-5.00) | |
| 3 | | | | | -1.20(-20.0) | |
| 4 | | | | | -1.91(-7.22) | |
| ***5*** | | | | | -4.40(-2.44) | |
| 6 | | | | | -1.25(-6.25) | |
| 7 | | | | | -2.08(-9.83) | |
| ***8*** | | | | | -2.57(-7.55) | |
| 9 | -.627(-4.10) | | .466(3.04) | | -2.88(-11.5) | |
| 10 | | | | 1.66(5.37) | -1.72(-7.58) | |
| 11 | | -.599(-2.28) | | | -2.46(-8.85) | |
| 12 | | | .947(5.24) | | -1.38(-7.94) | |
| 13 | | | | | -1.21(-7.05) | |
| ***14*** | | | | 1.50(2.88) | -1.39(-3.66) | |
| 15 | | | .317(2.66) | | -1.55(-12.0) | |
| 16 | | | | | -1.61(-10.5) | |
| ***17*** | | | | | | 1.49(2.29) |
| 18 | | | .670(.318) | | -.941(-2.83) | |
| 19 | | | | | -1.28(-7.36) | |
| ***20*** | | .562(4.11) | .376(2.22) | | -1.01(-6.33) | 2.30(3.00) |
| 21 | | | | | | |
| 22 | | | .468(6.69) | | -1.30(-17.5) | |
| 23 | | | | | -1.70(-7.20) | |
| 24 | | | .241(2.68) | | -1.01(-11.5) | |
| 25 | | | | | -1.74(-8.98) | |
| 26 | -1.58(-2.54) | | | | -1.90(-6.04) | |
| 27 | -1.01(-6.32) | | | | -2.06(-10.7) | |
| 28 | | | | | -1.31(-11.1) | |
| 29 | | .752(-2.06) | | | -1.21(-3.35) | |
| 30 | | | | | -1.93(-6.62) | |

…. – positive coefficient
…. – negative coefficient

---

"dark" – "dark time with no street lights" indicator variable
"day" – "daylight time" indicator variable
"precip" – "precipitation: rain OR snow OR sleet OR hail OR freezing rain" indicator variable
"snow" – "snowing weather" indicator variable
"dry" – "roadway surface is dry" indicator variable
"slush" – "roadway surface is covered by snow/slush" indicator variable

Table B.3 (Continued)

| # | Coefficient (t-ratio) | | | | | |
|---|---|---|---|---|---|---|
| | driv [$X_{17}$] | wall [$X_{17}$] | nojun [$X_{18}$] | ramp [$X_{18}$] | way4 [$X_{18}$] | T [$X_{18}$] |
| 1 | | | .482(4.03) | | | |
| 2 | | | | | | |
| 3 | | | -.300(-3.83) | | | |
| 4 | | | | | | |
| ***5*** | | | | | | |
| 6 | | | | | | |
| 7 | | | .976(3.15) | | | |
| ***8*** | | | | -1.90(-2.27) | | |
| 9 | | | | | | |
| 10 | | | | | | |
| 11 | | | | | | |
| 12 | -.698(-3.66) | | -.533(-3.03) | | | |
| 13 | | | .465(2.62) | | | |
| ***14*** | | | | | | .905(2.19) |
| 15 | | | -.540(-4.09) | | | |
| 16 | | | | | -.462(-2.76) | |
| ***17*** | | | | | | |
| 18 | | | -.560(-2.31) | | | |
| 19 | -.542(-2.99) | | | | | |
| ***20*** | | | | | | |
| 21 | | | | | | |
| 22 | | | | | -.210(-3.76) | |
| 23 | | 1.47(3.75) | | | | |
| 24 | | | | 1.22(5.13) | | |
| 25 | | | .691(3.49) | | | |
| 26 | | | | | | |
| 27 | | 1.31(5.24) | | | | |
| 28 | | | .383(3.17) | | | |
| 29 | | | | | -1.30(-2.85) | |
| 30 | | | | | | |

…. – positive coefficient
…. – negative coefficient

"driv" – "road median is a drivable" indicator variable
"wall" – "road median is a barrier wall" indicator variable
"nojun" – "no road junction at the accident location" indicator variable
"ramp" – "accident location is near or on a ramp" indicator variable
"way4" – "accident location is at a 4-way intersection" indicator variable
"T" – "accident location is at a T-intersection" indicator variable

Table B.3 (Continued)

| # | Coefficient (t-ratio) | | | | | |
|---|---|---|---|---|---|---|
| | curve [$X_{19}$] | sg [$X_{19}$] | sl [$X_{19}$] | str [$X_{19}$] | cl [$X_{19}$] | hl5 [$X_{22}$] |
| 1 | | | | | | |
| 2 | | | | | | |
| 3 | .892(14.9) | | | | | |
| 4 | | | -1.08(-4.10) | | | |
| ***5*** | | | | | | |
| 6 | | | | | .892(3.42) | |
| 7 | | | | | | .499(2.13) |
| ***8*** | | | | | | |
| 9 | | | -.482(-3.26) | | | .421(2.24) |
| 10 | | | | | | |
| 11 | | | | | | |
| 12 | | | | -.983(-5.73) | | |
| 13 | | | | | | |
| ***14*** | | | | | | |
| 15 | | | | -.959(-9.56) | | |
| 16 | | | | | | |
| ***17*** | | | | | | -1.92(-2.49) |
| 18 | .582(2.18) | | | | | |
| 19 | .865(3.47) | | | | | |
| ***20*** | | | -1.14(-8.21) | | | |
| 21 | | | | | | |
| 22 | .884(9.43) | | | | | |
| 23 | | | | | | |
| 24 | .896(11.5) | | | | | |
| 25 | | | | | | -.463(-2.05) |
| 26 | | | | | | -.856(-2.00) |
| 27 | | | | -1.35(-7.54) | | |
| 28 | | .608(3.18) | | | | -.283(-2.32) |
| 29 | | | | | | |
| 30 | .981(3.22) | | | | | |

…. – positive coefficient
…. – negative coefficient

"curve" – "road is at curve" indicator variable
"sg" – "road is straight AND at grade" indicator variable
"sl" – "road is straight AND level" indicator variable
"str" – "road is straight" indicator variable
"cl" – "road is at-curve AND level" indicator variable
"hl5" – "help arrived in 5 minutes or less after the crash" indicator variable

Table B.3 (Continued)

| # | Coefficient (t-ratio) | | | | | |
|---|---|---|---|---|---|---|
| | hl10 [$X_{22}$] | hg30 [$X_{22}$] | hg60 [$X_{22}$] | car [$X_{25}$] | heavy [$X_{25}$] | moto [$X_{25}$] |
| 1 | | | | | | |
| 2 | | | | | | |
| 3 | .141(2.41) | | | .220(3.87) | | |
| 4 | | | | | | |
| ***5*** | | | | | | |
| 6 | | | | | | |
| 7 | | | | | | |
| ***8*** | | | | | -2.32(-5.44) | |
| 9 | | | | | | |
| 10 | | -.736(-2.16) | | | | |
| 11 | | | | | -1.86(-5.15) | |
| 12 | | -.393(-2.12) | | | | |
| 13 | | | | | | |
| ***14*** | | | | | | |
| 15 | | | | | | |
| 16 | | | | | | |
| ***17*** | | | | | | |
| 18 | | | | | | |
| 19 | | | | | | |
| ***20*** | | | | | | |
| 21 | | | | | | |
| 22 | | | | | | |
| 23 | | | | | -.936(-3.74) | |
| 24 | | | | | -1.44(-4.96) | |
| 25 | | | | .411(2.06) | | |
| 26 | | | | | | |
| 27 | | | -.657(-2.03) | | | |
| 28 | | | | | | |
| 29 | | | | | -1.18(-2.88) | |
| 30 | | | | | | 2.49(5.38) |

…. – positive coefficient
…. – negative coefficient

---

"hl10" – "help arrived in 10 minutes or less after the crash" indicator variable
"hg30" – "help arrived in more than 30 minutes after the crash" indicator variable
"hg60" – "help arrived in more than 1 hour after the crash" indicator variable
"car" – "the vehicle at fault is a car" indicator variable
"heavy" – "the vehicle at fault is a truck or a tractor" indicator variable
"moto" – "the vehicle at fault is a motorcycle" indicator variable

Table B.3 (Continued)

| # | Coefficient (t-ratio) | | | | | |
|---|---|---|---|---|---|---|
| | pickup [$X_{25}$] | vage [$X_{26}$] | voldg [$X_{26}$] | v7g [$X_{26}$] | $X_{27}$ | Ind [$X_{28}$] |
| 1 | | | | | | |
| 2 | | | | | | |
| 3 | | | .256(4.66) | | .131(4.84) | -.353(-2.79) |
| 4 | | | | | | |
| ***5*** | | | | | -2.47(-2.15) | |
| 6 | | | | | | |
| 7 | | | | | | -.667(-3.22) |
| ***8*** | | | | | -.586(-3.35) | |
| 9 | | | | | | |
| 10 | | | | | | |
| 11 | | | | | .273(2.08) | |
| 12 | | | | | | |
| 13 | | | | | | |
| ***14*** | .942(2.34) | | | | | |
| 15 | | .0411(4.64) | | | | |
| 16 | | | | | | |
| ***17*** | | | | | | |
| 18 | | | | | | |
| 19 | | | | | | |
| ***20*** | | | | | | |
| 21 | | | | | | |
| 22 | | .0255(5.39) | | | | |
| 23 | | 0.0629(3.29) | | | | |
| 24 | | | | | | |
| 25 | | | | | -.302(-2.24) | |
| 26 | | | | | | |
| 27 | | | | | | |
| 28 | | | | | | |
| 29 | | | | | | |
| 30 | | | | -.724(-2.38) | .320(2.34) | |

…. – positive coefficient
…. – negative coefficient

"pickup" – "the vehicle at fault is a pickup" indicator variable
"vage" – "age (in years) of the vehicle at fault" quantitative variable
"voldg" – "the vehicle at fault is more than 7 years old" indicator variable
"v7g" – "age of the vehicle at fault is > 3 and ≤ 7 years" indicator variable
"$X_{27}$" – "number of occupants in the vehicle at fault" quantitative variable
"Ind" – "license state of the vehicle at fault is Indiana" indicator variable

Table B.3 (Continued)

| # | Coefficient (t-ratio) | | | | | | |
|---|---|---|---|---|---|---|---|
| | othUS [$X_{28}$] | neighs [$X_{28}$] | w2 [$X_{30}$] | ln2 [$X_{30}$] | r22 [$X_{30}$] | rmu2 [$X_{30}$] | rmu22 [$X_{30}$] |
| 1 | | | | | | | |
| 2 | | | | | | | |
| 3 | | | | | | | |
| 4 | | | | | | | |
| ***5*** | | | | | | 2.34(2.11) | |
| 6 | | | | | | 2.28(3.65) | |
| 7 | | | | | | | |
| ***8*** | | | | | | | |
| 9 | | | | | | | |
| 10 | | | | | | | |
| 11 | | | | | | | |
| 12 | | | | | | | |
| 13 | | | | | | | |
| ***14*** | | | | | | | |
| 15 | | | | | -.420(-3.41) | | |
| 16 | | | | | | | -2.07(-2.89) |
| ***17*** | | | | | | | |
| 18 | | | | | | | |
| 19 | .929(1.97) | | .720(2.27) | | | | |
| ***20*** | | | | | | | |
| 21 | | | | | | | |
| 22 | | | | .312(5.83) | | | |
| 23 | | | | | | | |
| 24 | | | | | | | |
| 25 | | | | | | | |
| 26 | | | | | | | |
| 27 | | | | | | | |
| 28 | | .653(3.66) | | | | .357(2.61) | |
| 29 | | | | | | | |
| 30 | | | | | | | |

…. – positive coefficient
…. – negative coefficient

"othUS" – "license state of the vehicle at fault is a U.S. state except Indiana and its neighboring states (IL, KY, OH, MI)" indicator variable
"neighs" – "license state of the vehicle at fault is an Indiana's neighboring state (IL, KY, OH, MI)" indicator variable
"w2" – "road traveled by the vehicle at fault is two-way" indicator variable
"ln2" – "road traveled by the vehicle at fault is two-lane" indicator variable
"r22" – "road traveled by the vehicle at fault is two-lane AND two-way" indicator variable
"rmu2" – "road traveled by the vehicle at fault is multi-lane AND undivided two-way" indicator variable
"rmu22" – "road traveled by the vehicle at fault is multi-lane AND undivided two-way left" indicator variable

Table B.3 (Continued)

| # | Coefficient (t-ratio) | | | | | |
|---|---|---|---|---|---|---|
| | priv [$X_{30}$] | stopsig [$X_{31}$] | nosig [$X_{31}$] | nopass [$X_{31}$] | lncontr [$X_{31}$] | sign [$X_{31}$] |
| 1 | | -.441(-2.98) | | | | |
| 2 | | | | | | |
| 3 | | | -.519(-9.34) | | | |
| 4 | | | | 1.55(4.73) | | |
| ***5*** | | | | | | |
| 6 | | | | .613(2.25) | | |
| 7 | | | | | | |
| ***8*** | | | | | | 1.30(3.88) |
| 9 | | | -.600(-2.97) | | | |
| 10 | | | | | .674(2.72) | |
| 11 | | | | | | |
| 12 | | | | | | |
| 13 | | -1.54(-3.25) | | | | |
| ***14*** | | | | | | |
| 15 | | .996(2.67) | | | | |
| 16 | | -1.32(-2.58) | | | | |
| ***17*** | | | | | | |
| 18 | | | -.665(-2.63) | | | |
| 19 | | | | | | |
| ***20*** | | | | | | |
| 21 | | | | | | |
| 22 | | -.619(-7.36) | | | | |
| 23 | | | | | | |
| 24 | -.599(-2.09) | | | | | .175(2.51) |
| 25 | | -1.66(-2.75) | | | | |
| 26 | | | | | | |
| 27 | | | -.351(-2.04) | | | |
| 28 | | | | | | |
| 29 | | | | | | |
| 30 | | | | | | |

…. – positive coefficient
…. – negative coefficient

"priv" – "road traveled by the vehicle at fault is a private drive" indicator variable
"stopsig" – "traffic control device for the vehicle at fault is a «stop sign»" indicator variable
"nosig" – "no any traffic control device for the vehicle at fault" indicator variable
"nopass" – "traffic control device for the vehicle at fault is a «no passing zone»" indicator variable
"lncontr" – "traffic control device for the vehicle at fault is a «lane control»" indicator variable
"sign" – "traffic control device for the vehicle at fault is any traffic sign" indicator Variable

Table B.3 (Continued)

| # | Coefficient (t-ratio) | | | | | |
|---|---|---|---|---|---|---|
| | $X_{33}$ | $X_{34}$ | age4 [$X_{34}$] | $X_{35}$ | maxpass [$X_{27}$] | mm [$X_{35}$] |
| 1 | | | -.0183(-5.71) | | .132(3.13) | |
| 2 | | -.0581(-8.00) | | | | |
| 3 | | -.0293(-13.1) | | -.248(-4.29) | | |
| 4 | | -.0321(-3.45) | | | | |
| ***5*** | | | | | 2.15(2.61) | |
| 6 | | -.0381(-8.19) | | | | |
| 7 | | | | | | |
| ***8*** | | | | | | |
| 9 | | -.0157(-4.01) | | | | |
| 10 | | | | | | .569(2.72) |
| 11 | | | | | | |
| 12 | | | | | | |
| 13 | 2.74(3.85) | -.0143(-2.75) | | | | |
| ***14*** | | | -1.84(-2.84) | | | |
| 15 | | -.0307(-9.37) | | | | |
| 16 | | | | | | |
| ***17*** | | | | | | |
| 18 | | -.0313(3.78) | | | | |
| 19 | | -.0201(-3.52) | | | | |
| ***20*** | | -.358(-7.93) | | -.563(-3.69) | | |
| 21 | | | | | | |
| 22 | | -.0241(-9.43) | | -.211(-3.99) | | |
| 23 | | -.0335(-3.79) | | | | |
| 24 | | -.0367(-14.3) | | -.422(-5.70) | | |
| 25 | | | | | | .498(2.58) |
| 26 | | | | | | |
| 27 | | -.0263(-4.90) | | .333(-2.01) | | |
| 28 | | -.0179(-4.69) | | | | |
| 29 | | | | | | |
| 30 | | -.0203(-2.25) | | | | |

…. – positive coefficient
…. – negative coefficient

---

"$X_{33}$" – "at least one of the vehicles involved was on fire" indicator variable
"$X_{34}$" – "age (in years) of the driver at fault" quantitative variable
"age4" – "age of the driver at fault is ≥ 40 and < 50" indicator variable
"$X_{35}$" – "gender of the driver at fault: 1 – female, 0 – male" indicator variable
"maxpass" – "the largest number of occupants in all vehicles involved" indicator variable
"mm" – "two male drivers involved into a two-vehicle accident" indicator variable

Table B.4 Binary logit models for 2006 accident causation

| # | Model name | | | Log-likelihood | | $R^2$ | Coefficient (t-ratio) | |
|---|---|---|---|---|---|---|---|---|
| | | | | model | restricted* | | $X_{29}$ | constant |
| 1** | County road | rural | (car/SUV)+(car/SUV) | -342.72 | -41.06 | .164 | | |
| 2a | County road | rural | (car)+(truck) | -18.849 | -26.763 | .296 | | -5.12(-5.10) |
| **2b** | County road | rural | (SUV)+(truck) | -5.0592 | -10.270 | .507 | | |
| 3 | County road | rural | one vehicle | -1204.3 | -1438.2 | .163 | | |
| 4 | County road | urban | (car/SUV)+(car/SUV) | -227.21 | -269.91 | .158 | | -2.63(-12.2) |
| **5** | County road | urban | (car/SUV)+(truck) | -15.788 | -18.893 | .164 | | -3.54(-4.94) |
| 6 | County road | urban | one vehicle | -256.10 | -300.36 | .147 | | |
| **7a** | Interstate | rural | (car)+(car) | -85.811 | -118.56 | .276 | | -1.50(-3.53) |
| **7b** | Interstate | rural | (car)+(SUV) | -127.43 | -178.37 | .286 | | -2.14(-3.06) |
| **7c** | Interstate | rural | (SUV)+(SUV) | -50.205 | -78.501 | .360 | | -1.98(-4.10) |
| **8** | Interstate | rural | (car/SUV)+(truck) | -222.35 | -287.72 | .227 | | -1.08(-2.67) |
| 9 | Interstate | rural | one vehicle | -918.40 | -1435.7 | .360 | -.0439(-5.72) | 3.59(6.96) |
| **10** | Interstate | urban | (car/SUV)+(car/SUV) | -508.16 | -603.72 | .158 | | -2.77(-6.43) |
| 11a | Interstate | urban | (car)+(truck) | -116.08 | -159.73 | .273 | | -1.31(-3.03) |
| **11b** | Interstate | urban | (SUV)+(truck) | -32.424 | -65.943 | .508 | | -4.48(-3.00) |
| 12 | Interstate | urban | one vehicle | -136.07 | -169.56 | .198 | | -1.04(-2.03) |
| 13 | State route | rural | (car/SUV)+(car/SUV) | -774.29 | -863.45 | .103 | | -1.68(-7.77) |
| **14** | State route | rural | (car/SUV)+(truck) | -92.951 | -114.16 | .186 | | -2.63(-6.19) |
| 15 | State route | rural | one vehicle | -1616.3 | -2084.9 | .225 | -.0373(-5.34) | 1.07(2.68) |
| 16 | State route | urban | (car/SUV)+(car/SUV) | -850.84 | -952.04 | .106 | .0277(3.47) | -2.50(-6.45) |
| **17** | State route | urban | (car/SUV)+(truck) | -81.158 | -93.656 | .133 | | -3.76(-11.3) |
| 18 | State route | urban | one vehicle | -358.35 | -441.38 | .188 | | .615(2.03) |
| **19** | City street | rural | (car/SUV)+(car/SUV) | -211.60 | -227.51 | .070 | | -3.41(-18.0) |
| 20a | City street | rural | (car)+(truck) | -13.637 | -16.670 | .182 | | -4.34(-4.32) |
| **20b** | City street | rural | (SUV)+(truck) | -11.753 | -11.753 | .000 | | -2.89(-4.87) |
| 21 | City street | rural | one vehicle | -597.08 | -694.78 | .141 | | -1.63(-6.62) |
| 22 | City street | urban | (car/SUV)+(car/SUV) | -2265.6 | -2584.8 | .123 | | -3.02(-12.6) |
| 23a | City street | urban | (car)+(truck) | -93.362 | -115.47 | .191 | | -5.49(-10.0) |
| **23b** | City street | urban | (SUV)+(truck) | -29.107 | -40.704 | .285 | | -6.80(-6.78) |
| 24 | City street | urban | one vehicle | -2597.7 | -3024.6 | .141 | | -.556(-4.78) |
| 25 | US route | rural | (car/SUV)+(car/SUV) | -456.42 | -499.97 | .087 | | |
| **26** | US route | rural | (car/SUV)+(truck) | -80.524 | -102.56 | .215 | | -4.55(-8.95) |
| 27 | US route | rural | one vehicle | -708.44 | -977.40 | .275 | | -1.60(-5.96) |
| **28** | US route | urban | (car/SUV)+(car/SUV) | -834.94 | -901.81 | .074 | | -1.68(-6.19) |
| 29a | US route | urban | (car)+(truck) | -46.549 | -51.986 | .105 | | -3.60(-10.6) |
| **29b** | US route | urban | (SUV)+(truck) | -11.446 | -11.446 | .000 | | -4.72(-6.64) |
| 30 | US route | urban | one vehicle | -250.31 | -315.51 | .207 | | -1.71(-5.00) |

.... – positive coefficient .... – negative coefficient

* – restricted log-likelihood found by setting all coefficients except intercepts to zero

** – models are estimated by using procedure A on page 32, except the models marked by bold numbers and estimated by using procedure B on page 33

"$X_{29}$" – "posted speed limit (if the same for all vehicles involved)" quantitative variable

"constant" – "constant term (intercept)" quantitative variable

Table B.4 (Continued)

| # | Coefficient (t-ratio) | | | | | |
|---|---|---|---|---|---|---|
| | wint [$X_3$] | fall [$X_3$] | tues [$X_4$] | wed[$X_4$] | thday [$X_4$] | peak [$X_5$] |
| ***1***** | | | | | | |
| 2a | | | | | | |
| ***2b*** | | | | | | |
| 3 | .460(4.29) | | | | | |
| 4 | .747(2.74) | | .932(3.21) | | | |
| ***5*** | | | | | | |
| 6 | | | | | | |
| ***7a*** | | | | | | |
| ***7b*** | .772(2.05) | | | | | |
| ***7c*** | | | | | 1.56(2.61) | |
| ***8*** | | | | | | |
| 9 | .590(4.59) | | | .345(2.22) | | |
| ***10*** | | | | | | -.775(-4.11) |
| 11a | | | | | | |
| ***11b*** | | | | | | |
| 12 | | | | | | |
| 13 | | | | | | |
| ***14*** | | | | | -1.66(-2.18) | |
| 15 | .540(5.17) | | | | | |
| 16 | | | | | | |
| ***17*** | | | | 1.08(2.32) | | |
| 18 | | | .729(3.00) | | | |
| ***19*** | | | | | | |
| 20a | 2.67(2.25) | | | | | |
| ***20b*** | | | | | | |
| 21 | | -.432(-2.34) | | | | |
| 22 | .229(2.39) | | | | | |
| 23a | | | | | | |
| ***23b*** | 2.38(3.01) | | | | | |
| 24 | .178(2.15) | | | | | |
| 25 | | | | | | |
| ***26*** | | | | | | |
| 27 | .650(4.32) | | | | | |
| ***28*** | | | | | | |
| 29a | | | | | | |
| ***29b*** | | | | | | |
| 30 | | -.727(-2.23) | | | | |

…. – positive coefficient …. – negative coefficient.

"wint" – "winter season" indicator variable
"fall" – "fall season" indicator variable
"tues" – "Tuesday" indicator variable
"wed" – "Wednesday" indicator variable
"thday" – "Thursday" indicator variable
"peak" – "rush hours: 7:00 to 9:00 OR 17:00 to 19:00" indicator variable

Table B.4 (Continued)

| # | Coefficient (t-ratio) | | | | | |
|---|---|---|---|---|---|---|
| | nigh [$X_5$] | dayt [$X_5$] | lunch [$X_5$] | light [$X_{14}$] | dark [$X_{14}$] | day [$X_{14}$] |
| ***1***** | | | | | | |
| 2a | | | | | | |
| ***2b*** | | | | | | |
| 3 | | | | | | .368(3.59) |
| 4 | | | | | | |
| ***5*** | | | | | | |
| 6 | | | | | | |
| ***7a*** | | | | | | |
| ***7b*** | | | | | | |
| ***7c*** | | | | | | |
| ***8*** | | | | | | |
| 9 | -.825(-4.07) | | | | | |
| ***10*** | | | | | | -.381(-2.15) |
| 11a | 1.48(2.73) | | | | | |
| ***11b*** | | | | | | -2.64(-3.47) |
| 12 | | | | | | |
| 13 | | | | | | |
| ***14*** | | | 1.06(2.36) | | | |
| 15 | | | | .792(8.08) | | |
| 16 | 1.41(3.56) | | | | | |
| ***17*** | | | | | | |
| 18 | | .511(2.37) | | | | |
| ***19*** | | | | | | |
| 20a | | | | | | |
| ***20b*** | | | | | | |
| 21 | | | | .527(3.37) | | |
| 22 | | -.216(-2.45) | | | | |
| 23a | | | | | | |
| ***23b*** | | | | | | |
| 24 | | | | | | |
| 25 | | | | | | |
| ***26*** | | | | | | |
| 27 | | | | | -.721(-4.92) | |
| ***28*** | | | | -.609(-2.91) | | |
| 29a | | | | | | |
| ***29b*** | | | | | | |
| 30 | | | | | | |

…. – positive coefficient …. – negative coefficient.

"nigh" – "late night hours: from 1:00 to 5:00" indicator variable
"dayt" – "day hours: from 9:00 to 17:00" indicator variable
"lunch" – "lunch hours: 11:00 to 14:00" indicator variable
"light" – "daylight time OR street lights lit up during dark time" indicator variable
"dark" – "dark time with no street lights" indicator variable
"day" – "daylight time" indicator variable

Table B.4 (Continued)

| # | Coefficient (t-ratio) | | | | | |
|---|---|---|---|---|---|---|
| | darklamp [$X_{14}$] | precip [$X_{15}$] | snow [$X_{15}$] | dry [$X_{16}$] | slush [$X_{16}$] | ice [$X_{16}$] |
| 1** | | | | -1.88(-8.85) | | |
| 2a | | 3.68(3.29) | | | | |
| ***2b*** | | | | | | |
| 3 | | | | -1.28(-12.6) | | |
| 4 | | | | -1.90(-6.62) | | |
| ***5*** | | | | | | |
| 6 | | | 1.34(2.95) | -.730(-3.18) | | |
| ***7a*** | | | | -2.77(-6.39) | | |
| ***7b*** | 1.42(3.10) | | | -2.34(-7.03) | | |
| ***7c*** | | | 3.49(3.00) | -2.09(-3.70) | | |
| ***8*** | | 1.02(2.34) | | -1.37(-3.04) | | |
| 9 | | .753(4.35) | | -2.57(-12.4) | | |
| ***10*** | | .718(2.30) | | -1.21(-3.91) | | |
| 11a | | | | -2.34(-6.59) | | |
| ***11b*** | | | 6.96(4.53) | | | |
| 12 | | 1.96(5.97) | | | | |
| 13 | | | | -1.70(-11.7) | | |
| ***14*** | | | | -1.44(-3.39) | | |
| 15 | | .342(2.56) | | -1.39(-10.2) | | |
| 16 | | | | -1.62(-11.4) | | |
| ***17*** | | | | | | 4.24(3.41) |
| 18 | | | | -1.46(-7.22) | | |
| ***19*** | | 1.13(3.93) | | | | |
| 20a | | | | | | |
| ***20b*** | | | | | | |
| 21 | | | .904(2.42) | | 1.22(3.29) | |
| 22 | | .305(2.44) | | -1.58(-12.0) | | |
| 23a | | | | | | |
| ***23b*** | | | | | | |
| 24 | | | | -.985(-13.3) | | |
| 25 | | | | -1.57(-8.64) | | |
| ***26*** | | 2.44(5.48) | | | | |
| 27 | | .468(2.48) | | -1.87(-8.86) | | |
| ***28*** | | | | -1.41(-9.68) | | |
| 29a | | | | | 2.76(2.62) | |
| ***29b*** | | | | | | |
| 30 | | 1.99(8.17) | | | | |

…. – positive coefficient    …. – negative coefficient.

"darklamp" – "dark time with street lights on" indicator variable
"precip" – "precipitation: rain OR snow OR sleet OR hail OR freezing rain" indicator variable
"snow" – "snowing weather" indicator variable
"dry" – "roadway surface is dry" indicator variable
"slush" – "roadway surface is covered by snow/slush" indicator variable
"ice" – "roadway surface is icy" indicator variable

Table B.4 (Continued)

| # | Coefficient (t-ratio) | | | | | |
|---|---|---|---|---|---|---|
| | driv [$X_{17}$] | wall [$X_{17}$] | nojun [$X_{18}$] | ramp [$X_{18}$] | way4 [$X_{18}$] | T [$X_{18}$] |
| ***1***** | | | | | -.786(-2.54) | |
| 2a | | | | | | |
| ***2b*** | | | | | | |
| 3 | | 1.64(2.35) | -.639(-5.53) | | | |
| 4 | | | | | | |
| ***5*** | | | | | | |
| 6 | | 2.33(2.29) | | | | .729(-3.18) |
| ***7a*** | | | | | | |
| ***7b*** | | | 1.66(2.82) | | | |
| ***7c*** | | | | | | |
| ***8*** | | | | | | |
| 9 | | .302(2.18) | -1.20(-5.30) | | | |
| ***10*** | | .637(2.64) | | | | |
| 11a | | | | | | |
| ***11b*** | | 3.16(2.10) | | | | |
| 12 | | | | | | |
| 13 | | | .357(2.40) | | | |
| ***14*** | 1.17(2.45) | | | | | |
| 15 | | | | .990(2.36) | | |
| 16 | | | | | | |
| ***17*** | | | | | | |
| 18 | | | | | | |
| ***19*** | | | | | | |
| 20a | | | | | | |
| ***20b*** | | | | | | |
| 21 | | | | 1.25(2.10) | | |
| 22 | .180(2.01) | | | | -.234(-2.52) | |
| 23a | | | | | | 1.31(2.72) |
| ***23b*** | | | | | | |
| 24 | | | | | | .401(4.14) |
| 25 | | | | | | |
| ***26*** | | | | | | |
| 27 | -.765(-3.56) | | | | | |
| ***28*** | | | .307(2.13) | | | |
| 29a | | | | | | |
| ***29b*** | | | | | | |
| 30 | | | | | | |

…. – positive coefficient      …. – negative coefficient.

"driv" – "road median is a drivable" indicator variable
"wall" – "road median is a barrier wall" indicator variable
"nojun" – "no road junction at the accident location" indicator variable
"ramp" – "accident location is near or on a ramp" indicator variable
"way4" – "accident location is at a 4-way intersection" indicator variable
"T" – "accident location is at a T-intersection" indicator variable

Table B.4 (Continued)

| # | Coefficient (t-ratio) | | | | | |
|---|---|---|---|---|---|---|
| | Y [$X_{18}$] | int [$X_{18}$] | curve [$X_{19}$] | sg [$X_{19}$] | sl [$X_{19}$] | str [$X_{19}$] |
| ***1**** | | | | | | |
| 2a | | | | | | |
| ***2b*** | | | | | | |
| 3 | | | 1.05(9.61) | | | |
| 4 | | | | | | |
| ***5*** | | | | | | |
| 6 | | | | | | -.857(-3.73) |
| ***7a*** | | | | | | |
| ***7b*** | | | | | | |
| ***7c*** | | | | | | |
| ***8*** | | | | | | |
| 9 | | | | | -.418(-3.33) | |
| ***10*** | | | | | | |
| 11a | | | | | | |
| ***11b*** | | | | | | |
| 12 | | | | | | -1.62(-4.84) |
| 13 | | | | | | |
| ***14*** | 3.27(3.48) | | | 1.17(2.71) | | |
| 15 | | | | | | -1.05(-10.6) |
| 16 | | | | | | |
| ***17*** | | | | | | |
| 18 | | | | | | -1.18(-5.70) |
| ***19*** | | | 1.34(3.18) | | | |
| 20a | | | | | | |
| ***20b*** | | | | | | |
| 21 | | | 1.15(7.43) | | | |
| 22 | | | .898(5.40) | | | |
| 23a | | | 1.90(2.84) | | | |
| ***23b*** | | | | | | |
| 24 | | | 1.11(13.9) | | | |
| 25 | | | | | | -.640(-2.84) |
| ***26*** | | | | | | |
| 27 | 2.15(3.85) | | 1.00(6.70) | | | |
| ***28*** | | | | | | |
| 29a | | | | | | |
| ***29b*** | | | | | | |
| 30 | | 1.45(2.24) | .813(2.91) | | | |

…. – positive coefficient …. – negative coefficient.

"Y" – "accident location is at a Y-intersection" indicator variable
"int" – "accident location is near or on an interchange" indicator variable
"curve" – "road is at curve" indicator variable
"sg" – "road is straight AND at grade" indicator variable
"sl" – "road is straight AND level" indicator variable
"str" – "road is straight" indicator variable

Table B.4 (Continued)

| # | Coefficient (t-ratio) | | | | | |
|---|---|---|---|---|---|---|
| | cl [$X_{19}$] | cg [$X_{19}$] | lev [$X_{19}$] | grd [$X_{19}$] | hl5 [$X_{22}$] | hl10 [$X_{22}$] |
| ***1**** | | | | | | |
| 2a | | | | | | |
| ***2b*** | | | | | | |
| 3 | | | | | | |
| 4 | 1.32(3.09) | | | | | |
| ***5*** | | | | | | |
| 6 | | | | | | |
| ***7a*** | 1.39(2.17) | | | | | |
| ***7b*** | 1.61(2.93) | | | | | |
| ***7c*** | | | | | | |
| ***8*** | | | | | | |
| 9 | | | | | .324(2.06) | |
| ***10*** | | | | | | |
| 11a | | | | | | |
| ***11b*** | | | | | | |
| 12 | | | | | | |
| 13 | | .937(3.86) | | | | |
| ***14*** | | | | | | |
| 15 | | | | | | |
| 16 | | | -.490(-2.92) | | | |
| ***17*** | | | | 1.11(2.28) | | |
| 18 | | | | | | |
| ***19*** | | | | | | |
| 20a | | | | | | |
| ***20b*** | | | | | | |
| 21 | | | | | | .403(2.61) |
| 22 | | | | | | |
| 23a | | | | | | |
| ***23b*** | | | | | | |
| 24 | | | | | | |
| 25 | | | | | | |
| ***26*** | | | | | | .957(2.04) |
| 27 | | | | | | .630(4.39) |
| ***28*** | | | | | | |
| 29a | | | | | | |
| ***29b*** | | | | | | |
| 30 | | | | | | |

…. – positive coefficient …. – negative coefficient.

"cl" – "road is at-curve AND level" indicator variable
"cg" – "road is at-curve AND grade" indicator variable
"lev" – "road is 'at-curve or straight' AND level" indicator variable
"grd" – "road is 'at-curve or straight' AND grade" indicator variable
"hl5" – "help arrived in 5 minutes or less after the crash" indicator variable
"hl10" – "help arrived in 10 minutes or less after the crash" indicator variable

Table B.4 (Continued)

| # | Coefficient (t-ratio) | | | | | |
|---|---|---|---|---|---|---|
| | car [$X_{25}$] | heavy [$X_{25}$] | van [$X_{25}$] | truck3 [$X_{25}$] | trac1 [$X_{25}$] | vage [$X_{26}$] |
| ***1***** | | | | | | |
| 2a | | | | | | |
| ***2b*** | | | | | | |
| 3 | | | | | | |
| 4 | | | | | | |
| ***5*** | | | 2.85(2.01) | | | |
| 6 | | | | | | |
| ***7a*** | | | | | | |
| ***7b*** | | | | | | |
| ***7c*** | | | | | | |
| ***8*** | | -1.18(-4.12) | | | | |
| 9 | | -.509(-2.61) | | | | |
| ***10*** | | | | | | |
| 11a | | | | | -1.28(-2.77) | |
| ***11b*** | | | | | | |
| 12 | | | | | | |
| 13 | | | | | | |
| ***14*** | | | | 1.33(1.98) | | |
| 15 | | | -.733(-2.65) | | | .0184(2.03) |
| 16 | | | | | | |
| ***17*** | | | | | | |
| 18 | | | | | | |
| ***19*** | | | | | | |
| 20a | | | | | | |
| ***20b*** | | | | | | |
| 21 | | | | | | |
| 22 | | | | | | .0277(3.44) |
| 23a | 1.71(3.17) | | | | | |
| ***23b*** | | | | | | |
| 24 | | -1.17(-3.74) | | | | |
| 25 | | | | | | |
| ***26*** | | | | | | |
| 27 | | | | | 1.04(3.77) | |
| ***28*** | | | | | | |
| 29a | | | | | | |
| ***29b*** | | | | | | |
| 30 | | | | | | |

…. – positive coefficient …. – negative coefficient.

"car" – “the vehicle at fault is a car” indicator variable
"heavy" – “the vehicle at fault is a truck or a tractor” indicator variable
"van" – “the vehicle at fault is a van” indicator variable
"track3" – “the vehicle at fault is a truck (single 3 or more axes)” indicator variable
"trac1" – “the vehicle at fault is a tractor/one semi trailer” indicator variable
"vage" – “age (in years) of the vehicle at fault” quantitative variable

Table B.4 (Continued)

| # | Coefficient (t-ratio) | | | | | |
|---|---|---|---|---|---|---|
| | voldg [$X_{26}$] | $X_{27}$ | Ind [$X_{28}$] | neighs [$X_{28}$] | neighc [$X_{28}$] | w2 [$X_{30}$] |
| ***1***** | | | | | | |
| 2a | | | | | | |
| ***2b*** | | | | 4.40(2.10) | | |
| 3 | | | | | | |
| 4 | | | | | | |
| ***5*** | | | | | | |
| 6 | .522(2.49) | | | | | |
| ***7a*** | | | | | | |
| ***7b*** | | -.527(-2.07) | | | | |
| ***7c*** | 1.33(2.42) | | | | | |
| ***8*** | | | | | | |
| 9 | | | | | | |
| ***10*** | | | | | | |
| 11a | | | .910(2.08) | | | |
| ***11b*** | | | | | | |
| 12 | | | | | | |
| 13 | | | | | | |
| ***14*** | | | | | | |
| 15 | | | | | | |
| 16 | | | | | | |
| ***17*** | | | | | | |
| 18 | | | | | | |
| ***19*** | | | | | | |
| 20a | | | | | | |
| ***20b*** | | | | | | |
| 21 | | | | | | |
| 22 | | | | | | .601(3.57) |
| 23a | | | | | | |
| ***23b*** | | | | | | |
| 24 | | .0805(3.26) | | | | |
| 25 | | | -.734(-3.35) | | | |
| ***26*** | | | | | | |
| 27 | | | | | | |
| ***28*** | | | | | | |
| 29a | | | | | | |
| ***29b*** | | | | | | |
| 30 | | | | .959(2.98) | | |

.... – positive coefficient .... – negative coefficient.

"voldg" – "the vehicle at fault is more than 7 years old" indicator variable
"$X_{27}$" – "number of occupants in the vehicle at fault" quantitative variable
"Ind" – "license state of the vehicle at fault is Indiana" indicator variable
"neighs" – "license state of the vehicle at fault is Indiana's neighboring state (IL, KY, OH, MI)" indicator variable
"neighc" – "license state of the vehicle at fault is from Canada, Mexico, or US territories" indicator variable
"w2" – "road traveled by the vehicle at fault is two-way" indicator variable

Table B.4 (Continued)

| # | Coefficient (t-ratio) | | | | | |
|---|---|---|---|---|---|---|
| | r11 [$X_{30}$] | r21 [$X_{30}$] | rmd2 [$X_{30}$] | priv [$X_{30}$] | w1 [$X_{30}$] | stopsig [$X_{31}$] |
| ***1**** | | | | | | -.658(-2.23) |
| 2a | | | | | | |
| ***2b*** | | | | | | |
| 3 | | | | | | |
| 4 | | | | | | |
| ***5*** | | | | | | |
| 6 | | | | | | |
| ***7a*** | | | | | -2.67(-2.49) | |
| ***7b*** | | | | | | |
| ***7c*** | | | | | | |
| ***8*** | | | | | -1.63(-2.59) | |
| 9 | | | | | | |
| ***10*** | | | .664(2.31) | | | |
| 11a | | | | | | |
| ***11b*** | 4.12(2.51) | | | | | |
| 12 | | | | | | |
| 13 | | | | | | |
| ***14*** | | | | | | |
| 15 | | | | | | |
| 16 | | | | | | |
| ***17*** | | | | | | |
| 18 | | | | | | |
| ***19*** | | | | | | -1.64(-2.26) |
| 20a | | | | | | |
| ***20b*** | | | | | | |
| 21 | | | | | | |
| 22 | | | | | | -.458(-3.12) |
| 23a | | | | | | |
| ***23b*** | | | | | | |
| 24 | | | | -.611(-3.14) | | |
| 25 | | | | | | |
| ***26*** | | | | | | |
| 27 | | | | | | |
| ***28*** | | | | | | |
| 29a | | 2.22(2.22) | | | | |
| ***29b*** | | | | | | |
| 30 | | | | | | |

…. – positive coefficient …. – negative coefficient.

"r11" – "road traveled by the vehicle at fault is one-lane & one-way" indicator var.
"r21" – "road traveled by the veh. at fault is two-lanes AND one-way" indic. var.
"rmd2" – "road traveled by the vehicle at fault is multy-lane divided 3 or more AND two-way" indicator variable
"priv" – "road traveled by the vehicle at fault is private drive" indicator variable
"w1" – "road traveled by the vehicle at fault is one-way" indicator variable
"stopsig" – "traffic control device for the vehicle at fault is «stop sign»" indicator var.

Table B.4 (Continued)

| # | Coefficient (t-ratio) | | | | | |
|---|---|---|---|---|---|---|
| | nopass [$X_{31}$] | lncontr [$X_{31}$] | fl [$X_{31}$] | $X_{33}$ | $X_{34}$ | age1 [$X_{34}$] |
| ***1***** | | | | | -.0213(-4.20) | |
| 2a | | | | | | |
| ***2b*** | | | | | | |
| 3 | .501(3.69) | | | | -.0391(-11.3) | |
| 4 | | | | | | |
| ***5*** | | | | | | |
| 6 | | | | | -.448(-5.98) | |
| ***7a*** | | | | | | |
| ***7b*** | | | | | | |
| ***7c*** | | | | | | |
| ***8*** | | | | | | |
| 9 | | | | | -.0182(-4.10) | |
| ***10*** | | | | | | |
| 11a | | | | | | |
| ***11b*** | | | | | | |
| 12 | | 1.09(3.01) | | | -.0229(-2.07) | |
| 13 | | | | | | |
| ***14*** | | | | | | |
| 15 | | | | | -.0346(-9.64) | |
| 16 | | | | | -.0201(-4.56) | |
| ***17*** | | | | | | |
| 18 | | | | | -.0525(-6.49) | |
| ***19*** | | | | | | |
| 20a | | | | | | |
| ***20b*** | | | | | | |
| 21 | | | | | -.0393(-6.33) | |
| 22 | | | | 1.38(3.29) | -.0169(-5.86) | |
| 23a | | | 4.18(4.23) | | | 1.05(2.24) |
| ***23b*** | | 1.56(2.01) | | | | |
| 24 | | | | | -.0400(-13.1) | |
| 25 | | | | | | |
| ***26*** | | | | | | 1.08(2.19) |
| 27 | | | | | -.0195(-3.99) | |
| ***28*** | | | | 2.06(3.21) | -.0149(-3.42) | |
| 29a | | | | | | |
| ***29b*** | | | | | | |
| 30 | | | | | -.0481(-5.01) | |

…. – positive coefficient  …. – negative coefficient.

"nopass" – "traffic control device for the vehicle at fault is a «no passing zone»" indicator variable

"lncontr" – "traffic control device for the vehicle at fault is a «lane control»" indicator variable

"fl" – "traffic control device for the vehicle at fault is flashing signal" indicator

"$X_{33}$" – "at least one of the vehicles involved was on fire" indicator variable

"$X_{34}$" – "age (in years) of the driver at fault" quantitative variable

Table B.4 (Continued)

| # | Coefficient (t-ratio) | | | | | |
|---|---|---|---|---|---|---|
| | $X_{35}$ | maxpass [$X_{27}$] | youngdrv [$X_{34}$] | olddrv [$X_{34}$] | mm [$X_{35}$] | |
| ***1***** | | -.383(-3.91) | | | | |
| 2a | | | | | | |
| ***2b*** | | | | -1.32(-3.34) | | |
| 3 | -.305(-2.74) | | | | | |
| 4 | | | | | | |
| ***5*** | 2.29(2.13) | | | | | |
| 6 | | | | | | |
| ***7a*** | | .496(2.78) | | | | |
| ***7b*** | .818(2.46) | | | | | |
| ***7c*** | | | | | | |
| ***8*** | | | | | | |
| 9 | | | | | | |
| ***10*** | | .189(2.82) | | | | |
| 11a | | | | | | |
| ***11b*** | | | | | | |
| 12 | | | | | | |
| 13 | | | -.0186(-2.88) | | | |
| ***14*** | | | | | | |
| 15 | | | | | | |
| 16 | | | | | .457(3.14) | |
| ***17*** | | | | | | |
| 18 | | | | | | |
| ***19*** | | | | | | |
| 20a | | | | | | |
| ***20b*** | | | | | | |
| 21 | -.322(-1.97) | | | | | |
| 22 | | | | | .254(2.72) | |
| 23a | | | | | | |
| ***23b*** | 2.66(3.08) | | | | | |
| 24 | -.456(-5.71) | | | | | |
| 25 | | | -.0218(-3.14) | | | |
| ***26*** | | | | | | |
| 27 | | | | | | |
| ***28*** | | | | | .493(3.33) | |
| 29a | | | | | | |
| ***29b*** | | | | | | |
| 30 | | | | | | |

…. – positive coefficient …. – negative coefficient.

"$X_{35}$" – "gender of the driver at fault: 1 – female, 0 – male" indicator variable
"maxpass" – "the largest number of occupants in all vehicles involved" indicator variable
"youngdrv" – "the driver at fault is younger than the other driver involved" indicator variable
"olddrv" – "the driver at fault is older than the other driver involved" indicator var.
"mm" – "two male drivers involved into a two-vehicle accident" indicator var.

Table B.5 Tests of car-SUV separation in 2004 accident causation study[24]

| # | Model name | | | $M$ | $K$ | $LL(\boldsymbol{\beta}_{\mathbf{m}})$ | $\sum LL(\boldsymbol{\beta}_{\mathbf{m}})$ | df | p-value | conclusion* |
|---|---|---|---|---|---|---|---|---|---|---|
| 1 | County road | rural | (car/SUV)+(car/SUV) | 3 | 7 | -1426.61 | -1418.48 | 14 | 0.30 | Car = SUV |
| 2 | | | (car/SUV)+(truck) | 2 | 2 | -87.04 | -86.24 | 2 | 0.45 | Car = SUV |
| 4 | | urban | (car/SUV)+(car/SUV) | 3 | 6 | -246.94 | -240.33 | 12 | 0.35 | Car = SUV |
| 5 | | | (car/SUV)+(truck) | 2 | 5 | -16.14 | -15.11 | 5 | 0.84 | Car = SUV |
| 7 | Interstate | rural | (car/SUV)+(car/SUV) | 3 | 7 | -327.44 | -317.43 | 14 | 0.13 | Car = SUV |
| 8 | | | (car/SUV)+(truck) | 2 | 6 | -128.79 | -127.07 | 6 | 0.75 | Car = SUV |
| 10 | | urban | (car/SUV)+(car/SUV) | 3 | 7 | -339.99 | -333.09 | 14 | 0.46 | Car = SUV |
| 11 | | | (car/SUV)+(truck) | 2 | 7 | -203.66 | -201.12 | 7 | 0.65 | Car = SUV |
| 13 | State route | rural | (car/SUV)+(car/SUV) | 3 | 7 | -537.67 | -530.30 | 14 | 0.40 | Car = SUV |
| 14 | | | (car/SUV)+(truck) | 2 | 6 | -128.69 | -123.18 | 6 | 0.88 | Car = SUV |
| 16 | | urban | (car/SUV)+(car/SUV) | 3 | 7 | -716.67 | -707.19 | 14 | 0.17 | Car = SUV |
| 17 | | | (car/SUV)+(truck) | 2 | 4 | -57.96 | -57.47 | 4 | 0.91 | Car = SUV |
| 19 | City street | rural | (car/SUV)+(car/SUV) | 3 | 11 | -525.56 | -519.80 | 22 | 0.97 | Car = SUV |
| 20 | | | (car/SUV)+(truck) | 2 | 2 | -35.71 | -35.33 | 2 | 0.68 | Car = SUV |
| 22 | | urban | (car/SUV)+(car/SUV) | 3 | 12 | -6259.14 | -6246.99 | 24 | 0.44 | Car = SUV |
| 23 | | | (car/SUV)+(truck) | 2 | 7 | -323.90 | -322.02 | 7 | 0.81 | Car = SUV |
| 25 | US route | rural | (car/SUV)+(car/SUV) | 3 | 7 | -404.95 | -399.41 | 14 | 0.68 | Car = SUV |
| 26 | | | (car/-SUV)+(truck) | 2 | 5 | -147.16 | -143.83 | 5 | 0.25 | Car = SUV |
| 28 | | urban | (car/SUV)+(car/SUV) | 3 | 11 | -1088.61 | -1081.30 | 22 | 0.88 | Car = SUV |
| 29 | | | (car/SUV)+(truck) | 2 | 5 | -117.72 | -115.99 | 5 | 0.63 | Car = SUV |

* For all models 1–29 we find that "Car = SUV", which means that in 2004 unsafe-speed-related accident causation study cars and SUVs can be considered together.

[24] These tests are intended for testing whether in two-vehicle accidents cars and SUVs can be considered together or must be considered separately. The testing is done for all two-vehicle accident best final models by using the likelihood ratio test given in Equation (2.5). Please refer to Equation (2.5) for explanation of the quantities reported in the table. The p-values given in the next to last column are the probability values of the test statistic under the zero hypothesis (which is that cars and SUVs can be considered together).

Table B.6 Tests of car-SUV separation in 2006 accident causation study

| # | Model name | | | $M$ | $K$ | $LL(\boldsymbol{\beta_m})$ | $\sum LL(\boldsymbol{\beta_m})$ | df | p-value | conclusion* |
|---|---|---|---|---|---|---|---|---|---|---|
| 1 | County road | rural | (car/SUV)+(car/SUV) | 3 | 5 | -342.72 | -335.58 | 10 | 0.16 | Car = SUV |
| 2 | | | (car/SUV)+(truck) | 2 | 4 | -191.86 | -131.52 | 4 | 0.02 | Car ≠ SUV |
| 4 | | urban | (car/SUV)+(car/SUV) | 3 | 5 | -227.21 | -223.57 | 10 | 0.70 | Car = SUV |
| 5 | | | (car/SUV)+(truck) | 2 | 3 | -157.00 | -154.17 | 3 | 0.90 | Car = SUV |
| 7 | Interstate | rural | (car/SUV)+(car/SUV) | 3 | 3 | -50.625 | -44.218 | 6 | 0.05 | Car ≠ SUV |
| 8 | | | (car/SUV)+(truck) | 2 | 5 | -222.35 | -219.80 | 5 | 0.40 | Car = SUV |
| 10 | | urban | (car/SUV)+(car/SUV) | 3 | 8 | -508.16 | -501.60 | 16 | 0.66 | Car = SUV |
| 11 | | | (car/SUV)+(truck) | 2 | 6 | -137.34 | -129.39 | 6 | 0.01 | Car ≠ SUV |
| 13 | State route | rural | (car/SUV)+(car/SUV) | 3 | 5 | -774.29 | -770.94 | 10 | 0.75 | Car = SUV |
| 14 | | | (car/SUV)+(truck) | 2 | 8 | -92.951 | -89.165 | 8 | 0.48 | Car = SUV |
| 16 | | urban | (car/SUV)+(car/SUV) | 3 | 6 | -850.84 | -841.88 | 12 | 0.12 | Car = SUV |
| 17 | | | (car/SUV)+(truck) | 2 | 4 | -81.158 | -79.467 | 4 | 0.50 | Car = SUV |
| 19 | City street | rural | (car/SUV)+(car/SUV) | 3 | 4 | -211.60 | -209.63 | 8 | 0.86 | Car = SUV |
| 20 | | | (car/SUV)+(truck) | 2 | 3 | -20.366 | -16.153 | 3 | 0.04 | Car ≠ SUV |
| 22 | | urban | (car/SUV)+(car/SUV) | 3 | 14 | -2265.6 | -2246.7 | 28 | 0.102 | Car = SUV |
| 23 | | | (car/SUV)+(truck) | 2 | 6 | -143.81 | -137.52 | 6 | 0.05 | Car ≠ SUV |
| 25 | US route | rural | (car/SUV)+(car/SUV) | 3 | 4 | -456.42 | -454.21 | 8 | 0.82 | Car = SUV |
| 26 | | | (car/-SUV)+(truck) | 2 | 5 | -80.524 | -77.449 | 5 | 0.29 | Car = SUV |
| 28 | | urban | (car/SUV)+(car/SUV) | 3 | 7 | -834.94 | -829.20 | 14 | 0.65 | Car = SUV |
| 29 | | | (car/SUV)+(truck) | 2 | 4 | -56.753 | -51.514 | 4 | 0.03 | Car ≠ SUV |

* For all models 2, 7, 11, 20, 23 and 29 we find that "Car ≠ SUV", which means that for these models cars and SUVs must be considered separately in 2006 accident causation study. For all other models we find that "Car = SUV", which means that cars and SUVs can be considered together.

# Appendix C.

Table C.1 Road classes and accident types in 2004 accident severity study

| # | Road-class-accident-type combination | | | Number of observations | | | | |
|---|---|---|---|---|---|---|---|---|
| | | | | all | available for the models* | | | |
| | | | | | total | fatal | injury | PDO |
| 1 | County road | rural | (car/SUV**)+(car/SUV) | 7260 | 2788 | 22 | 741 | 2025 |
| 2 | | | (car/SUV)+(truck***) | 649 | 615 | 5 | 143 | 467 |
| 3 | | | one vehicle | 18121 | 9433 | 115 | 2289 | 7029 |
| 4 | | urban | (car/SUV)+(car/SUV) | 1861 | 1400 | 1 | 293 | 1106 |
| 5 | | | (car/SUV)+(truck) | 143 | 120 | 0 | 21 | 99 |
| 6 | | | one vehicle | 980 | 713 | 3 | 165 | 545 |
| 7 | Interstate | rural | (car/SUV)+(car/SUV) | 1044 | 955 | 3 | 157 | 795 |
| 8a | | | (car)+(truck) | 516 | 470 | 6 | 78 | 386 |
| 8b | | | (SUV)+(truck) | 295 | 244 | 1 | 58 | 185 |
| 9 | | | one vehicle | 3351 | 3284 | 23 | 510 | 2751 |
| 10 | | urban | (car/SUV)+(car/SUV) | 2234 | 1698 | 4 | 258 | 1436 |
| 11 | | | (car/SUV)+(truck) | 1085 | 756 | 5 | 122 | 629 |
| 12 | | | one vehicle | 1614 | 1526 | 14 | 345 | 1167 |
| 13 | State route | rural | (car/SUV)+(car/SUV) | 4788 | 1979 | 23 | 638 | 1318 |
| 14 | | | (car/SUV)+(truck) | 683 | 653 | 22 | 195 | 436 |
| 15 | | | one vehicle | 9798 | 6368 | 50 | 1200 | 5118 |
| 16a | | urban | (car)+(car) | 2701 | 2318 | 2 | 581 | 1735 |
| 16b | | | (car)+(SUV) | 3872 | 3344 | 4 | 783 | 2557 |
| 16c | | | (SUV)+(SUV) | 1473 | 1084 | 6 | 231 | 847 |
| 17 | | | (car/SUV)+(truck) | 641 | 339 | 2 | 69 | 268 |
| 18 | | | one vehicle | 1502 | 1381 | 9 | 308 | 1064 |
| 19 | City street | rural | (car/SUV)+(car/SUV) | 3820 | 2610 | 4 | 591 | 2015 |
| 20 | | | (car/SUV)+(truck) | 263 | 175 | 1 | 35 | 739 |
| 21 | | | one vehicle | 2412 | 2033 | 13 | 495 | 1525 |
| 22 | | urban | (car/SUV)+(car/SUV) | 63367 | 33141 | 28 | 7290 | 25823 |
| 23a | | | (car)+(truck) | 2321 | 1421 | 3 | 236 | 1182 |
| 23b | | | (SUV)+(truck) | 1273 | 990 | 3 | 145 | 842 |
| 24 | | | one vehicle | 12549 | 6431 | 64 | 2249 | 4118 |
| 25 | US route | rural | (car/SUV)+(car/SUV) | 2592 | 1499 | 17 | 484 | 998 |
| 26 | | | (car/SUV)+(truck) | 566 | 413 | 21 | 154 | 238 |
| 27 | | | one vehicle | 4211 | 2982 | 24 | 508 | 2450 |
| 28 | | urban | (car/SUV)+(car/SUV) | 6931 | 4839 | 6 | 1219 | 3614 |
| 29 | | | (car/SUV)+(truck) | 752 | 704 | 3 | 163 | 538 |
| 30 | | | one vehicle | 1073 | 1063 | 13 | 292 | 758 |

* – observations available for the best final estimated statistical models after exclusion of all missing observations

** – "SUV" includes sport utility vehicles, pickups and vans

*** – "truck" includes any possible kind of truck or tractor

Table C.2 Road classes and accident types in 2006 accident severity study

| # | Road-class-accident-type combination | | | Number of observations | | | | |
|---|---|---|---|---|---|---|---|---|
| | | | | all | available for the models* | | | |
| | | | | | total | fatal | injury | PDO |
| 1 | County road | rural | (car/SUV**)+(car/SUV) | 5966 | 4323 | 14 | 369 | 3940 |
| 2 | | | (car/SUV)+(truck***) | 345 | 286 | 3 | 42 | 241 |
| 3 | | | one vehicle | 16165 | 14733 | 143 | 3313 | 11277 |
| 4a | | urban | (car)+(car) | 536 | 489 | 0 | 107 | 275 |
| 4b | | | (car)+(SUV) | 691 | 232 | 0 | 40 | 192 |
| 4c | | | (SUV)+(SUV) | 261 | 225 | 0 | 37 | 188 |
| 5 | | | (car/SUV)+(truck) | 80 | 80 | 1 | 5 | 74 |
| 6 | | | one vehicle | 800 | 745 | 2 | 174 | 569 |
| 7 | Interstate | rural | (car/SUV)+(car/SUV) | 1124 | 994 | 6 | 167 | 821 |
| 8 | | | (car/SUV)+(truck) | 758 | 649 | 11 | 79 | 559 |
| 9 | | | one vehicle | 3736 | 3676 | 22 | 571 | 3083 |
| 10 | | urban | (car/SUV)+(car/SUV) | 2395 | 2178 | 1 | 303 | 1874 |
| 11 | | | (car/SUV)+(truck) | 850 | 692 | 2 | 89 | 601 |
| 12 | | | one vehicle | 1884 | 1834 | 13 | 397 | 1424 |
| 13 | State route | rural | (car/SUV)+(car/SUV) | 4582 | 763 | 28 | 274 | 461 |
| 14 | | | (car/SUV)+(truck) | 524 | 125 | 10 | 18 | 97 |
| 15 | | | one vehicle | 10172 | 9611 | 81 | 1659 | 7871 |
| 16 | | urban | (car/SUV)+(car/SUV) | 7483 | 6224 | 7 | 1398 | 4819 |
| 17 | | | (car/SUV)+(truck) | 510 | 482 | 2 | 65 | 415 |
| 18 | | | one vehicle | 1715 | 1579 | 16 | 404 | 1159 |
| 19 | City street | rural | (car/SUV)+(car/SUV) | 2926 | 2103 | 1 | 473 | 1629 |
| 20 | | | (car/SUV)+(truck) | 155 | 84 | 0 | 9 | 75 |
| 21 | | | one vehicle | 2115 | 624 | 4 | 164 | 456 |
| 22a | | urban | (car)+(car) | 20109 | 13499 | 2 | 2983 | 10514 |
| 22b | | | (car)+(SUV) | 23000 | 8282 | 8 | 1866 | 6408 |
| 22c | | | (SUV)+(SUV) | 7216 | 2544 | 3 | 601 | 1940 |
| 23 | | | (car/SUV)+(truck) | 2323 | 1570 | 2 | 175 | 1393 |
| 24 | | | one vehicle | 10869 | 4152 | 40 | 1561 | 2551 |
| 25 | US route | rural | (car/SUV)+(car/SUV) | 2481 | 541 | 16 | 185 | 340 |
| 26 | | | (car/SUV)+(truck) | 566 | 386 | 9 | 108 | 269 |
| 27 | | | one vehicle | 4257 | 4019 | 26 | 642 | 3351 |
| 28 | | urban | (car/SUV)+(car/SUV) | 6941 | 1680 | 2 | 463 | 1215 |
| 29 | | | (car/SUV)+(truck) | 572 | 467 | 5 | 62 | 400 |
| 30 | | | one vehicle | 1211 | 285 | 4 | 98 | 183 |

* – observations available for the best final estimated statistical models after exclusion of all missing observations

** – "SUV" includes sport utility vehicles, pickups and vans

*** – "truck" includes any possible kind of truck or tractor

Table C.3 Speed limit data bins chosen in 2004 accident severity study

<table>
<tr><th rowspan="2">#</th><th rowspan="2">Road type</th><th colspan="13">Speed limit intervals (in mph)</th></tr>
<tr><th>[5,10)*</th><th>[10,15)</th><th>[15,20)</th><th>[20,25)</th><th>[25,30)</th><th>[30,35)</th><th>[35,40)</th><th>[40,45)</th><th>[45,50)</th><th>[50,55)</th><th>[55,60)</th><th>[60,65)</th><th>≥65</th></tr>
<tr><td>1</td><td rowspan="6">County road</td><td colspan="6">533**</td><td>305</td><td>623</td><td>576</td><td>156</td><td>604</td><td colspan="2">-</td></tr>
<tr><td>2</td><td colspan="6">109</td><td>56</td><td>78</td><td colspan="2">140</td><td>134</td><td colspan="2">-</td></tr>
<tr><td>3</td><td colspan="5">129</td><td>856</td><td>794</td><td>1345</td><td>1584</td><td>603</td><td colspan="3">4122</td></tr>
<tr><td>4</td><td colspan="5">101</td><td>376</td><td>242</td><td>323</td><td>280</td><td>33</td><td>45</td><td colspan="2">-</td></tr>
<tr><td>5</td><td colspan="6">40</td><td colspan="2">33</td><td colspan="3">34</td><td colspan="2">-</td></tr>
<tr><td>6</td><td colspan="6">200</td><td>87</td><td colspan="3">284</td><td colspan="2">99</td><td>-</td></tr>
<tr><td>7</td><td rowspan="7">Interstate</td><td colspan="8">57</td><td colspan="2">111</td><td colspan="2">205</td><td>549</td></tr>
<tr><td>8a</td><td colspan="9">57</td><td colspan="2">64</td><td colspan="2">81</td></tr>
<tr><td>8b</td><td colspan="11">65</td><td colspan="2">52</td></tr>
<tr><td>9</td><td colspan="9">164</td><td colspan="2">340</td><td>318</td><td>2444</td></tr>
<tr><td>10</td><td colspan="10">470</td><td>959</td><td colspan="2">215</td></tr>
<tr><td>11</td><td colspan="11">643</td><td colspan="2">45</td></tr>
<tr><td>12</td><td colspan="8">62</td><td>173</td><td>90</td><td>929</td><td colspan="2">261</td></tr>
<tr><td>13</td><td rowspan="8">State route</td><td colspan="9">603</td><td colspan="2">1376</td><td colspan="2">-</td></tr>
<tr><td>14</td><td colspan="8">65</td><td>82</td><td>55</td><td>376</td><td colspan="2">-</td></tr>
<tr><td>15</td><td colspan="6">81</td><td>51</td><td>146</td><td>581</td><td>607</td><td colspan="3">4792</td></tr>
<tr><td>16a</td><td colspan="4">35</td><td>132</td><td>473</td><td>545</td><td>332</td><td>306</td><td>109</td><td colspan="3">136</td></tr>
<tr><td>16b</td><td colspan="5">230</td><td>659</td><td>751</td><td>401</td><td>486</td><td>196</td><td colspan="3">235</td></tr>
<tr><td>16c</td><td colspan="6">261</td><td colspan="2">370</td><td colspan="5">328</td></tr>
<tr><td>17</td><td colspan="5">26</td><td>58</td><td>83</td><td>41</td><td colspan="2">82</td><td colspan="3">49</td></tr>
<tr><td>18</td><td colspan="5">67</td><td>214</td><td>165</td><td>162</td><td>242</td><td>125</td><td colspan="3">323</td></tr>
<tr><td>19</td><td rowspan="7">City street</td><td colspan="3">74</td><td>70</td><td>129</td><td>579</td><td>486</td><td>717</td><td>241</td><td>710</td><td>194</td><td colspan="2">49</td></tr>
<tr><td>20</td><td colspan="7">68</td><td>46</td><td colspan="2">30</td><td colspan="3">31</td></tr>
<tr><td>21</td><td colspan="3">38</td><td>53</td><td>60</td><td colspan="2">516</td><td colspan="3">598</td><td colspan="3">606</td></tr>
<tr><td>22</td><td colspan="2">242</td><td>319</td><td>574</td><td>2022</td><td>9862</td><td>8379</td><td>5219</td><td>1464</td><td>163</td><td colspan="3">179</td></tr>
<tr><td>23a</td><td colspan="6">560</td><td>345</td><td colspan="2">287</td><td colspan="4">36</td></tr>
<tr><td>23b</td><td colspan="5">93</td><td>333</td><td>225</td><td>133</td><td>54</td><td colspan="4">22</td></tr>
<tr><td>24</td><td colspan="5">965</td><td>2271</td><td>1256</td><td colspan="3">1053</td><td colspan="3">197</td></tr>
<tr><td>25</td><td rowspan="6">US route</td><td colspan="6">42</td><td>59</td><td>76</td><td>206</td><td>153</td><td colspan="3">963</td></tr>
<tr><td>26</td><td colspan="10">97</td><td>276</td><td colspan="2">-</td></tr>
<tr><td>27</td><td colspan="6">30</td><td colspan="2">66</td><td>149</td><td>159</td><td colspan="3">2553</td></tr>
<tr><td>28</td><td colspan="6">919</td><td>949</td><td colspan="2">1825</td><td>358</td><td colspan="3">272</td></tr>
<tr><td>29</td><td colspan="3">-</td><td colspan="2">30</td><td>87</td><td>120</td><td>75</td><td>187</td><td>65</td><td colspan="3">70</td></tr>
<tr><td>30</td><td colspan="5">53</td><td>73</td><td>135</td><td>86</td><td>258</td><td>120</td><td>242</td><td colspan="2">-</td></tr>
</table>

* – Interval [5,10) includes speed limits larger or equal to 5 mph and smaller than 10 mph. All other intervals are similarly defined.

** – Numbers printed on top the speed limit data bins inside the table, give data sample sizes in the corresponding bins.

Table C.4 Speed limit data bins chosen in 2006 accident severity study

<table>
<tr><th rowspan="2">#</th><th rowspan="2">Road type</th><th colspan="15">Speed limit intervals (in mph)</th></tr>
<tr><th>[5,10)*</th><th>[10,15)</th><th>[15,20)</th><th>[20,25)</th><th>[25,30)</th><th>[30,35)</th><th>[35,40)</th><th>[40,45)</th><th>[45,50)</th><th>[50,55)</th><th>[55,60)</th><th>[60,65)</th><th>[65,70)</th><th>[70,75)</th><th>≥75</th></tr>
<tr><td>1</td><td rowspan="8">County road</td><td colspan="8">807**</td><td>345</td><td>98</td><td colspan="2">223</td><td colspan="3">-</td></tr>
<tr><td>2</td><td colspan="6">45</td><td>35</td><td>54</td><td colspan="2">70</td><td>82</td><td colspan="4">-</td></tr>
<tr><td>3</td><td colspan="6">1127</td><td colspan="2">3233</td><td>2993</td><td>1021</td><td colspan="4">6359</td><td>-</td></tr>
<tr><td>4a</td><td colspan="5">26</td><td>83</td><td>72</td><td>110</td><td>111</td><td>9</td><td>11</td><td colspan="4">-</td></tr>
<tr><td>4b</td><td colspan="6">84</td><td>51</td><td>44</td><td colspan="4">53</td><td colspan="3">-</td></tr>
<tr><td>4c</td><td colspan="6">63</td><td>43</td><td>47</td><td colspan="3">47</td><td colspan="4">-</td></tr>
<tr><td>5</td><td colspan="6">21</td><td colspan="2">25</td><td colspan="3">16</td><td colspan="4">-</td></tr>
<tr><td>6</td><td colspan="5">44</td><td>143</td><td>128</td><td>108</td><td colspan="3">295</td><td colspan="4">-</td></tr>
<tr><td>7</td><td rowspan="6">Interstate</td><td colspan="10">173</td><td colspan="2">225</td><td>177</td><td>336</td><td>-</td></tr>
<tr><td>8</td><td colspan="10">56</td><td colspan="4">265</td><td>-</td></tr>
<tr><td>9</td><td colspan="8">75</td><td colspan="2">74</td><td colspan="2">433</td><td>1019</td><td>2054</td><td>-</td></tr>
<tr><td>10</td><td colspan="5">53</td><td>50</td><td>74</td><td>93</td><td>146</td><td>206</td><td colspan="2">1165</td><td colspan="2">331</td><td>-</td></tr>
<tr><td>11</td><td colspan="10">121</td><td colspan="4">511</td><td>-</td></tr>
<tr><td>12</td><td colspan="8">102</td><td colspan="2">207</td><td>1131</td><td colspan="2">259</td><td>119</td><td>-</td></tr>
<tr><td>13</td><td rowspan="6">State route</td><td colspan="9">252</td><td colspan="4">511</td><td colspan="2">-</td></tr>
<tr><td>14</td><td colspan="10">57</td><td colspan="3">68</td><td colspan="2">-</td></tr>
<tr><td>15</td><td colspan="8">365</td><td>935</td><td>748</td><td>7267</td><td colspan="3">296</td><td>-</td></tr>
<tr><td>16</td><td colspan="4">147</td><td>339</td><td>1470</td><td>1508</td><td>970</td><td>981</td><td>367</td><td colspan="3">442</td><td colspan="2">-</td></tr>
<tr><td>17</td><td colspan="7">212</td><td>580</td><td>71</td><td>46</td><td colspan="3">45</td><td colspan="2">-</td></tr>
<tr><td>18</td><td colspan="6">277</td><td>205</td><td>172</td><td>282</td><td>143</td><td colspan="3">368</td><td colspan="2">-</td></tr>
<tr><td>19</td><td rowspan="8">City street</td><td colspan="4">140</td><td>100</td><td>523</td><td>438</td><td>577</td><td>208</td><td colspan="5">117</td><td>-</td></tr>
<tr><td>20</td><td colspan="7">1665</td><td>660</td><td colspan="3">381</td><td colspan="4">-</td></tr>
<tr><td>21</td><td colspan="8">338</td><td colspan="6">228</td><td>-</td></tr>
<tr><td>22a</td><td colspan="4">850</td><td>985</td><td>5440</td><td>3629</td><td>1928</td><td colspan="3">667</td><td colspan="4">-</td></tr>
<tr><td>22b</td><td colspan="4">335</td><td>572</td><td>2901</td><td>2313</td><td>1620</td><td colspan="5">541</td><td colspan="2">-</td></tr>
<tr><td>22c</td><td colspan="6">1211</td><td colspan="7">1333</td><td colspan="2">-</td></tr>
<tr><td>23</td><td colspan="4">86</td><td>134</td><td>561</td><td>442</td><td>227</td><td colspan="5">120</td><td colspan="2">-</td></tr>
<tr><td>24</td><td colspan="5">593</td><td>1598</td><td>818</td><td colspan="7">723</td><td>-</td></tr>
<tr><td>25</td><td rowspan="6">US route</td><td colspan="8">68</td><td colspan="2">121</td><td>244</td><td colspan="2">34</td><td colspan="2">-</td></tr>
<tr><td>26</td><td colspan="9">84</td><td colspan="2">240</td><td colspan="2">62</td><td colspan="2">-</td></tr>
<tr><td>27</td><td colspan="11">3134</td><td colspan="3">869</td><td>-</td></tr>
<tr><td>28</td><td colspan="8">943</td><td colspan="5">737</td><td colspan="2">-</td></tr>
<tr><td>29</td><td colspan="9">391</td><td colspan="4">76</td><td colspan="2">-</td></tr>
<tr><td>30</td><td colspan="8">100</td><td colspan="2">116</td><td colspan="3">61</td><td colspan="2">-</td></tr>
</table>

* – Interval [5,10) includes speed limits larger or equal to 5 mph and smaller than 10 mph. All other intervals are similarly defined.

** – Numbers printed on top the speed limit data bins inside the table, give data sample sizes in the corresponding bins.

Table C.5 Multinomial logit models for 2004 accident severity[25]

| # | Model name | | | Log-likelihood | | $R^2$ | Coefficient (t-ratio) X29 | |
|---|---|---|---|---|---|---|---|---|
| | | | | model | restricted* | | fatality [ $\beta_1$ ] | injury [ $\beta_2$ ] |
| 1 | County road | rural | (car/SUV)+(car/SUV) | -1636.3 | -1735.9 | .057 | .108 (3.61) | .0255 (5.15) |
| ***2*** | County road | rural | (car/SUV)+(truck) | -250.47 | -287.90 | .130 | | .0414 (3.42) |
| 3 | County road | rural | one vehicle | -5060.1 | -5816.0 | .130 | .0382 (3.47) | |
| 4 | County road | urban | (car/SUV)+(car/SUV) | -683.95 | -726.22 | .058 | .0323 (3.75) | .0323 (3.75) |
| ***5**** | County road | urban | (car/SUV)+(truck) | -104.80 | -131.83 | .205 | | |
| 6 | County road | urban | one vehicle | -359.79 | -404.07 | .110 | | |
| 7 | Interstate | rural | (car/SUV)+(car/SUV) | -456.77 | -473.11 | .035 | | |
| ***8a*** | Interstate | rural | (car)+(truck) | -198.37 | -219.34 | .095 | | |
| ***8b*** | Interstate | rural | (SUV)+(truck) | -114.91 | -140.04 | .179 | | |
| 9 | Interstate | rural | one vehicle | -1360.9 | -1550.4 | .122 | | |
| 10 | Interstate | urban | (car/SUV)+(car/SUV) | -710.09 | -751.00 | .054 | | |
| 11 | Interstate | urban | (car/SUV)+(truck) | -336.87 | -363.30 | .073 | | |
| 12 | Interstate | urban | one vehicle | -772.32 | -887.87 | .130 | | |
| 13 | State route | rural | (car/SUV)+(car/SUV) | -1302.1 | -1360.4 | .043 | | .0306 (3.90) |
| ***14*** | State route | rural | (car/SUV)+(truck) | -362.86 | -405.25 | .105 | | |
| 15 | State route | rural | one vehicle | -2188.0 | -2714.2 | .194 | | |
| 16a | State route | urban | (car)+(car) | -1116.2 | -1157.9 | .036 | .0340 (5.14) | .0340 (5.14) |
| 16b | State route | urban | (car)+(SUV) | -1547.8 | -1608.4 | .038 | .0225 (4.36) | .0225 (4.36) |
| ***16c*** | State route | urban | (SUV)+(SUV) | -492.07 | -517.33 | .049 | .0315 (3.39) | .0315 (3.39) |
| ***17*** | State route | urban | (car/SUV)+(truck) | -173.40 | -183.09 | .053 | .0418 (2.89) | .0418 (2.89) |
| 18 | State route | urban | one vehicle | -677.08 | -784.90 | .137 | | |
| 19 | City street | rural | (car/SUV)+(car/SUV) | -1358.7 | -1425.1 | .047 | .114 (2.53) | .0273 (5.40) |
| ***20*** | City street | rural | (car/SUV)+(truck) | -68.07 | -93.51 | .272 | | .0676 (3.17) |
| 21 | City street | rural | one vehicle | -1025.6 | -1203.4 | .148 | | |
| 22 | City street | urban | (car/SUV)+(car/SUV) | -14547 | -15236 | .045 | .0938 (2.76) | .0304 (11.8) |
| ***23a*** | City street | urban | (car)+(truck) | -526.00 | -580.96 | .095 | .0469 (3.99) | .0469 (3.99) |
| ***23b*** | City street | urban | (SUV)+(truck) | -344.88 | -366.77 | .060 | .0640 (4.41) | .0640 (4.41) |
| 24 | City street | urban | one vehicle | -4004.8 | -4493.6 | .109 | | |
| 25 | US route | rural | (car/SUV)+(car/SUV) | -978.44 | -1029.3 | .049 | .340 (2.48) | .0409 (4.54) |
| ***26*** | US route | rural | (car/SUV)+(truck) | -270.09 | -309.90 | .128 | | .0720 (3.02) |
| 27 | US route | rural | one vehicle | -1251.3 | -1496.3 | .164 | | |
| 28 | US route | urban | (car/SUV)+(car/SUV) | -2345.1 | -2457.6 | .046 | .0263 (5.76) | .0263 (5.76) |
| ***29*** | US route | urban | (car/SUV)+(truck) | -311.62 | -336.81 | .075 | .0307 (2.52) | .0307 (2.52) |
| 30 | US route | urban | one vehicle | -549.94 | -639.11 | .140 | | |

.... – positive coefficient .... – negative coefficient

* – restricted log-likelihood found by setting all coefficients except intercepts to zero (with the exception of model 5, in case of which intercepts are also set to zero)

** – models are estimated by using procedure A on page 32, except the models marked by bold numbers and estimated by using procedure B on page 33

"$X_{29}$" – "posted speed limit (if the same for all vehicles involved)" quantitative variable

[25] See Equation (2.4), where outcomes "1", "2", "3" correspond to "fatality", "injury", "PDO". Only statistically significant coefficients, which are components of vectors $\beta_1$ and $\beta_2$, are given.

Table C.5 (Continued)

| # | Coefficient (t-ratio) | | | | | |
|---|---|---|---|---|---|---|
| | constant | | wint [$X_3$] | | sum [$X_3$] | |
| | fatality | injury | fatality | injury | fatality | injury |
| 1 | -12.0 (-6.27) | -3.30 (-12.4) | -.260 (-2.61) | -.260 (-2.61) | | |
| ***2*** | -7.41 (-5.59) | -4.07 (-6.90) | | | | |
| 3 | -6.43 (-10.7) | -1.45 (-13.7) | | | .232 (3.51) | .232 (3.51) |
| 4 | -9.13 (-8.46) | -3.57 (-8.23) | | | | |
| ***5*** | | -2.78 (-3.72) | | | | |
| 6 | -6.90 (-9.19) | -2.50 (-9.49) | | | | |
| 7 | -6.25 (-10.2) | -2.33 (-10.7) | | | | |
| ***8a*** | -5.97 (-5.85) | -1.87 (-7.79) | | | | |
| ***8b*** | -6.76 (-5.65) | -3.35 (-4.72) | | | | |
| 9 | -4.84 (-20.4) | -3.04 (-12.7) | | | | |
| 10 | -8.01 (-8.12) | -2.40 (-11.2) | | | | |
| 11 | -6.23 (-10.0) | -2.53 (-10.4) | | | | |
| 12 | -5.88 (-12.5) | -3.25 (-9.55) | | | 1.13 (2.09) | |
| 13 | -5.24 (-13.2) | -3.47 (-7.44) | | | | |
| ***14*** | -4.30 (-9.91) | -2.88 (-7.34) | -.619 (-2.58) | -.619 (-2.58) | | |
| 15 | -4.31 (-16.9) | -1.64 (-11.7) | | | | |
| 16a | -9.26 (-8.86) | -3.02 (-3.09) | | | | |
| 16b | -8.88 (-11.6) | -3.14 (-10.3) | | | | |
| ***16c*** | -7.13 (-8.91) | -3.22 (-7.13) | | | | .412 (2.37) |
| ***17*** | -7.42 (-6.33) | -3.22 (-5.16) | | | | |
| 18 | -6.31 (-10.7) | -1.60 (-11.3) | | | 1.96 (2.76) | |
| 19 | -11.3 (-5.04) | -2.25 (-8.63) | -.276 (-2.34) | -.276 (-2.34) | | |
| ***20*** | -6.22 (-5.18) | -5.50 (-4.69) | | | | |
| 21 | -5.62 (-17.1) | -3.16 (-14.6) | -.317 (-2.24) | -.317 (-2.24) | | |
| 22 | -14.0 (-8.80) | -2.93 (-26.5) | | -.0872 (-2.45) | | |
| ***23a*** | -9.67 (-8.74) | -3.87 (-8.47) | | | | |
| ***23b*** | -8.24 (-9.27) | -4.38 (-7.74) | | | | |
| 24 | -7.38 (-16.7) | -4.30 (-13.9) | -1.09 (-2.65) | -.410 (-5.95) | | |
| 25 | -22.6 (-3.00) | -3.11 (-6.50) | -.287 (-2.12) | -.287 (-2.12) | | |
| ***26*** | -4.73 (-6.79) | -5.49 (-4.06) | | | | |
| 27 | -5.56 (-22.6) | -1.86 (-11.2) | | | | |
| 28 | -8.57 (-15.6) | -3.12 (-9.77) | | | | |
| ***29*** | -8.67 (-7.46) | -4.01 (-6.27) | | | | |
| 30 | -5.20 (-10.1) | -1.61 (-9.51) | | | .429 (2.43) | .429 (2.43) |

…. – positive coefficient
…. – negative coefficient

---

“constant” – “constant term (intercept)” quantitative variable
"wint" – “winter season” indicator variable
"sum" – “summer season” indicator variable

Table C.5 (Continued)

| # | Coefficient (t-ratio) | | | | | |
|---|---|---|---|---|---|---|
| | fall [$X_3$] | | mon [$X_4$] | | tues [$X_4$] | |
| | fatality | injury | fatality | injury | fatality | injury |
| 1 | | | | | | |
| ***2*** | | | 2.67 (2.11) | .933 (3.29) | | |
| 3 | | | | | | |
| 4 | -.445 (-2.68) | -.445 (-2.68) | | | | |
| ***5*** | | | | | | |
| 6 | | | | | | |
| 7 | | | | | | |
| ***8a*** | | | | | | |
| ***8b*** | | | | | | 1.10 (2.57) |
| 9 | | -.264 (-2.04) | | | | |
| 10 | | | | | -.457 (-2.14) | -.457 (-2.14) |
| 11 | | | | | | |
| 12 | | | | | | |
| 13 | | | | | | |
| ***14*** | | | | .524 (1.99) | | |
| 15 | | | | | | |
| 16a | | | | | | |
| 16b | | | | | | |
| ***16c*** | | | | | | |
| ***17*** | | | | | .823 (2.50) | .823 (2.50) |
| 18 | | | | | | |
| 19 | | | | | | |
| ***20*** | | | | | | |
| 21 | | | -.374 (-2.09) | -.374 (-2.09) | | |
| 22 | | | | | 1.03 (2.23) | |
| ***23a*** | | | | | | |
| ***23b*** | | | | | | |
| 24 | | | | | | |
| 25 | | | | | | |
| ***26*** | 1.08 (2.04) | | | | | |
| 27 | | | | | | |
| 28 | | | | | | |
| ***29*** | | -.679 (-2.54) | | | | |
| 30 | | | | | | |

…. – positive coefficient
…. – negative coefficient

"fall" – "fall season" indicator variable
"mon" – "Monday" indicator variable
"tues" – "Tuesday" indicator variable

Table C.5 (Continued)

| # | Coefficient (t-ratio) | | | | | |
|---|---|---|---|---|---|---|
| | sund [$X_4$] | | sat [$X_4$] | | wed [$X_4$] | |
| | fatality | injury | fatality | injury | fatality | injury |
| 1 | | | | | | |
| ***2*** | | | | | | |
| 3 | | | | | | |
| 4 | | | | | | |
| ***5*** | | | | | | |
| 6 | | | | | | |
| 7 | | | | | | |
| ***8a*** | | | | | | |
| ***8b*** | | | | | | |
| 9 | | | | | | |
| 10 | | | | | | |
| 11 | | | | | | |
| 12 | | | | | | |
| 13 | 1.20 (2.50) | | | | | |
| ***14*** | | | | | | |
| 15 | | | | | | |
| 16a | | | | | | |
| 16b | | | | | | |
| ***16c*** | | | | | | |
| ***17*** | | | | | | |
| 18 | | | | | | |
| 19 | | | | | | |
| ***20*** | | | 2.08 (2.64) | 2.08 (2.64) | | |
| 21 | | | | | | |
| 22 | | | | | | |
| ***23a*** | | | | | | |
| ***23b*** | | | | | | |
| 24 | | | | | | |
| 25 | | | | | | |
| ***26*** | | | | | | |
| 27 | | | | | -.397 (-2.21) | -.397 (-2.21) |
| 28 | | | | | | |
| ***29*** | | | | | | |
| 30 | | | | | | |

…. – positive coefficient
…. – negative coefficient

"sund" – "Sunday" indicator variable
"sat" – "Saturday" indicator variable
"wed" – "Wednesday" indicator variable

Table C.5 (Continued)

| # | Coefficient (t-ratio) | | | | | |
|---|---|---|---|---|---|---|
| | thday [$X_4$] | | nigh [$X_5$] | | morn [$X_5$] | |
| | fatality | injury | fatality | injury | fatality | injury |
| 1 | | | | | | |
| ***2*** | | | | | | |
| 3 | | | | | | |
| 4 | | | | | | |
| ***5*** | | | | | | |
| 6 | | | | | | |
| 7 | | | | | | |
| ***8a*** | | | | | | |
| ***8b*** | | | | | | |
| 9 | | | | | | |
| 10 | | | | | -.414 (-2.33) | -.414 (-2.33) |
| 11 | | | | | | |
| 12 | -.517 (-2.56) | -.517 (-2.56) | | | | |
| 13 | | | | | | |
| ***14*** | | | | | | |
| 15 | | | | | | |
| 16a | | | | | | |
| 16b | | | | | | |
| ***16c*** | | | | | | |
| ***17*** | | | | | | |
| 18 | | | | | | |
| 19 | | | | | | |
| ***20*** | | | | | | |
| 21 | | | | | | |
| 22 | | | | | | |
| ***23a*** | | | 3.30 (2.30) | | | |
| ***23b*** | | | | | | |
| 24 | | | | | | |
| 25 | | | | | | |
| ***26*** | | | | | | |
| 27 | | | | | | |
| 28 | | | | | | |
| ***29*** | | | | | | |
| 30 | 1.73 (2.71) | | | | | |

…. – positive coefficient
…. – negative coefficient

"thday" – "Thursday" indicator variable
"nigh" – "late night hours: 1:00 to 5:00" indicator variable[26]
"morn" – "morning hours: 5:00 to 9:00" indicator variable

[26] We use military 24-hour time everywhere in our research.

Table C.5 (Continued)

| # | Coefficient (t-ratio) | | | | | |
|---|---|---|---|---|---|---|
| | dayt [$X_5$] | | nocons [$X_{13}$] | | cons [$X_{13}$] | |
| | fatality | injury | fatality | injury | fatality | injury |
| 1 | | | | | | |
| ***2*** | | | | | | |
| 3 | | | | | | |
| 4 | | | | | | |
| ***5*** | | | | | | |
| 6 | | | | | | |
| 7 | | | | | | |
| ***8a*** | | | | | | |
| ***8b*** | | | | | | |
| 9 | | | | | | |
| 10 | | | | | | |
| 11 | | | | | | |
| 12 | | | .510 (2.19) | .510 (2.19) | | |
| 13 | | | | | | |
| ***14*** | | | | | | |
| 15 | | | | | | |
| 16a | -.228 (-2.15) | -.228 (-2.15) | | | | |
| 16b | | | | | | |
| ***16c*** | | | | | 2.58 (2.09) | |
| ***17*** | | | | | | |
| 18 | .324 (2.21) | .324 (2.21) | | | | |
| 19 | | | | | | |
| ***20*** | | | | | | |
| 21 | | | | | | |
| 22 | | | | | | |
| ***23a*** | | | | | | |
| ***23b*** | | | | | | |
| 24 | | | .672 (2.73) | .672 (2.73) | | |
| 25 | | | | | | |
| ***26*** | | | | | | |
| 27 | .282 (2.40) | .282 (2.40) | | | | |
| 28 | | | | | | |
| ***29*** | | | | | | |
| 30 | | | | | | |

…. – positive coefficient
…. – negative coefficient

"dayt" – "day hours: 9:00 to 17:00" indicator variable
"nocons" – "no construction at the accident location" indicator variable
"cons" – "construction at the accident location" indicator variable

Table C.5 (Continued)

| # | Coefficient (t-ratio) | | | | | |
|---|---|---|---|---|---|---|
| | light [$X_{14}$] | | dark [$X_{14}$] | | day [$X_{14}$] | |
| | fatality | injury | fatality | injury | fatality | injury |
| 1 | | | | | | |
| ***2*** | | | | | | |
| 3 | | | | | | .199 (3.74) |
| 4 | | | | | | |
| ***5*** | | | | | | |
| 6 | | | | | | |
| 7 | | | .845 (4.25) | .845 (4.25) | | |
| ***8a*** | | | | | | |
| ***8b*** | | | 1.22 (3.07) | 1.22 (3.07) | | |
| 9 | | | | | | |
| 10 | | | | | | |
| 11 | | | | | | |
| 12 | | | | | | |
| 13 | | | 1.06 (2.35) | | | |
| ***14*** | | | 1.24 (2.44) | | | |
| 15 | | .261 (3.10) | | | | |
| 16a | | | | | | |
| 16b | | | | | | |
| ***16c*** | | | | | | |
| ***17*** | | | | | | |
| 18 | | | | | | |
| 19 | | | | | | -.297 (-2.75) |
| ***20*** | | | | | | |
| 21 | | | | | | |
| 22 | | | | | | |
| ***23a*** | | | | | | |
| ***23b*** | | | | | | |
| 24 | | | | | -.824 (-3.15) | |
| 25 | | | 1.29 (2.55) | | | |
| ***26*** | | | | | | |
| 27 | | | | | | |
| 28 | | | | | | |
| ***29*** | | | | | | |
| 30 | | | | | | .642 (4.04) |

…. – positive coefficient
…. – negative coefficient

---

"light" – "daylight time OR street lights lit up during dark time" indicator variable
"dark" – "dark time with no street lights" indicator variable
"day" – "daylight time" indicator variable

Table C.5 (Continued)

| # | Coefficient (t-ratio) | | | | | |
|---|---|---|---|---|---|---|
| | dawn [$X_{14}$] | | darklamp [$X_{14}$] | | precip [$X_{15}$] | |
| | fatality | injury | fatality | injury | fatality | injury |
| 1 | | | | | | |
| ***2*** | | | | | | |
| 3 | | | | | | |
| 4 | | | | | | |
| ***5*** | | | | | | 1.46 (2.39) |
| 6 | | | | | | |
| 7 | | | | | | |
| ***8a*** | | | | | | |
| ***8b*** | | | | | | |
| 9 | | | | | | |
| 10 | | | | | | |
| 11 | | | | | | |
| 12 | | | | | -1.02 (-6.41) | -1.02 (-6.41) |
| 13 | | | | | | |
| ***14*** | | | | | | |
| 15 | | | | | | |
| 16a | | | | | | |
| 16b | | | | | | |
| ***16c*** | | | | | | |
| ***17*** | | | | | | |
| 18 | | | | | | |
| 19 | | | | | | |
| ***20*** | | 2.78 (2.53) | | | | |
| 21 | | | | | | |
| 22 | | | .188 (4.76) | .188 (4.76) | | |
| ***23a*** | | | .545 (2.55) | .545 (2.55) | | |
| ***23b*** | | | .641 (2.00) | | | |
| 24 | | | | | | |
| 25 | | | | | | |
| ***26*** | | | | | | |
| 27 | | | | | | |
| 28 | | | | | | |
| ***29*** | | | .643 (2.17) | .643 (2.17) | | |
| 30 | | | | | | |

…. – positive coefficient
…. – negative coefficient

---

"dawn" – “dawn OR dask” indicator variable
"darklamp" – “dark AND street lights on” indicator variable
"precip" – “precipitation: rain OR snow OR sleet OR hail OR freezing rain” indicator variable

Table C.5 (Continued)

| # | Coefficient (t-ratio) | | | | | |
|---|---|---|---|---|---|---|
| | snow [$X_{15}$] | | clear [$X_{15}$] | | clo [$X_{15}$] | |
| | fatality | injury | fatality | injury | fatality | injury |
| 1 | | | | | | |
| ***2*** | | | | | | |
| 3 | | | | | | |
| 4 | | | | | | |
| ***5*** | | | | | | |
| 6 | | | | | | |
| 7 | | | | | | |
| ***8a*** | | | | | | |
| ***8b*** | | | | | | |
| 9 | | -1.14 (-6.01) | | | | |
| 10 | | | | | | |
| 11 | | | | | | |
| 12 | | | | | | |
| 13 | | | | | | |
| ***14*** | | | | | | |
| 15 | | | | | | |
| 16a | | | | | | |
| 16b | | | | | | |
| ***16c*** | | | | | | |
| ***17*** | | | | | | |
| 18 | -.779 (-2.39) | -.779 (-2.39) | | | | |
| 19 | | | | | | |
| ***20*** | | | | | | |
| 21 | | | | | | |
| 22 | | | | | | |
| ***23a*** | | | | | | |
| ***23b*** | | | | | | |
| 24 | | | | | | |
| 25 | | | | | | |
| ***26*** | | | | | 1.64 (2.99) | |
| 27 | | | | | | |
| 28 | | -1.01 (-2.82) | | | | |
| ***29*** | | | .615 (2.88) | .615 (2.88) | | |
| 30 | | | | | | |

…. – positive coefficient
…. – negative coefficient

"snow" – "snowing weather" indicator variable
"clear" – "clear weather" indicator variable
"clo" – "cloudy weather" indicator variable

Table C.5 (Continued)

| # | Coefficient (t-ratio) | | | | | |
|---|---|---|---|---|---|---|
| | rain [$X_{15}$] | | soil [$X_{15}$] | | dry [$X_{16}$] | |
| | fatality | injury | fatality | injury | fatality | injury |
| 1 | | | | | | |
| ***2*** | 3.23 (2.57) | | | | | |
| 3 | | | | | | |
| 4 | | | | | | |
| ***5*** | | | | | | |
| 6 | | | | | | |
| 7 | | .644 (2.58) | | | | |
| ***8a*** | | | 1.89 (1.99) | 1.89 (1.99) | | |
| ***8b*** | | | | | | |
| 9 | | | | | | |
| 10 | | | | | | |
| 11 | | | | | | |
| 12 | | | | | | |
| 13 | | | | | | |
| ***14*** | | | | | | |
| 15 | | | | | | |
| 16a | | | | | | |
| 16b | | | | | | |
| ***16c*** | | | | | | |
| ***17*** | | | | | | |
| 18 | | | | | | |
| 19 | | | | | | |
| ***20*** | | | | | | |
| 21 | | | | | | |
| 22 | | | | | | |
| ***23a*** | | | | | | |
| ***23b*** | | | | | | |
| 24 | | | | | .861 (2.52) | .324 (5.05) |
| 25 | | | | | | |
| ***26*** | | | | | | |
| 27 | | | | | | |
| 28 | | | | | | |
| ***29*** | | | | | | |
| 30 | | | | | | |

…. – positive coefficient
…. – negative coefficient

"rain" – “rainy weather” indicator variable
"soil" – “blowing sand OR soil OR snow” indicator variable
"dry" – “roadway surface is dry” indicator variable

Table C.5 (Continued)

| # | Coefficient (t-ratio) | | | | | |
|---|---|---|---|---|---|---|
| | slush [$X_{16}$] | | driv [$X_{17}$] | | nomed [$X_{17}$] | |
| | fatality | injury | fatality | injury | fatality | injury |
| 1 | | | | | | .304 (2.60) |
| ***2*** | | | | | | |
| 3 | -1.46 (-2.48) | -.430 (-4.63) | | | | |
| 4 | | | | | | |
| ***5*** | | | | | | |
| 6 | | | | | | |
| 7 | | | | | | |
| ***8a*** | 2.70 (2.18) | | | | | |
| ***8b*** | | | | | | |
| 9 | | | | | | |
| 10 | | | | | | |
| 11 | | | | | | |
| 12 | | | | | | |
| 13 | | | | | | |
| ***14*** | | | | | | |
| 15 | | | | | | |
| 16a | | | | -.387 (-3.09) | | |
| 16b | | | | | | |
| ***16c*** | | | | | | |
| ***17*** | | | | | | |
| 18 | | | | | | |
| 19 | | | | | | |
| ***20*** | | | -1.14 (-2.38) | -1.14 (-2.38) | | |
| 21 | -.713 (-2.67) | -.713 (-2.67) | | | | |
| 22 | -.236 (-2.84) | -.236 (-2.84) | | | | -.112 (-3.54) |
| ***23a*** | | | | | | |
| ***23b*** | | | | | | |
| 24 | | | .361 (6.27) | .361 (6.27) | | |
| 25 | | | -1.66 (-2.19) | | | |
| ***26*** | | | | | | |
| 27 | | | | | | |
| 28 | | -.870 (-2.97) | | | | |
| ***29*** | | | | | | |
| 30 | -.779 (-2.16) | -.779 (-2.16) | | | -1.62 (-2.03) | |

…. – positive coefficient
…. – negative coefficient

---

"slush" – "roadway surface is covered by snow/slush" indicator variable
"driv" – "road median is a drivable" indicator variable
"nomed" – "no median" indicator variable

Table C.5 (Continued)

| # | Coefficient (t-ratio) | | | | | |
|---|---|---|---|---|---|---|
| | curb [$X_{17}$] | | nojun [$X_{18}$] | | way4 [$X_{18}$] | |
| | fatality | injury | fatality | injury | fatality | injury |
| 1 | | | | | | .240 (2.29) |
| ***2*** | | | | | | .699 (2.81) |
| 3 | | | | | | |
| 4 | | | | | .542 (3.93) | .542 (3.93) |
| ***5*** | | | | | | |
| 6 | | | | | | |
| 7 | | | .498 (2.11) | .498 (2.11) | | |
| ***8a*** | | | | | | |
| ***8b*** | | | | | | |
| 9 | | | | -.367 (-2.09) | | |
| 10 | | | | | | |
| 11 | | | | | | |
| 12 | | | | | | |
| 13 | | | | | | |
| ***14*** | | | | | | |
| 15 | | | | | | |
| 16a | | | | | | |
| 16b | .407 (2.67) | .407 (2.67) | | | | |
| ***16c*** | | | | | | |
| ***17*** | | | | | | |
| 18 | | | | | | |
| 19 | | | -.278 (-2.82) | | | |
| ***20*** | | | | | | |
| 21 | -1.21 (-3.07) | -1.21 (-3.07) | | | | |
| 22 | | | -.371 (-11.3) | -.371 (-11.3) | | |
| ***23a*** | | | -.472 (-2.75) | | | |
| ***23b*** | | | | | | |
| 24 | | | | | .298 (3.41) | .298 (3.41) |
| 25 | | | | | | |
| ***26*** | | | | | .517 (2.16) | .517 (2.16) |
| 27 | | | | | | |
| 28 | | | -.182 (-2.44) | -.182 (-2.44) | | |
| ***29*** | | | | | .441 (2.17) | .441 (2.17) |
| 30 | | | | | | |

…. – positive coefficient
…. – negative coefficient

---

"curb" – "road median is curbed" indicator variable
"nojun" – "no road junction at the accident location" indicator variable
"way4" – "accident location is at a 4-way intersection" indicator variable

Table C.5 (Continued)

| # | Coefficient (t-ratio) | | | | | |
|---|---|---|---|---|---|---|
| | T [$X_{18}$] | | ramp [$X_{18}$] | | curve [$X_{19}$] | |
| | fatality | injury | fatality | injury | fatality | injury |
| 1 | | | | | | |
| ***2*** | | | | | | |
| 3 | | | | | | |
| 4 | | | | | | |
| ***5*** | | | | | | |
| 6 | | | | | | |
| 7 | | | | | | |
| ***8a*** | | | | | | |
| ***8b*** | | | | | | |
| 9 | | | | | | |
| 10 | | | -.394 (-2.28) | -.394 (-2.28) | | |
| 11 | | | | | | |
| 12 | | | | | | |
| 13 | | | | | | |
| ***14*** | | | | | | |
| 15 | | | | | | |
| 16a | | | | | | |
| 16b | | | | | | |
| ***16c*** | | | | | | |
| ***17*** | | | | | | |
| 18 | | | | | | |
| 19 | | | | | | |
| ***20*** | | 1.82 (2.83) | | | | |
| 21 | | | | | .311 (2.30) | .311 (2.30) |
| 22 | | | | | | |
| ***23a*** | | | | | | |
| ***23b*** | | | | | | |
| 24 | | | | | | |
| 25 | | | | | | |
| ***26*** | | | | | | |
| 27 | | | | | | .278 (2.01) |
| 28 | | | | | | |
| ***29*** | | | | | | |
| 30 | | | | | | |

…. – positive coefficient
…. – negative coefficient

"T" – "accident location is at a T-intersection" indicator variable
"ramp" – "accident location is near or on a ramp" indicator variable
"curve" – "road is at curve" indicator variable

Table C.5 (Continued)

| # | Coefficient (t-ratio) | | | | | |
|---|---|---|---|---|---|---|
| | sg [$X_{19}$] | | sl [$X_{19}$] | | str [$X_{19}$] | |
| | fatality | injury | fatality | injury | fatality | injury |
| 1 | | | | | | |
| ***2*** | | | | | | |
| 3 | | | | | -.176 (-2.93) | -.176 (-2.93) |
| 4 | | | | | | |
| ***5*** | | | | | | |
| 6 | | | | | | |
| 7 | | | | | | |
| ***8a*** | | | | | | |
| ***8b*** | -1.39 (-2.36) | -1.39 (-2.36) | | | | |
| 9 | | | -.256 (-2.33) | -.256 (-2.33) | | |
| 10 | | | | | | |
| 11 | | | | | | |
| 12 | | | | | | |
| 13 | | | | | | |
| ***14*** | | | | | | |
| 15 | | | | | -1.20 (-4.13) | -.291 (-3.12) |
| 16a | | | | | | |
| 16b | | | | | | |
| ***16c*** | | | | | | |
| ***17*** | | | | | | |
| 18 | | | | | | |
| 19 | | | | | | |
| ***20*** | | | | | | |
| 21 | | | | | | |
| 22 | | | | | | |
| ***23a*** | | | | | | |
| ***23b*** | | | | | | |
| 24 | | | .173 (2.68) | .173 (2.68) | | |
| 25 | | | | | | |
| ***26*** | | | | | | |
| 27 | | | | | | |
| 28 | | | | | | |
| ***29*** | | | | | | |
| 30 | | | | | | |

…. – positive coefficient
…. – negative coefficient

---

"sg" – "road is straight AND at grade" indicator variable
"sl" – "road is straight AND level" indicator variable
"str" – "road is straight" indicator variable

Table C.5 (Continued)

| # | Coefficient (t-ratio) | | | | | |
|---|---|---|---|---|---|---|
| | cl [$X_{19}$] | | sh [$X_{19}$] | | cg [$X_{19}$] | |
| | fatality | injury | fatality | injury | fatality | injury |
| 1 | | | .329 (2.05) | .329 (2.05) | | |
| ***2*** | | | | | | |
| 3 | | | | | | |
| 4 | | | | | | |
| ***5*** | | | | | | |
| 6 | | | | | | |
| 7 | | | | | | |
| ***8a*** | | | | | | |
| ***8b*** | | | | | | |
| 9 | | | | | | |
| 10 | | | | | | |
| 11 | | | | | .999 (2.40) | .999 (2.40) |
| 12 | | | | | | |
| 13 | | | | | | |
| ***14*** | | | | | | |
| 15 | | | | | | |
| 16a | | | | | | |
| 16b | | | | | | |
| ***16c*** | | | | | | |
| ***17*** | | | | | | |
| 18 | | | | | | |
| 19 | | | | | | |
| ***20*** | | | | | | |
| 21 | | | | | | |
| 22 | | | | | | |
| ***23a*** | | | | | | |
| ***23b*** | | | | | | |
| 24 | | | | | | |
| 25 | | | | | | |
| ***26*** | 1.62 (2.39) | | | | | |
| 27 | | | | | | |
| 28 | | | | | | |
| ***29*** | | | | | | |
| 30 | 2.73 (4.20) | | | | | |

…. – positive coefficient
…. – negative coefficient

---

"cl" – "road is at-curve AND level" indicator variable
"sh" – "road is straight AND hillcrest" indicator variable
"cg" – "road is at-curve AND at grade" indicator variable

Table C.5 (Continued)

| # | Coefficient (t-ratio) | | | | | |
|---|---|---|---|---|---|---|
| | lev [$X_{19}$] | | driver [$X_{20}$] | | veh [$X_{20}$] | |
| | fatality | injury | fatality | injury | fatality | injury |
| 1 | | | | | | |
| ***2*** | | | | | | |
| 3 | | | | | | |
| 4 | | | | | | |
| ***5*** | | | | | | |
| 6 | | | 1.27 (4.94) | 1.27 (4.94) | | |
| 7 | | | | | | |
| ***8a*** | | | | | | |
| ***8b*** | | | 1.96 (2.89) | 1.96 (2.89) | | |
| 9 | | | | 1.46 (11.6) | | |
| 10 | | | | | | |
| 11 | | | | | | |
| 12 | | -.397 (-2.76) | | 1.44 (6.83) | | |
| 13 | | | .573 (2.66) | .573 (2.66) | | |
| ***14*** | | | | | -1.58 (-2.32) | -1.58 (-2.32) |
| 15 | | | | | | |
| 16a | | | | | | |
| 16b | | | | | | |
| ***16c*** | | | | | | |
| ***17*** | | | | | | |
| 18 | | | | | | |
| 19 | | | | | | |
| ***20*** | | | | | | |
| 21 | | | | 1.45 (9.71) | | |
| 22 | | | | | | |
| ***23a*** | | | | | | |
| ***23b*** | | | | | | |
| 24 | | | 1.04 (10.5) | 1.04 (10.5) | | |
| 25 | | | | | | |
| ***26*** | | | 1.01 (2.30) | 1.01 (2.30) | | |
| 27 | | | | | | |
| 28 | | | .569 (2.37) | .569 (2.37) | | |
| ***29*** | | | | | | |
| 30 | | | | | | |

…. – positive coefficient
…. – negative coefficient

"lev" – "road is at level" indicator variable
"driver" – "primary cause of accident is driver-related" indicator variable
"veh" – "primary cause of accident is vehicle-related" indicator variable

Table C.5 (Continued)

| # | Coefficient (t-ratio) | | | | | |
|---|---|---|---|---|---|---|
| | env [$X_{20}$] | | hl5 [$X_{22}$] | | hl10 [$X_{22}$] | |
| | fatality | injury | fatality | injury | fatality | injury |
| 1 | | | | | | |
| ***2*** | | | | | 1.16 (5.04) | 1.16 (5.04) |
| 3 | -2.90 (-6.30) | -1.27 (-19.7) | | | | |
| 4 | | | | | | |
| ***5*** | | | | | | |
| 6 | | | | | .804 (4.26) | .804 (4.26) |
| 7 | | | | | | |
| ***8a*** | | | | | | |
| ***8b*** | | | | | | 1.10 (3.20) |
| 9 | | | | | | |
| 10 | | | | | | |
| 11 | | | | | | |
| 12 | | | | | | |
| 13 | | | | | .631 (6.37) | .631 (6.37) |
| ***14*** | | | | | | |
| 15 | | -1.78 (-18.5) | | | | |
| 16a | | | .700 (6.33) | .700 (6.33) | | |
| 16b | | | .662 (7.07) | .662 (7.07) | | |
| ***16c*** | | | | .442 (2.64) | | |
| ***17*** | | | | | | |
| 18 | | -1.62 (-7.63) | .887 (6.27) | .887 (6.27) | | |
| 19 | | | | | .742 (7.20) | .742 (7.20) |
| ***20*** | | | | | | |
| 21 | | | | | | |
| 22 | | | | | 1.47 (2.35) | .777 (22.8) |
| ***23a*** | | | .969 (6.00) | .969 (6.00) | | |
| ***23b*** | | | | .629 (3.08) | | |
| 24 | | | | | .872 (13.6) | .872 (13.6) |
| 25 | | | | | .647 (5.65) | |
| ***26*** | | | | | | |
| 27 | | -1.59 (-12.7) | | | | |
| 28 | | | .690 (9.24) | .690 (9.24) | | |
| ***29*** | | | .974 (4.62) | .974 (4.62) | | |
| 30 | | -1.34 (-6.07) | .965 (6.09) | | | |

…. – positive coefficient
…. – negative coefficient

"env" – "primary cause of accident is environment-related" indicator variable
"hl5" – "help arrived in 5 minutes or less after the crash" indicator variable
"hl10" – "help arrived in 10 minutes or less after the crash" indicator variable

Table C.5 (Continued)

| # | Coefficient (t-ratio) | | | | | |
|---|---|---|---|---|---|---|
| | hl20 [$X_{22}$] | | hg30 [$X_{22}$] | | car [$X_{25}$] | |
| | fatality | injury | fatality | injury | fatality | injury |
| 1 | 2.27 (2.21) | .740 (6.37) | | | | |
| ***2*** | | | | | | |
| 3 | 1.38 (5.19) | .705 (12.3) | | | | |
| 4 | .758 (3.78) | .758 (3.78) | | | | |
| ***5*** | | | | | | |
| 6 | | | | | | |
| 7 | | | | | | |
| ***8a*** | | | | | | |
| ***8b*** | | | | | | |
| 9 | | 1.06 (7.99) | | | | |
| 10 | .886 (4.97) | .886 (4.97) | | | | |
| 11 | 1.27 (4.82) | 1.27 (4.82) | | | | |
| 12 | .938 (5.52) | .938 (5.52) | | | | |
| 13 | | | | | | |
| ***14*** | 1.55 (5.23) | 1.55 (5.23) | | | | |
| 15 | .872 (9.39) | .872 (9.39) | | | | |
| 16a | | | | | | |
| 16b | | | | | | |
| ***16c*** | | | | | | |
| ***17*** | | | | | | |
| 18 | | | | | | |
| 19 | | | | | | |
| ***20*** | 1.88 (2.80) | 1.88 (2.80) | | | | |
| 21 | .881 (6.04) | .881 (6.04) | | | | |
| 22 | | | | | | |
| ***23a*** | | | | | .575 (2.87) | .575 (2.87) |
| ***23b*** | | | | | | |
| 24 | | | | | | |
| 25 | | | | | | |
| ***26*** | | | -2.40 (-3.22) | -2.40 (-3.22) | | |
| 27 | .851 (6.24) | .851 (6.24) | | | | |
| 28 | | | | | | |
| ***29*** | | | | | | |
| 30 | | | | | | |

…. – positive coefficient
…. – negative coefficient

---

"hl20" – "help arrived in 20 minutes or less after the crash" indicator variable
"hg30" – "help arrived in more than 30 minutes after the crash" indicator variable
"car" – "the vehicle at fault is a car" indicator variable

Table C.5 (Continued)

| # | Coefficient (t-ratio) | | | | | |
|---|---|---|---|---|---|---|
| | SUV [$X_{25}$] | | heavy [$X_{25}$] | | moto [$X_{25}$] | |
| | fatality | injury | fatality | injury | fatality | injury |
| 1 | | | | | | |
| ***2*** | | | | | | |
| 3 | | | | | 2.93 (12.2) | 2.93 (12.2) |
| 4 | | | | | | |
| ***5*** | | | | | | |
| 6 | | | | | 2.32 (3.69) | 2.32 (3.69) |
| 7 | | | | | | |
| ***8a*** | | | | | | |
| ***8b*** | | | | | | |
| 9 | | | | | | |
| 10 | | | | | | |
| 11 | | | | | | |
| 12 | | | | | 2.86 (4.51) | 2.86 (4.51) |
| 13 | | | | | | |
| ***14*** | | | -1.19 (-2.07) | -1.19 (-2.07) | | |
| 15 | | | | | 3.09 (9.82) | 3.09 (9.82) |
| 16a | | | | | | |
| 16b | | | | | | |
| ***16c*** | | | | | | |
| ***17*** | | | | | | |
| 18 | | | | | 1.94 (4.97) | 1.94 (4.97) |
| 19 | | | | | | |
| ***20*** | | | | | | |
| 21 | | | | | | 3.46 (8.00) |
| 22 | | | | | | |
| ***23a*** | | | | | | |
| ***23b*** | | | | | | |
| 24 | | | | | 2.31 (11.5) | 2.31 (11.5) |
| 25 | | -.550 (-2.78) | | | | |
| ***26*** | | | | | | |
| 27 | | | | | 4.19 (5.67) | 4.19 (5.67) |
| 28 | | | | | | |
| ***29*** | | | | | | |
| 30 | | | | | | |

.... – positive coefficient
.... – negative coefficient

---

"SUV" – "the vehicle at fault is a SUV" indicator variable
"heavy" – "the vehicle at fault is a truck or a tractor" indicator variable
"moto" – "the vehicle at fault is a motorcycle" indicator variable

Table C.5 (Continued)

| # | Coefficient (t-ratio) | | | | | |
|---|---|---|---|---|---|---|
| | pickup [$X_{25}$] | | van [$X_{25}$] | | trac1 [$X_{25}$] | |
| | fatality | injury | fatality | injury | fatality | injury |
| 1 | | | | | | |
| ***2*** | | | | | | |
| 3 | | | | | | |
| 4 | | | | | | |
| ***5*** | | | | | | |
| 6 | | | | | | |
| 7 | | | | | | |
| ***8a*** | | | | | | |
| ***8b*** | | | | | | |
| 9 | | | | | | |
| 10 | | | | | | |
| 11 | | | | | | |
| 12 | | | | | | |
| 13 | | | | | | |
| ***14*** | | | | | | |
| 15 | | | | | | |
| 16a | | | | | | |
| 16b | | | .300 (2.08) | .300 (2.08) | | |
| ***16c*** | | -.351 (-2.11) | | | | |
| ***17*** | | | | | | |
| 18 | | | | | | |
| 19 | | | -.428 (-2.20) | -.428 (-2.20) | | |
| ***20*** | | | | | | |
| 21 | | | | | | |
| 22 | | | | | | |
| ***23a*** | | | | | | |
| ***23b*** | | | | | -.905 (-2.71) | -.905 (-2.71) |
| 24 | | | | | | |
| 25 | | | | | | |
| ***26*** | | | | | | |
| 27 | | | | | | |
| 28 | | -.291 (-2.83) | | | | |
| ***29*** | | | | | | |
| 30 | | | | | | |

…. – positive coefficient
…. – negative coefficient

"pickup" – "the vehicle at fault is a pickup" indicator variable
"van" – "the vehicle at fault is a van" indicator variable
"trac1" – "the vehicle at fault is a tractor OR one semi-trailer" indicator variable

Table C.5 (Continued)

| # | Coefficient (t-ratio) | | | | | |
|---|---|---|---|---|---|---|
| | vage [$X_{26}$] | | voldg [$X_{26}$] | | v7g [$X_{26}$] | |
| | fatality | injury | fatality | injury | fatality | injury |
| 1 | .128 (3.91) | | | | | |
| ***2*** | | | | | | |
| 3 | .0311 (6.59) | .0311 (6.59) | | | | |
| 4 | | .0420 (3.14) | | | | |
| ***5*** | | | -1.25 (-3.74) | -1.25 (-3.74) | | |
| 6 | | | | | -.532 (-2.47) | -.532 (-2.47) |
| 7 | | | | | | |
| ***8a*** | | | | | -.797 (-2.55) | -.797 (-2.55) |
| ***8b*** | | | | | | |
| 9 | | | | .228 (2.04) | | |
| 10 | | | | | | |
| 11 | | | | | | |
| 12 | | | | | | |
| 13 | | | | | | |
| ***14*** | .0601 (3.35) | .0601 (3.35) | | | | |
| 15 | .0332 (4.37) | .0332 (4.37) | | | | |
| 16a | | | | | | |
| 16b | | | | | | |
| ***16c*** | | | | | | |
| ***17*** | | | | | | |
| 18 | | | | | | |
| 19 | | | | | | |
| ***20*** | | | | | | |
| 21 | .0216 (2.02) | .0216 (2.02) | | | | |
| 22 | | | 1.40 (2.74) | .110 (3.72) | | |
| ***23a*** | | | | | | |
| ***23b*** | | | | | | |
| 24 | | | | .315 (5.53) | | |
| 25 | | | | | | |
| ***26*** | | | | | | |
| 27 | .0329 (3.27) | .0329 (3.27) | | | | |
| 28 | | | | | | |
| ***29*** | | | | | | |
| 30 | | | | | | |

…. – positive coefficient
…. – negative coefficient

"vage" – "age (in years) of the vehicle at fault" quantitative variable
"voldg" – "the vehicle at fault is more than 7 years old" indicator variable
"v7g" – "age of the vehicle at fault is ≥ 3 and ≤ 7 years" indicator variable

Table C.5 (Continued)

| # | Coefficient (t-ratio) | | | | | |
|---|---|---|---|---|---|---|
| | v3g [$X_{26}$] | | $X_{27}$ | | Ind [$X_{28}$] | |
| | fatality | injury | fatality | injury | fatality | injury |
| 1 | | | | | | |
| ***2*** | | | | | | |
| 3 | | | .0382 (3.47) | .0382 (3.47) | | |
| 4 | | | | | | |
| ***5*** | | | | | | |
| 6 | | | | | | |
| 7 | | | | | | |
| ***8a*** | | | .303 (2.37) | .303 (2.37) | | |
| ***8b*** | | | | | | |
| 9 | | | .110 (2.60) | .110 (2.60) | | |
| 10 | | | | | | |
| 11 | | | | | | |
| 12 | | | | | | |
| 13 | | | | | | |
| ***14*** | | | | | | |
| 15 | | | | | | |
| 16a | | | | | | |
| 16b | -.406 (-2.71) | -.406 (-2.71) | -.142 (-2.15) | -.142 (-2.15) | .404 (2.04) | .404 (2.04) |
| ***16c*** | | | | | | |
| ***17*** | | | | | | |
| 18 | | | | | | |
| 19 | | | -.238 (-3.14) | -.238 (-3.14) | | |
| ***20*** | | | | | | |
| 21 | | | .133 (2.26) | .133 (2.26) | | |
| 22 | | | -.0628 (-2.89) | -.0628 (-2.89) | | |
| ***23a*** | | | | | | |
| ***23b*** | | | | | | |
| 24 | | | .0817 (3.11) | .0817 (3.11) | .477 (4.10) | .477 (4.10) |
| 25 | | | | | | |
| ***26*** | | | | | | |
| 27 | | | | | | |
| 28 | | | | | | |
| ***29*** | | | | | | |
| 30 | | | | | | |

…. – positive coefficient
…. – negative coefficient

---

"v3g" – "age of the vehicle at fault is > 1 and ≤ 3 years" indicator variable
"$X_{27}$" – "number of occupants in the vehicle at fault" quantitative variable
"Ind" – "license state of the vehicle at fault is Indiana" indicator variable

Table C.5 (Continued)

| # | Coefficient (t-ratio) | | | | | |
|---|---|---|---|---|---|---|
| | othUS [$X_{28}$] | | lnm [$X_{30}$] | | r22 [$X_{30}$] | |
| | fatality | injury | fatality | injury | fatality | injury |
| 1 | | | | | | |
| ***2*** | | | | | | |
| 3 | | | | | | |
| 4 | | | | | | |
| ***5*** | | | | | | |
| 6 | 3.60 (2.82) | 3.60 (2.82) | | | | |
| 7 | | | | | | |
| ***8a*** | | | | | | |
| ***8b*** | | | | | | |
| 9 | | | | | | |
| 10 | -.970 (-2.20) | -.970 (-2.20) | | | | |
| 11 | -.695 (-2.28) | -.695 (-2.28) | | | | |
| 12 | | | | | | |
| 13 | | | | | | .369 (3.18) |
| ***14*** | | | | | | |
| 15 | | | | | | |
| 16a | | | | | | |
| 16b | | | | | | |
| ***16c*** | | | | | | |
| ***17*** | | | | | | |
| 18 | -1.12 (-2.26) | -1.12 (-2.26) | .414 (2.85) | .414 (2.85) | | |
| 19 | | | | | | |
| ***20*** | | | | | | |
| 21 | | | | | | |
| 22 | | | | | | |
| ***23a*** | | | | | | |
| ***23b*** | | | | | | |
| 24 | | | | | | |
| 25 | | | | | | |
| ***26*** | | | | | | |
| 27 | | | | | | |
| 28 | | | | | | |
| ***29*** | | | | | | |
| 30 | | | | | | |

…. – positive coefficient
…. – negative coefficient

"othUS" – "license state of the vehicle at fault is a U.S. state except Indiana and its neighboring states (IL, KY, OH, MI)" indicator variable
"lnm" – "road traveled by the vehicle at fault is multi-lane" indicator variable
"r22" – "road traveled by the vehicle at fault is two-lane AND two-way" indicator variable

Table C.5 (Continued)

| # | Coefficient (t-ratio) | | | | | |
|---|---|---|---|---|---|---|
| | rmu22 [$X_{30}$] | | rmd2 [$X_{30}$] | | priv [$X_{30}$] | |
| | fatality | injury | fatality | injury | fatality | injury |
| 1 | | | | | | |
| ***2*** | | | | | | |
| 3 | | | | | | |
| 4 | | | | | | |
| ***5*** | | | | 2.06 (2.99) | | |
| 6 | | | | | | |
| 7 | | | | | | |
| ***8a*** | | | | | | |
| ***8b*** | | | | | | |
| 9 | | | | | | |
| 10 | 3.89 (3.15) | | | | | |
| 11 | | | | | | |
| 12 | | | | | | |
| 13 | | | | | | |
| ***14*** | | | | | | |
| 15 | | | | | | |
| 16a | | | | | | |
| 16b | | | | | | |
| ***16c*** | | | | | | |
| ***17*** | | | | | | |
| 18 | | | | | | |
| 19 | | | | | | |
| ***20*** | | | | | | |
| 21 | | | | | -1.14 (-3.10) | -1.14 (-3.10) |
| 22 | | | | | -.489 (-3.05) | -.489 (-3.05) |
| ***23a*** | | | .374 (1.98) | .374 (1.98) | | |
| ***23b*** | | | | | | |
| 24 | | | | | | -.490 (-4.09) |
| 25 | | | | | | |
| ***26*** | | | | | | |
| 27 | | | | | | |
| 28 | | | | | | |
| ***29*** | | | | | | |
| 30 | | | | | | |

…. – positive coefficient
…. – negative coefficient

"rmu22" – "road traveled by the vehicle at fault is multi-lane AND undivided two-way left" indicator variable
"rmd2" – "road traveled by the vehicle at fault is multi-lane AND divided three or more" indicator variable
"priv" – "road traveled by the vehicle at fault is a private drive" indicator variable

Table C.5 (Continued)

| # | Coefficient (t-ratio) | | | | | |
|---|---|---|---|---|---|---|
| | stopsig [$X_{31}$] | | nosig [$X_{31}$] | | nopass [$X_{31}$] | |
| | fatality | injury | fatality | injury | fatality | injury |
| 1 | .260 (2.50) | .260 (2.50) | | | | |
| ***2*** | | | | | | |
| 3 | | | | | | |
| 4 | | | | | | |
| ***5*** | | | | | | |
| 6 | | | | | | |
| 7 | | | | | | |
| ***8a*** | | | | | | |
| ***8b*** | | | | | | |
| 9 | | | | | | |
| 10 | | | | | | |
| 11 | | | | | | |
| 12 | | | | | | |
| 13 | 1.24 (2.48) | .444 (3.02) | | | | |
| ***14*** | | | | | | |
| 15 | | | | | .198 (2.02) | .198 (2.02) |
| 16a | | | | | | |
| 16b | | | | | | |
| ***16c*** | | | | | | |
| ***17*** | | | | | 3.51 (2.41) | |
| 18 | | | | | | |
| 19 | | | | | | |
| ***20*** | | | | | | |
| 21 | | | | | | |
| 22 | | | -.106 (-2.71) | -.106 (-2.71) | | |
| ***23a*** | | | | | | |
| ***23b*** | | | | | | |
| 24 | | | | | | |
| 25 | .505 (2.82) | .505 (2.82) | | | | |
| ***26*** | | | | | | |
| 27 | | | | -.286 (-2.45) | | |
| 28 | | | | | | |
| ***29*** | | | | | | |
| 30 | | | | | | |

…. – positive coefficient
…. – negative coefficient

---

"stopsig" – "traffic control device for the vehicle at fault is a «stop sign»" indicator variable
"nosig" – "no any traffic control device for the vehicle at fault" indicator variable
"nopass" – "traffic control device for the vehicle at fault is a «no passing zone»" indicator variable

Table C.5 (Continued)

| # | Coefficient (t-ratio) | | | | | |
|---|---|---|---|---|---|---|
| | sig [$X_{31}$] | | other [$X_{31}$] | | $X_{33}$ | |
| | fatality | injury | fatality | injury | fatality | injury |
| 1 | | | | | 2.36 (1.99) | 1.85 (3.11) |
| ***2*** | | | | | | |
| 3 | | | 1.56 (2.85) | .631 (2.52) | 2.24 (5.03) | 1.00 (3.88) |
| 4 | | | | | | |
| ***5*** | | | | | | |
| 6 | | | | | 2.81 (2.36) | 2.81 (2.36) |
| 7 | | | | | | |
| ***8a*** | | | | | 6.11 (4.48) | |
| ***8b*** | | | | | | |
| 9 | | | | | | |
| 10 | | | | | 1.78 (2.03) | 1.78 (2.03) |
| 11 | | | | | 4.51 (4.33) | |
| 12 | | | | | | |
| 13 | | | | | 2.62 (2.82) | |
| ***14*** | | | | | 2.81 (2.74) | |
| 15 | | | | | 1.96 (3.13) | |
| 16a | | | | | | |
| 16b | | | | | | |
| ***16c*** | | | | | 5.07 (3.73) | |
| ***17*** | | | | | | |
| 18 | | | | | | |
| 19 | | | | | | |
| ***20*** | | | | | | |
| 21 | | | | | | |
| 22 | | | | | 1.95 (8.94) | 1.95 (8.94) |
| ***23a*** | | | | | | 2.67 (3.01) |
| ***23b*** | | | | | | |
| 24 | .268 (2.44) | .268 (2.44) | | | 1.16 (4.19) | 1.16 (4.19) |
| 25 | | | | | | |
| ***26*** | | | | | 3.21 (3.57) | |
| 27 | | | | | | |
| 28 | | | | | 1.91 (3.60) | 1.91 (3.60) |
| ***29*** | | | | | | |
| 30 | | | | | | |

…. – positive coefficient
…. – negative coefficient

"sig" – "traffic control device for the vehicle at fault is a signal" indicator variable
"other" – "traffic control device for the vehicle at fault is an «other regulatory sign or marking»" indicator variable
"$X_{33}$" – "at least one of the vehicles involved was on fire" indicator variable

Table C.5 (Continued)

| # | Coefficient (t-ratio) | | | | | |
|---|---|---|---|---|---|---|
| | $X_{34}$ | | age2 [$X_{34}$] | | age3 [$X_{34}$] | |
| | fatality | injury | fatality | injury | fatality | injury |
| 1 | | | | | | |
| ***2*** | | | | | | |
| 3 | -.0071 (3.93) | | | | | |
| 4 | | | | | | |
| ***5*** | | | | | | |
| 6 | | | | | | |
| 7 | | | | | | |
| ***8a*** | | | | | | |
| ***8b*** | | | | | | |
| 9 | | | -.576 (-3.67) | -.576 (-3.67) | | |
| 10 | | | | | -.377 (-2.09) | -.377 (-2.09) |
| 11 | | | | | | |
| 12 | | | | | | |
| 13 | | | | | | |
| ***14*** | | | | | | |
| 15 | | | | | | |
| 16a | | | | | | |
| 16b | | | | | | |
| ***16c*** | | | | | | |
| ***17*** | | | | | | |
| 18 | | | | | | |
| 19 | | | | | | |
| ***20*** | | | | | | |
| 21 | | | | | | |
| 22 | .0346 (3.54) | .00180 (2.20) | | | | |
| ***23a*** | | | | | | |
| ***23b*** | | | | | | |
| 24 | | .381 (2.20) | | | | |
| 25 | | | | | | |
| ***26*** | | | | | | |
| 27 | | | | | | |
| 28 | | | | | | |
| ***29*** | | | | | | |
| 30 | | | | | | |

…. – positive coefficient
…. – negative coefficient

---

"$X_{34}$" – "age (in years) of the driver at fault" quantitative variable
"age2" – "age of the driver at fault is ≥ 24 and < 30" indicator variable
"age3" – "age of the driver at fault is ≥ 30 and < 40" indicator variable

Table C.5 (Continued)

| # | Coefficient (t-ratio) | | | | | |
|---|---|---|---|---|---|---|
| | age5 [$X_{34}$] | | $X_{35}$ | | oldvage [$X_{26}$] | |
| | fatality | injury | fatality | injury | fatality | injury |
| 1 | 1.09 (2.37) | | | | | |
| ***2*** | | | | .540 (2.11) | | |
| 3 | | | | .290 (5.39) | | |
| 4 | | | -.403 (-2.51) | -.403 (-2.51) | | |
| ***5*** | | | | | | |
| 6 | | | | | | |
| 7 | | | | | | |
| ***8a*** | | | | | | |
| ***8b*** | | | | | | |
| 9 | | | .365 (3.44) | .365 (3.44) | | |
| 10 | | | | | | |
| 11 | | | | | | |
| 12 | | | | .533 (3.94) | | |
| 13 | | | -2.02 (-2.69) | | | |
| ***14*** | | | | | | |
| 15 | | | -.900 (-2.42) | .229 (2.74) | | |
| 16a | | | | | | |
| 16b | | | | | | |
| ***16c*** | | | | | | .0428 (2.73) |
| ***17*** | | | | | | |
| 18 | | | | | | |
| 19 | | | | | | |
| ***20*** | | -1.33 (-2.11) | | | | |
| 21 | | | | | | |
| 22 | | | | | | |
| ***23a*** | | -.556 (-2.82) | | -.600 (-2.47) | | |
| ***23b*** | | | | | | |
| 24 | | | | .412 (7.07) | | |
| 25 | | | | | | |
| ***26*** | | | | | | |
| 27 | | | | | | |
| 28 | | | | | | |
| ***29*** | | | | | | |
| 30 | | | | | | |

…. – positive coefficient
…. – negative coefficient

"age5" – "age of the driver at fault is ≥ 50 years" indicator variable
"$X_{35}$" – "gender of the driver at fault: 1 – female, 0 – male" indicator variable
"oldvage" – "age (in years) of the oldest vehicle involved" quantitative variable

Table C.5 (Continued)

| # | Coefficient (t-ratio) | | | | | |
|---|---|---|---|---|---|---|
| | voldo [$X_{26}$] | | maxpass [$X_{27}$] | | age5y [$X_{34}$] | |
| | fatality | injury | fatality | injury | fatality | injury |
| 1 | | | -.691 (-1.99) | .133 (3.60) | | |
| ***2*** | | | | | | |
| 3 | | | | | | |
| 4 | | | .186 (2.71) | .186 (2.71) | | |
| ***5*** | | | | .863 (2.60) | | |
| 6 | | | | | | |
| 7 | | | | | | |
| ***8a*** | | | | | | |
| ***8b*** | | | | | -1.50 (-3.39) | -1.50 (-3.39) |
| 9 | | | | | | |
| 10 | | | .600 (2.64) | .141 (2.42) | | |
| 11 | | | | | | |
| 12 | | | | | | |
| 13 | | | | | | |
| ***14*** | | | | .282 (2.34) | | |
| 15 | | | | | | |
| 16a | | | .182 (3.32) | .182 (3.32) | | |
| 16b | | | .207 (4.04) | .207 (4.04) | | |
| ***16c*** | | | | | | |
| ***17*** | | | | | | |
| 18 | | | | | | |
| 19 | | | .198 (3.40) | .198 (3.40) | | |
| ***20*** | | | | | | |
| 21 | | | | | | |
| 22 | | | .190 (10.8) | .190 (10.8) | | |
| ***23a*** | | | | | | |
| ***23b*** | .415 (2.07) | .415 (2.07) | | | | |
| 24 | | | | | | |
| 25 | | | | | | |
| ***26*** | .592 (2.60) | .592 (2.60) | | | | |
| 27 | | | | | | |
| 28 | | | .132 (4.10) | .132 (4.10) | | |
| ***29*** | | | | .244 (2.10) | | |
| 30 | | | | | | |

…. – positive coefficient
…. – negative coefficient

---

"voldo" – "age of the oldest vehicle involved is > 7 years" indicator variable
"maxpass" – "the largest number of occupants in all vehicles involved" quantitative variable
"age5y" – "age of the youngest driver is ≥ 50 years" indicator variable

Table C.5 (Continued)

| # | Coefficient (t-ratio) | | | | | |
|---|---|---|---|---|---|---|
| | ff [$X_{35}$] | | mm [$X_{35}$] | | mf [$X_{35}$] | |
| | fatality | injury | fatality | injury | fatality | injury |
| 1 | | | | | | |
| ***2*** | | | | | | |
| 3 | | | | | | |
| 4 | | | | -.676 (-3.68) | | |
| ***5*** | | | | | | |
| 6 | | | | | | |
| 7 | | | | | | |
| ***8a*** | | | | | | |
| ***8b*** | | | | | | |
| 9 | | | | | | |
| 10 | .448 (2.66) | .448 (2.66) | | | | |
| 11 | | | | | | |
| 12 | | | | | | |
| 13 | | | | | | |
| ***14*** | | | | | | |
| 15 | | | | | | |
| 16a | | .280 (2.53) | | | | |
| 16b | | .462 (4.21) | | | | |
| ***16c*** | | | | | | |
| ***17*** | | | | | | |
| 18 | | | | | | |
| 19 | | | -.245 (-2.27) | -.245 (-2.27) | | |
| ***20*** | | | | | | |
| 21 | | | | | | |
| 22 | | | | -.262 (-7.73) | | |
| ***23a*** | | | | | .473 (2.49) | .473 (2.49) |
| ***23b*** | | | | | | |
| 24 | | | | | | |
| 25 | | | | | | |
| ***26*** | | | | | -1.97 (-2.50) | |
| 27 | | | | | | |
| 28 | | | -.306 (-3.73) | -.306 (-3.73) | | |
| ***29*** | | | | | | |
| 30 | | | | | | |

…. – positive coefficient
…. – negative coefficient

---

"ff" – "two female drivers involved into a two-vehicle accident" indicator variable
"mm" – "two male drivers involved into a two-vehicle accident" indicator variable
"mf" – "male and female drivers involved into a two-vehicle accident" indicator variable

Table C.6 Multinomial logit models for 2006 accident severity

| # | Model name | | | Log-likelihood | | $R^2$ | Coefficient (t-ratio) | |
|---|---|---|---|---|---|---|---|---|
| | | | | | | | $X_{29}$ | |
| | | | | model | restricted* | | fatality [ $\beta_1$ ] | injury [ $\beta_2$ ] |
| 1 | County road | rural | (car/SUV)+(car/SUV) | -2345.1 | -2457.6 | .046 | .0396 (5.48) | .0396 (5.48) |
| ***2*** | County road | rural | (car/SUV)+(truck) | -113.99 | -135.50 | .159 | .0648 (3.06) | .0648 (3.06) |
| 3 | County road | rural | one vehicle | -7152.1 | -8621.2 | .170 | .00506 (2.04) | .00506 (2.04) |
| ***4a*** | County road | urban | (car)+(car) | -356.34 | -256.92 | -.387 | | |
| ***4b*** | County road | urban | (car)+(SUV) | -232.26 | -106.65 | n/a | | .0613 (2.43) |
| ***4c*** | County road | urban | (SUV)+(SUV) | -183.12 | -100.57 | -.821 | | |
| ***5*** | County road | urban | (car/SUV)+(truck) | -16.716 | -24.014 | .304 | | |
| 6 | County road | urban | one vehicle | -367.48 | -418.24 | .121 | | |
| 7 | Interstate | rural | (car/SUV)+(car/SUV) | -459.31 | -485.53 | .054 | | |
| ***8*** | Interstate | rural | (car/SUV)+(truck) | -271.78 | -294.68 | .078 | | |
| 9 | Interstate | rural | one vehicle | -1491.0 | -1718.3 | .132 | | |
| 10 | Interstate | urban | (car/SUV)+(car/SUV) | -853.29 | -887.05 | .038 | | |
| ***11*** | Interstate | urban | (car/SUV)+(truck) | -252.55 | -278.96 | .097 | | |
| 12 | Interstate | urban | one vehicle | -900.79 | -1032.2 | .127 | | |
| 13 | State route | rural | (car/SUV)+(car/SUV) | -560.40 | -605.43 | .074 | .248 (3.48) | .0416 (3.25) |
| ***14*** | State route | rural | (car/SUV)+(truck) | -72.403 | -84.740 | .146 | .127 (2.50) | .127 (2.50) |
| 15 | State route | rural | one vehicle | -3704.3 | -4873.2 | .240 | .0636 (2.34) | |
| 16 | State route | urban | (car/SUV)+(car/SUV) | -3225.9 | -3368.2 | .042 | .252 (3.35) | .0290 (7.95) |
| ***17*** | State route | urban | (car/SUV)+(truck) | -192.97 | -203.31 | .051 | | |
| 18 | State route | urban | one vehicle | -899.36 | -982.58 | .085 | | |
| 19 | City street | rural | (car/SUV)+(car/SUV) | -1068.42 | -1129.4 | .054 | .0414 (6.13) | .0414 (6.13) |
| ***20*** | City street | rural | (car/SUV)+(truck) | -77.136 | -28.602 | n/a | | |
| 21 | City street | rural | one vehicle | -317.09 | -382.38 | .171 | | |
| 22a | City street | urban | (car)+(car) | -6828.1 | -7148.6 | .045 | .0251 (6.33) | .0251 (6.33) |
| 22b | City street | urban | (car)+(SUV) | -4289.5 | -4480.3 | .043 | .0218 (4.65) | .0218 (4.65) |
| 22c | City street | urban | (SUV)+(SUV) | -1341.6 | -1413.2 | .051 | .0343 (4.20) | .0343 (4.20) |
| 23 | City street | urban | (car/SUV)+(truck) | -520.63 | -563.91 | .077 | .0284 (2.34) | .0284 (2.34) |
| 24 | City street | urban | one vehicle | -2681.6 | -2955.4 | .093 | | |
| 25 | US route | rural | (car/SUV)+(car/SUV) | -380.51 | -412.77 | .078 | | |
| ***26*** | US route | rural | (car/SUV)+(truck) | -237.97 | -268.53 | .113 | .0608 (3.07) | .0608 (3.07) |
| 27 | US route | rural | one vehicle | -1491.0 | -1917.7 | .223 | | |
| 28 | US route | urban | (car/SUV)+(car/SUV) | -967.22 | -1003.9 | .037 | .0154 (2.14) | .0154 (2.14) |
| ***29*** | US route | urban | (car/SUV)+(truck) | -185.87 | -209.82 | .114 | .0586 (3.60) | .0586 (3.60) |
| 30 | US route | urban | one vehicle | -176.49 | -202.75 | .129 | | |

.... – positive coefficient  .... – negative coefficient

* – restricted log-likelihood found by setting all coefficients except intercepts to zero

** – models are estimated by using procedure A on page 32, except the models marked by bold numbers and estimated by using procedure B on page 33

---

$X_{29}$” – “posted speed limit (if the same for all vehicles involved)” quantitative variable

Table C.6 (Continued)

| # | Coefficient (t-ratio) | | | | | |
|---|---|---|---|---|---|---|
| | constant | | wint [$X_3$] | | sum [$X_3$] | |
| | fatality | injury | fatality | injury | fatality | injury |
| 1 | -9.88 (-8.23) | -3.45 (-10.0) | | | | |
| ***2*** | -6.57 (-4.36) | -5.00 (-4.82) | | | | |
| 3 | -5.31 (-21.2) | -2.25 (-14.9) | | | | |
| ***4a*** | | .954 (5.64) | | | | |
| ***4b*** | | -4.09 (-3.78) | | | | |
| ***4c*** | | | | | | |
| ***5*** | -8.86 (-4.10) | -7.25 (-3.69) | | | | |
| 6 | -7.00 (-9.20) | -2.54 (-8.65) | | | | |
| 7 | -6.27 (-11.2) | -2.89 (-10.2) | | | | |
| ***8*** | -5.01 (-10.8) | -3.24 (-8.56) | | | | |
| 9 | -10.2 (-7.26) | -3.76 (-22.9) | | | | |
| 10 | -8.42 (-8.23) | -2.90 (-12.5) | | | | |
| ***11*** | -6.05 (-8.32) | -2.70 (-10.9) | | | | |
| 12 | -6.62 (-15.0) | -3.84 (-15.8) | | -.414 (-2.69) | | |
| 13 | -18.5 (-4.74) | -5.00 (-5.88) | | | | |
| ***14*** | -9.24 (-3.42) | -8.38 (-3.12) | | | | |
| 15 | -7.83 (-5.28) | -1.78 (-16.9) | | | .198 (2.71) | .198 (2.71) |
| 16 | -19.1 (-4.58) | -2.77 (-16.6) | | | | |
| ***17*** | -3.43 (-3.37) | -2.22 (-12.6) | | | | |
| 18 | -4.93 (-14.3) | -1.99 (-8.01) | | | | |
| 19 | -9.35 (-9.03) | -3.58 (-11.8) | | | | |
| ***20*** | | -2.08 (-4.04) | | | | |
| 21 | -4.65 (-8.25) | -1.07 (-3.87) | | | .582 (2.55) | .582 (2.55) |
| 22a | -9.47 (-13.1) | -2.58 (-15.9) | | | | |
| 22b | -8.25 (-15.5) | -2.27 (-11.9) | | | | |
| 22c | -8.34 (-12.6) | -3.04 (-9.33) | | | | |
| 23 | -8.69 (-7.93) | -4.19 (-8.75) | | | | |
| 24 | -5.63 (-21.9) | -3.42 (-15.1) | | -.207 (-2.48) | | |
| 25 | -4.44 (-9.52) | -2.04 (4.11) | | | | |
| ***26*** | -7.48 (-6.35) | -4.32 (-3.99) | | | | |
| 27 | -4.93 (-10.5) | -1.24 (-3.40) | | | | |
| 28 | -8.27 (-7.86) | -2.13 (-6.45) | | | | |
| ***29*** | -11.4 (-4.75) | -3.56 (-5.04) | | | | |
| 30 | -5.39 (-7.12) | -2.32 (-4.48) | | | | |

…. – positive coefficient
…. – negative coefficient

"constant" – "constant term (intercept)" quantitative variable
"wint" – "winter season" indicator variable
"sum" – "summer season" indicator variable

Table C.6 (Continued)

| # | Coefficient (t-ratio) | | | | | |
|---|---|---|---|---|---|---|
| | fall [$X_3$] | | mon [$X_4$] | | sund [$X_4$] | |
| | fatality | injury | fatality | injury | fatality | injury |
| 1 | 2.28 (4.01) | | | | | |
| ***2*** | | | | | | |
| 3 | | | | | | |
| ***4a*** | | | | | | |
| ***4b*** | | | | | | |
| ***4c*** | | | | | | |
| ***5*** | | | 3.49 (2.53) | | | |
| 6 | | | | | | |
| 7 | | | | | | |
| ***8*** | | | | | | |
| 9 | | | | | | |
| 10 | | | | | | |
| ***11*** | -1.18 (-3.64) | -1.18 (-3.64) | | | | |
| 12 | | | | | | |
| 13 | | | | | | |
| ***14*** | | | | | | |
| 15 | | | | | | |
| 16 | | | | | | |
| ***17*** | | | | | | |
| 18 | | | | | | |
| 19 | | | | | | |
| ***20*** | | | | | | |
| 21 | | | | | | |
| 22a | | | | | | |
| 22b | | | 1.79 (2.52) | | | |
| 22c | | | | | | |
| 23 | | | | | .876 (2.19) | .876 (2.19) |
| 24 | | | | | | |
| 25 | 2.14 (4.11) | | | | | |
| ***26*** | | -1.49 (-3.92) | | | 2.99 (4.08) | |
| 27 | | | | | | |
| 28 | | | | | | |
| ***29*** | -.714 (-2.00) | -.714 (-2.00) | | | | |
| 30 | | | | | | |

…. – positive coefficient
…. – negative coefficient

"fall" – "fall season" indicator variable
"mon" – "Monday" indicator variable
"sund" – "Sunday" indicator variable

Table C.6 (Continued)

| # | Coefficient (t-ratio) | | | | | |
|---|---|---|---|---|---|---|
| | wday [$X_4$] | | wed [$X_4$] | | jobend [$X_5$] | |
| | fatality | injury | fatality | injury | fatality | injury |
| 1 | | | | | | |
| ***2*** | | | | | | |
| 3 | -.159 (-3.41) | -.159 (-3.41) | | | | |
| ***4a*** | | | | | | |
| ***4b*** | | | | | | |
| ***4c*** | | | | | | |
| ***5*** | | | | | | |
| 6 | | | | | | |
| 7 | | | | | | |
| ***8*** | | | | | | |
| 9 | | | | | | |
| 10 | | | | | | |
| ***11*** | | | | | | |
| 12 | | | | | | |
| 13 | | | | | | |
| ***14*** | | | | | | |
| 15 | | | | | | |
| 16 | | | | | | |
| ***17*** | -2.81 (-1.97) | | | | | |
| 18 | | | | | | |
| 19 | | | | | | |
| ***20*** | | | | | | |
| 21 | | | | | | |
| 22a | | | | | | |
| 22b | | | | | | |
| 22c | | | | | | |
| 23 | | | | | | |
| 24 | | | | -.264 (-2.58) | .214 (2.45) | .214 (2.45) |
| 25 | | | | | | |
| ***26*** | | | | | | |
| 27 | | | | | | |
| 28 | | | | | | |
| ***29*** | | | | | | |
| 30 | | | | | | |

…. – positive coefficient
…. – negative coefficient

"wday" – "any weekday except Saturday and Sunday" indicator variable
"wed" – "Wednesday" indicator variable
"jobend" – "end of job hours: from 16:00 to 19:00" indicator variable

Table C.6 (Continued)

| # | Coefficient (t-ratio) | | | | | |
|---|---|---|---|---|---|---|
| | peak $[X_5]$ | | nigh $[X_5]$ | | dayt $[X_5]$ | |
| | fatality | injury | fatality | injury | fatality | injury |
| 1 | | | 2.15 (2.32) | | | |
| ***2*** | | | | | | |
| 3 | | | | | | |
| ***4a*** | | | | | | |
| ***4b*** | | | | | | |
| ***4c*** | | | | | | |
| ***5*** | | | | | | |
| 6 | | | | | | |
| 7 | | | 1.06 (2.27) | 1.06 (2.27) | | |
| ***8*** | | | | | -.641 (-2.62) | -.641 (-2.62) |
| 9 | | | | | | .252 (2.46) |
| 10 | | | | | | |
| ***11*** | | | | | | |
| 12 | | | | | | |
| 13 | | | | | | |
| ***14*** | | | | | | |
| 15 | | | | | | .178 (2.59) |
| 16 | | | | | | |
| ***17*** | | | | | | |
| 18 | | | | | | |
| 19 | | | | | | |
| ***20*** | | | | | | |
| 21 | | | | | | .477 (2.11) |
| 22a | -.117 (-2.39) | -.117 (-2.39) | | | | |
| 22b | | -.137 (-2.20) | | | | |
| 22c | | | | | | |
| 23 | | | | | | |
| 24 | | | | | | |
| 25 | | | | | | |
| ***26*** | | | | | | |
| 27 | | | | | | |
| 28 | | | | | | |
| ***29*** | | | | | | |
| 30 | | | | | | |

.... – positive coefficient
.... – negative coefficient

"peak" – "rush hours: 7:00 to 9:00 OR 17:00 to 19:00" indicator variable
"nigh" – "late night hours: from 1:00 to 5:00" indicator variable
"dayt" – "day hours: from 9:00 to 17:00" indicator variable

Table C.6 (Continued)

| # | Coefficient (t-ratio) | | | | | |
|---|---|---|---|---|---|---|
| | lunch [$X_5$] | | even [$X_5$] | | nocons [$X_{13}$] | |
| | fatality | injury | fatality | injury | fatality | injury |
| 1 | | | | | | |
| ***2*** | -1.25 (-2.44) | -1.25 (-2.44) | | | | |
| 3 | | | | | | |
| ***4a*** | | | | | | |
| ***4b*** | | | | | | |
| ***4c*** | | | -3.10 (-5.13) | -3.10 (-5.13) | | |
| ***5*** | | | | | | |
| 6 | | | | | | |
| 7 | | | | | | |
| ***8*** | | | | | | |
| 9 | | | | | | |
| 10 | .581 (3.23) | .581 (3.23) | | | | |
| ***11*** | | | | | | |
| 12 | | | | | | |
| 13 | | | | | | |
| ***14*** | | | | | | |
| 15 | | | | | | |
| 16 | | | | | | |
| ***17*** | | | | | | |
| 18 | | | | | | |
| 19 | | | | | | |
| ***20*** | | | | | | |
| 21 | | | | | | |
| 22a | | | | | | |
| 22b | | | | | | |
| 22c | | | | | | |
| 23 | | | | | | |
| 24 | | | | | | |
| 25 | | | | | | |
| ***26*** | | | | | | |
| 27 | | | | | -.677 (-2.06) | -.677 (-2.06) |
| 28 | | | | | | |
| ***29*** | | | | | | |
| 30 | | | | | | |

…. – positive coefficient
…. – negative coefficient

---

“lunch” – “lunch hours: 11:00 to 14:00” indicator variable
“even” – “evening hours: 17:00 to 22:00” indicator variable
"nocons" – “no construction at the accident location” indicator variable

Table C.6 (Continued)

| # | Coefficient (t-ratio) | | | | | |
|---|---|---|---|---|---|---|
| | cons [$X_{13}$] | | dark [$X_{14}$] | | day [$X_{14}$] | |
| | fatality | injury | fatality | injury | fatality | injury |
| 1 | | | | | | |
| ***2*** | | | | | | |
| 3 | | | | | -.738 (-4.06) | |
| ***4a*** | | | | | | |
| ***4b*** | | | | | | |
| ***4c*** | | | | | | |
| ***5*** | | | | | | |
| 6 | | | | | | |
| 7 | | | .482 (2.12) | .482 (2.12) | | |
| ***8*** | | | | | | |
| 9 | | | | | | |
| 10 | | | | | -.425 (-2.95) | -.425 (-2.95) |
| ***11*** | | | | | | |
| 12 | | | 1.40 (2.33) | 1.40 (2.33) | | |
| 13 | | | 1.24 (2.85) | | | |
| ***14*** | | | | | | |
| 15 | | | | | -.931 (-3.80) | |
| 16 | | | | | | |
| ***17*** | | | | 1.52 (3.16) | | |
| 18 | | | | | .529 (4.35) | .529 (4.35) |
| 19 | -1.51 (-2.06) | -1.51 (-2.06) | | | | |
| ***20*** | | | | | | |
| 21 | | | | | | |
| 22a | | | | | | |
| 22b | | | | | -.150 (-2.26) | -.150 (-2.26) |
| 22c | | | | | | |
| 23 | | | | | | |
| 24 | | | | | -1.20 (-3.37) | |
| 25 | | | | | | |
| ***26*** | | | | | | |
| 27 | | | | | .235 (2.33) | .235 (2.33) |
| 28 | | | | | | |
| ***29*** | | | | | | |
| 30 | | | | | | |

…. – positive coefficient
…. – negative coefficient

"cons" – "construction at the accident location" indicator variable
"dark" – "dark time with no street lights" indicator variable
"day" – "daylight time" indicator variable

Table C.6 (Continued)

| # | Coefficient (t-ratio) | | | | | |
|---|---|---|---|---|---|---|
| | dawn [$X_{14}$] | | darklamp [$X_{14}$] | | precip [$X_{15}$] | |
| | fatality | injury | fatality | injury | fatality | injury |
| 1 | | | | | | |
| ***2*** | | | | | | |
| 3 | | | | | | |
| ***4a*** | | | | | | |
| ***4b*** | | | | | | |
| ***4c*** | | | | | | |
| ***5*** | | | | | | |
| 6 | | | | | | |
| 7 | | | | | | |
| ***8*** | | | | | | |
| 9 | | | | | | |
| 10 | | | | | | |
| ***11*** | | | .722 (2.45) | .722 (2.45) | | |
| 12 | | | | | | |
| 13 | | | | | | |
| ***14*** | | | 2.97 (2.42) | 2.97 (2.42) | | |
| 15 | | | | | -1.09 (-2.56) | |
| 16 | | | | | | |
| ***17*** | | | | | | |
| 18 | | | | | | |
| 19 | | | | | | |
| ***20*** | | | | | | |
| 21 | | | | | | |
| 22a | | | | | | |
| 22b | | | | | | |
| 22c | | | | | | |
| 23 | 3.14 (2.21) | | | | | |
| 24 | | | | | | |
| 25 | | | | | | |
| ***26*** | | | | | | |
| 27 | | | | | | |
| 28 | | | | | | |
| ***29*** | | | | | | |
| 30 | | | | | | |

…. – positive coefficient
…. – negative coefficient

"dawn" – "dawn OR dask" indicator variable
"darklamp" – "dark AND street lights on" indicator variable
"precip" – "precipitation: rain OR snow OR sleet OR hail OR freezing rain" indicator variable

Table C.6 (Continued)

| # | Coefficient (t-ratio) | | | | | |
|---|---|---|---|---|---|---|
| | snow [$X_{15}$] | | clear [$X_{15}$] | | dry [$X_{16}$] | |
| | fatality | injury | fatality | injury | fatality | injury |
| 1 | | | | | | |
| ***2*** | | | | | | |
| 3 | | | | .0929 (2.11) | | |
| ***4a*** | | | | | | |
| ***4b*** | | | | | | |
| ***4c*** | | | | | | |
| ***5*** | | | | | | |
| 6 | | | | | | |
| 7 | | | | | | |
| ***8*** | | | | | | |
| 9 | | | | | | |
| 10 | | | | | | |
| ***11*** | | | | | | |
| 12 | | | | | .574 (4.44) | .574 (4.44) |
| 13 | | | | | | |
| ***14*** | | | | | | |
| 15 | | | | | | |
| 16 | | | | | | |
| ***17*** | | | | | | |
| 18 | | | | | | |
| 19 | | | | | | |
| ***20*** | | | | | | |
| 21 | | | .518 (2.46) | .518 (2.46) | | |
| 22a | | | | -.143 (-3.36) | | |
| 22b | | | | | | |
| 22c | | | | | | |
| 23 | | | | | | |
| 24 | -.651 (-3.07) | -.651 (-3.07) | | | .385 (4.78) | .385 (4.78) |
| 25 | | | | | | |
| ***26*** | | | | | | |
| 27 | | | | | | |
| 28 | | | | | | |
| ***29*** | | | | | | |
| 30 | | | | | | |

…. – positive coefficient
…. – negative coefficient

"snow" – "snowing weather" indicator variable
"clear" – "clear weather" indicator variable
"dry" – "roadway surface is dry" indicator variable

Table C.6 (Continued)

| # | Coefficient (t-ratio) | | | | | |
|---|---|---|---|---|---|---|
| | wet [$X_{16}$] | | ice [$X_{16}$] | | lose [$X_{16}$] | |
| | fatality | injury | fatality | injury | fatality | injury |
| 1 | | | | | | |
| ***2*** | | | | | | |
| 3 | | | | | .428 (3.30) | .428 (3.30) |
| ***4a*** | | | | | | |
| ***4b*** | | | | | | |
| ***4c*** | | | | | | |
| ***5*** | | | | | | |
| 6 | | | | | | |
| 7 | | | | | | |
| ***8*** | | | | | | |
| 9 | | | | | | |
| 10 | | | | | | |
| ***11*** | | | | | | |
| 12 | | | | | | |
| 13 | | | | | | |
| ***14*** | | | | | | |
| 15 | | | | | | |
| 16 | | | | | | |
| ***17*** | | | | | | |
| 18 | | | | | | |
| 19 | | | | | | |
| ***20*** | | | | | | |
| 21 | | | -2.16 (-2.10) | -2.16 (-2.10) | | |
| 22a | | | | | | |
| 22b | | | | | | |
| 22c | | | | | | |
| 23 | .584 (3.01) | .584 (3.01) | | | | |
| 24 | | | | | | |
| 25 | | | | | | |
| ***26*** | | | | | | |
| 27 | | | | | | |
| 28 | | | | | | |
| ***29*** | | | | | | |
| 30 | | | | | | |

…. – positive coefficient
…. – negative coefficient

---

"wet" – "roadway surface is wet" indicator variable
"ice" – "roadway surface is icy" indicator variable
"lose" – "roadway surface has loose material on it" indicator variable

Table C.6 (Continued)

| # | Coefficient (t-ratio) | | | | | |
|---|---|---|---|---|---|---|
| | water [$X_{16}$] | | driv [$X_{17}$] | | wall [$X_{17}$] | |
| | fatality | injury | fatality | injury | fatality | injury |
| 1 | | | | | | |
| ***2*** | | | | | | |
| 3 | | | | | | |
| ***4a*** | | | | | | |
| ***4b*** | | | | | | |
| ***4c*** | | | -1.13 (-2.41) | | | |
| ***5*** | | | | | | |
| 6 | | | | | | |
| 7 | | | .467 (2.55) | .467 (2.55) | | |
| ***8*** | | | | | | |
| 9 | | | | | | |
| 10 | | | .439 (2.44) | .439 (2.44) | | |
| ***11*** | | 2.30 (2.40) | | | | |
| 12 | | | | | | |
| 13 | | | | | | |
| ***14*** | | | | | | |
| 15 | | | | | | |
| 16 | | | | | | |
| ***17*** | | | | | | |
| 18 | | | | | | |
| 19 | | | | | | |
| ***20*** | | | | | | |
| 21 | | | | | | |
| 22a | | | | | | |
| 22b | | | | | | |
| 22c | | | | | | |
| 23 | | | | | | |
| 24 | | | .281 (3.83) | .281 (3.83) | | |
| 25 | | | | | | |
| ***26*** | | | | | -2.10 (-1.97) | |
| 27 | | | | | | |
| 28 | | | | | | |
| ***29*** | | | | | | |
| 30 | | | | | | |

.... – positive coefficient
.... – negative coefficient

---

"water" – “roadway surface has water on it” indicator variable
"driv" – “road median is a drivable” indicator variable
"wall" – “road median is a wall” indicator variable

Table C.6 (Continued)

| # | Coefficient (t-ratio) | | | | | |
|---|---|---|---|---|---|---|
| | nomed [$X_{17}$] | | curb [$X_{17}$] | | nojun [$X_{18}$] | |
| | fatality | injury | fatality | injury | fatality | injury |
| 1 | | | | | | |
| ***2*** | -2.73 (-2.10) | | | | | |
| 3 | | | -1.71 (-2.57) | | .167 (2.35) | .167 (2.35) |
| ***4a*** | | | | | | |
| ***4b*** | | | | | | |
| ***4c*** | | | | | | |
| ***5*** | | | | | | |
| 6 | | | | | | |
| 7 | | | | | .591 (2.19) | .591 (2.19) |
| ***8*** | | | | | | |
| 9 | | | | | | |
| 10 | | | | | .536 (3.73) | .536 (3.73) |
| ***11*** | | | | | | |
| 12 | | | | | | |
| 13 | | | | | | |
| ***14*** | | | | | | |
| 15 | | | | | | |
| 16 | | | | | -.367 (-5.86) | -.367 (-5.86) |
| ***17*** | | | | | | |
| 18 | | | | | -.548 (-4.14) | -.548 (-4.14) |
| 19 | | -.236 (-2.13) | | | | |
| ***20*** | | | | | | |
| 21 | | | | | | |
| 22a | | -.112 (-2.37) | | | -.389 (-8.68) | -.389 (-8.68) |
| 22b | | | | | -.252 (-4.18) | -.252 (-4.18) |
| 22c | | | | | | |
| 23 | | | | | | |
| 24 | | | | | | |
| 25 | | | -1.21 (-2.38) | -1.21 (-2.38) | | |
| ***26*** | | | | | | |
| 27 | | | | | | |
| 28 | | | | | | |
| ***29*** | | | | | | |
| 30 | | | | | | |

…. – positive coefficient
…. – negative coefficient

"nomed" – "no median" indicator variable
"curb" – "road median is curbed" indicator variable
"nojun" – "no road junction at the accident location" indicator variable

Table C.6 (Continued)

| # | Coefficient (t-ratio) | | | | | |
|---|---|---|---|---|---|---|
| | way4 [$X_{18}$] | | T [$X_{18}$] | | str [$X_{19}$] | |
| | fatality | injury | fatality | injury | fatality | injury |
| 1 | | | | | | |
| ***2*** | 1.13 (3.18) | 1.13 (3.18) | | | | |
| 3 | | | | | -.194 (-3.95) | -.194 (-3.95) |
| ***4a*** | | | | | | |
| ***4b*** | | | | | | |
| ***4c*** | | | | | | |
| ***5*** | | | | | | |
| 6 | | | | | | |
| 7 | | | | | | |
| ***8*** | | | 3.37 (2.65) | | | |
| 9 | | | | | | |
| 10 | | | | | | |
| ***11*** | | | | | | |
| 12 | | | | | | |
| 13 | | | | | | |
| ***14*** | | | | | | |
| 15 | | | | | -.322 (-4.80) | -.322 (-4.80) |
| 16 | | | | | | |
| ***17*** | | | | | | |
| 18 | | | | | | |
| 19 | .328 (2.95) | .328 (2.95) | | | | |
| ***20*** | | | | | | |
| 21 | | | | | | |
| 22a | | | | | | |
| 22b | | | | | | |
| 22c | | | | | | |
| 23 | | .682 (4.11) | | | | |
| 24 | | | | | | |
| 25 | | .436 (2.03) | | | | |
| ***26*** | | | | | | |
| 27 | | | | | -.227 (-2.02) | -.227 (-2.02) |
| 28 | | | | | | |
| ***29*** | | | | | | |
| 30 | | | | | | |

…. – positive coefficient
…. – negative coefficient

"way4" – "accident location is at a 4-way intersection" indicator variable
"T" – "accident location is at a T-intersection" indicator variable
"str" – "road is straight" indicator variable

Table C.6 (Continued)

| # | Coefficient (t-ratio) | | | | | |
|---|---|---|---|---|---|---|
| | hill [$X_{19}$] | | driver [$X_{20}$] | | env [$X_{20}$] | |
| | fatality | injury | fatality | injury | fatality | injury |
| 1 | | | | | | |
| ***2*** | | | | | | |
| 3 | | | | | -3.70 (-7.24) | -1.75 (-30.4) |
| ***4a*** | | | | | | |
| ***4b*** | | | | | | |
| ***4c*** | | | | | | |
| ***5*** | | | | | | |
| 6 | | | | | -1.01 (-4.10) | -1.01 (-4.10) |
| 7 | | | | | | |
| ***8*** | | | | | | |
| 9 | | | 3.64 (3.12) | 1.67 (14.5) | | |
| 10 | | | | | | |
| ***11*** | | 1.25 (2.18) | | | | |
| 12 | | | | 1.52 (8.49) | | |
| 13 | | | 1.08 (2.43) | 1.08 (2.43) | | |
| ***14*** | | | | | | |
| 15 | | | | | -4.07 (-7.81) | -2.04 (-27.3) |
| 16 | | | -2.50 (-2.76) | | | |
| ***17*** | | | | | | |
| 18 | | | | | | |
| 19 | | | | | | |
| ***20*** | | | | | | |
| 21 | | | | | -1.39 (-3.55) | -1.39 (-3.55) |
| 22a | | | | | | |
| 22b | | | | | | |
| 22c | | | | | | |
| 23 | | | | | | |
| 24 | | | | 1.01 (7.15) | | |
| 25 | | | | | | |
| ***26*** | | | | | | |
| 27 | | | | | -3.91 (-3.79) | -1.98 (-17.1) |
| 28 | | | | | | |
| ***29*** | | | | | | |
| 30 | | | .817 (2.06) | .817 (2.06) | | |

…. – positive coefficient
…. – negative coefficient

"hill" – “road is at hill” indicator variable
"driver" – “primary cause of accident is driver-related” indicator variable
"env" – “primary cause of accident is environment-related” indicator variable

Table C.6 (Continued)

| # | Coefficient (t-ratio) | | | | | |
|---|---|---|---|---|---|---|
| | hl5 [$X_{22}$] | | help [$X_{22}$] | | hl10 [$X_{22}$] | |
| | fatality | injury | fatality | injury | fatality | injury |
| 1 | | | | | 1.00 (8.01) | |
| ***2*** | | | | | | |
| 3 | | | | | | |
| ***4a*** | | | | | | |
| ***4b*** | | | | | | .805 (2.12) |
| ***4c*** | | | | -.0809 (-3.77) | | |
| ***5*** | | | | | | |
| 6 | | | | | | |
| 7 | | | | | | .658 (3.75) |
| ***8*** | | | | | | |
| 9 | | | | | | |
| 10 | | | | | | |
| ***11*** | | | | | .905 (3.76) | .905 (3.76) |
| 12 | | | | | | |
| 13 | | | | | | |
| ***14*** | | | | | | |
| 15 | | | | | | |
| 16 | .676 (10.5) | .676 (10.5) | | | | |
| ***17*** | | | | | | |
| 18 | .752 (6.14) | .752 (6.14) | | | | |
| 19 | | | | | .567 (4.93) | .567 (4.93) |
| ***20*** | | | | | | |
| 21 | | | | | | |
| 22a | | | | | .692 (13.2) | .692 (13.2) |
| 22b | .671 (12.3) | .671 (12.3) | | | | |
| 22c | | | | | .770 (.687) | .770 (.687) |
| 23 | | .784 (4.69) | | | | |
| 24 | | | | | .867 (10.5) | .867 (10.5) |
| 25 | | | | | | 1.01 (5.04) |
| ***26*** | | .756 (2.84) | | | | |
| 27 | | | | | | |
| 28 | .647 (5.72) | .647 (5.72) | | | | |
| ***29*** | | | -.0871 (-2.93) | -.0871 (-2.93) | | |
| 30 | | .677 (2.42) | | | | |

…. – positive coefficient
…. – negative coefficient

"hl5" – "help arrived in 5 minutes or less after the crash" indicator variable
"help" – "time when help arrived after the crash" indicator variable
"hl10" – "help arrived in 10 minutes or less after the crash" indicator variable

Table C.6 (Continued)

| # | Coefficient (t-ratio) | | | | | |
|---|---|---|---|---|---|---|
| | hl20 [$X_{22}$] | | car [$X_{25}$] | | moto [$X_{25}$] | |
| | fatality | injury | fatality | injury | fatality | injury |
| 1 | | | | | | |
| ***2*** | | | | | | |
| 3 | .910 (18.7) | | | | 3.36 (9.64) | 3.13 (19.4) |
| ***4a*** | -1.19 (-6.06) | -1.19 (-6.06) | | | | |
| ***4b*** | | | | | | |
| ***4c*** | | | | | | |
| ***5*** | | | | | | |
| 6 | 1.06 (4.31) | 1.06 (4.31) | | | 2.46 (5.05) | 2.46 (5.05) |
| 7 | | | | | | |
| ***8*** | 1.57 (4.29) | 1.57 (4.29) | | | | |
| 9 | 1.71 (2.19) | .651 (5.74) | | | 4.58 (5.63) | 2.89 (5.44) |
| 10 | .578 (3.80) | .578 (3.80) | | | | |
| ***11*** | | | | .560 (2.35) | | |
| 12 | .929 (5.83) | .929 (5.83) | | | 4.41 (6.05) | 2.65 (5.76) |
| 13 | 1.10 (4.48) | 1.10 (4.48) | | | | |
| ***14*** | | | | | | |
| 15 | .952 (12.3) | .952 (12.3) | | | 3.25 (16.4) | 3.25 (16.4) |
| 16 | | | | | | |
| ***17*** | | | | | | |
| 18 | | | | | | 2.37 (6.54) |
| 19 | | | | | | |
| ***20*** | | | | | | |
| 21 | | | | | 4.63 (4.32) | 4.63 (4.32) |
| 22a | | | | | | |
| 22b | | | | | | |
| 22c | | | | | | |
| 23 | | | | .519 (3.06) | | |
| 24 | | | | | 2.35 (9.76) | 2.35 (9.76) |
| 25 | | | | | | |
| ***26*** | | | | | | |
| 27 | .950 (7.15) | .950 (7.15) | | | | 3.47 (8.12) |
| 28 | | | | | | |
| ***29*** | | | | | | |
| 30 | | | | | 3.38 (3.05) | 3.38 (3.05) |

.... – positive coefficient
.... – negative coefficient

"hl20" – "help arrived in 20 minutes or less after the crash" indicator variable
"car" – "the vehicle at fault is a car" indicator variable
"moto" – "the vehicle at fault is a motorcycle" indicator variable

Table C.6 (Continued)

| # | Coefficient (t-ratio) | | | | | |
|---|---|---|---|---|---|---|
| | pickup [$X_{25}$] | | vage [$X_{26}$] | | voldg [$X_{26}$] | |
| | fatality | injury | fatality | injury | fatality | injury |
| 1 | | | | | | |
| ***2*** | 2.76 (2.11) | 1.02 (2.20) | | | | |
| 3 | | | .0376 (9.76) | .0376 (9.76) | | |
| ***4a*** | | | | | | |
| ***4b*** | | | | | | |
| ***4c*** | | | | | | |
| ***5*** | | | | | | |
| 6 | | | .0543 (3.24) | .0543 (3.24) | | |
| 7 | | | | | | |
| ***8*** | | | | | | |
| 9 | | | .0372 (3.67) | .0372 (3.67) | | |
| 10 | | | | | | |
| ***11*** | | | | | | |
| 12 | | | | | .318 (2.55) | .318 (2.55) |
| 13 | | | | | | |
| ***14*** | | | | | | |
| 15 | | | .0373 (6.55) | .0373 (6.55) | | |
| 16 | | | | | | |
| ***17*** | | | | | | |
| 18 | | | | | | |
| 19 | | | | | | |
| ***20*** | | | | | | |
| 21 | | | | | | |
| 22a | | | | | | .137 (3.19) |
| 22b | | | | | | .142 (2.63) |
| 22c | | | | | .237 (2.46) | .237 (2.46) |
| 23 | | | .0333 (2.31) | .0333 (2.31) | | |
| 24 | | | .0145 (2.35) | .0145 (2.35) | | |
| 25 | | | | | | |
| ***26*** | | | | | | |
| 27 | | | .0375 (4.20) | .0375 (4.20) | | |
| 28 | | | | | | |
| ***29*** | | | | | | |
| 30 | | | .0566 (2.14) | .0566 (2.14) | | |

…. – positive coefficient
…. – negative coefficient

"pickup" – "the vehicle at fault is a pickup" indicator variable
"vage" – "age (in years) of the vehicle at fault" quantitative variable
"voldg" – "the vehicle at fault is more than 7 years old" indicator variable

Table C.6 (Continued)

| # | Coefficient (t-ratio) | | | | | |
|---|---|---|---|---|---|---|
| | $X_{27}$ | | Ind [$X_{28}$] | | othUS [$X_{28}$] | |
| | fatality | injury | fatality | injury | fatality | injury |
| 1 | | | | | | |
| ***2*** | | | | | | |
| 3 | .200 (8.23) | .200 (8.23) | | | | |
| ***4a*** | | | | | | |
| ***4b*** | | | | | | |
| ***4c*** | | | | | | |
| ***5*** | 2.15 (2.74) | 2.15 (2.74) | | | | |
| 6 | | | | | | |
| 7 | | | | | | |
| ***8*** | | | | | | |
| 9 | .385 (4.20) | 2.89 (5.44) | | | | |
| 10 | | | | | | |
| ***11*** | | | | | | |
| 12 | | | | | | |
| 13 | | | | | | |
| ***14*** | | | | | | |
| 15 | | | | | | |
| 16 | | | | | | |
| ***17*** | | | | | | |
| 18 | | | .433 (2.05) | .433 (2.05) | | |
| 19 | | | | | | |
| ***20*** | | | | | | |
| 21 | | | | | | |
| 22a | -.127 (-3.76) | -.127 (-3.76) | | | | |
| 22b | | | | | | |
| 22c | | | | | .617 (2.08) | .617 (2.08) |
| 23 | | | | | | |
| 24 | .105 (2.66) | .105 (2.66) | .585 (4.28) | .585 (4.28) | | |
| 25 | | | | | | |
| ***26*** | | | | | | |
| 27 | .415 (3.83) | | | | .621 (2.54) | .621 (2.54) |
| 28 | | | | | | |
| ***29*** | | | | | | |
| 30 | | | | | | |

…. – positive coefficient
…. – negative coefficient

---

"$X_{27}$" – “number of occupants in the vehicle at fault” quantitative variable
"Ind" – “license state of the vehicle at fault is Indiana” indicator variable
"othUS" – “license state of the vehicle at fault is a U.S. state except Indiana and its neighboring states (IL, KY, OH, MI)” indicator variable

Table C.6 (Continued)

| # | Coefficient (t-ratio) | | | | | |
|---|---|---|---|---|---|---|
| | neighs [$X_{28}$] | | lnm [$X_{30}$] | | priv [$X_{30}$] | |
| | fatality | injury | fatality | injury | fatality | injury |
| 1 | | | | | | |
| ***2*** | | | | | | |
| 3 | | | | | | |
| ***4a*** | | | | | | |
| ***4b*** | | | | | | |
| ***4c*** | | | | | | |
| ***5*** | | | | | | |
| 6 | | | | | | |
| 7 | | | | | | |
| ***8*** | | | | | | |
| 9 | | | | | | |
| 10 | | | | | | |
| ***11*** | | | | | | |
| 12 | | | | | | |
| 13 | -.858 (-2.44) | -.858 (-2.44) | | | | |
| ***14*** | | | | | | |
| 15 | | | | | | |
| 16 | | | | | | |
| ***17*** | | | | | | |
| 18 | | | | | | |
| 19 | | | | | | |
| ***20*** | | | | | | |
| 21 | | | | | | |
| 22a | -.266 (-2.64) | -.266 (-2.64) | | | -1.41 (-5.03) | -1.41 (-5.03) |
| 22b | | | | | -.973 (-3.14) | -.973 (-3.14) |
| 22c | | | | | -1.49 (-2.02) | -1.49 (-2.02) |
| 23 | | | | | | |
| 24 | | | | | | -.616 (-4.35) |
| 25 | | | | | | |
| ***26*** | | | | | | |
| 27 | | | | | | |
| 28 | | | | | | |
| ***29*** | | | | | | |
| 30 | | | .587 (2.01) | | | |

…. – positive coefficient
…. – negative coefficient

"neighs" – "license state of the vehicle at fault is Indiana's neighboring states (IL, KY, OH, MI)" indicator variable
"lnm" – "road traveled by the vehicle at fault is multi-lane" indicator variable
"priv" – "road traveled by the vehicle at fault is a private drive" indicator variable

Table C.6 (Continued)

| # | Coefficient (t-ratio) | | | | | |
|---|---|---|---|---|---|---|
| | rmu2 [$X_{30}$] | | stopsig [$X_{31}$] | | nosig [$X_{31}$] | |
| | fatality | injury | fatality | injury | fatality | injury |
| 1 | | | | | | |
| ***2*** | | | | | | |
| 3 | | | | | | |
| ***4a*** | | | | | | |
| ***4b*** | | | | | | |
| ***4c*** | | | | | | |
| ***5*** | | | | | | |
| 6 | .588 (1.97) | .588 (1.97) | | | | |
| 7 | | | | | | |
| ***8*** | | | | | | |
| 9 | | | | | | |
| 10 | | | | | | |
| ***11*** | | | | | | |
| 12 | | | | | | |
| 13 | | | .557 (2.44) | .557 (2.44) | | |
| ***14*** | | | 2.29 (2.93) | | | |
| 15 | | | | | | |
| 16 | | | | | | |
| ***17*** | | | | | | |
| 18 | | | | | | |
| 19 | | | | | | |
| ***20*** | | | | | | |
| 21 | | | | | | |
| 22a | | | | | | |
| 22b | | | | | | -.141 (-2.00) |
| 22c | | | | | -.488 (-3.97) | -.488 (-3.97) |
| 23 | | | | | | |
| 24 | | | | | | |
| 25 | | | | | | |
| ***26*** | | | | | | |
| 27 | | | | | | |
| 28 | | | | | | |
| ***29*** | | | | | | |
| 30 | | | | | -.845 (-2.91) | -.845 (-2.91) |

…. – positive coefficient
…. – negative coefficient

---

"rmu2" – "road traveled by the vehicle at fault is multi-lane AND undivided two-way" indicator variable
"stopsig" – "traffic control device for the vehicle at fault is a «stop sign»" indicator variable
"nosig" – "no any traffic control device for the vehicle at fault" indicator variable

Table C.6 (Continued)

| # | Coefficient (t-ratio) | | | | | |
|---|---|---|---|---|---|---|
| | sig [$X_{31}$] | | other [$X_{31}$] | | sign [$X_{31}$] | |
| | fatality | injury | fatality | injury | fatality | injury |
| 1 | | | | | 2.14 (2.04) | |
| ***2*** | | | | | | |
| 3 | | | | | | |
| ***4a*** | | | | | | |
| ***4b*** | | | | | | .848 (2.28) |
| ***4c*** | | | | | | |
| ***5*** | | | | | | |
| 6 | | | | | | |
| 7 | | | | | | |
| ***8*** | | | | | | |
| 9 | | | | | | |
| 10 | | | | | | |
| ***11*** | | | | | | |
| 12 | | | | | | |
| 13 | | | | | | |
| ***14*** | | | | | | |
| 15 | | | | | | |
| 16 | | | | | | |
| ***17*** | | | | | | |
| 18 | | | | | | |
| 19 | | | | | | |
| ***20*** | | 1.98 (2.66) | | | | |
| 21 | 1.03 (2.12) | 1.03 (2.12) | | | | |
| 22a | | | | | | |
| 22b | | | | | | |
| 22c | | | | | | |
| 23 | | | | | | |
| 24 | | .487 (4.25) | | | | |
| 25 | | | | | | |
| ***26*** | | | | | | |
| 27 | | | | | | |
| 28 | | | 7.18 (4.14) | | | |
| ***29*** | | | | | | |
| 30 | | | | | | |

…. – positive coefficient
…. – negative coefficient

---

"sig" – "traffic control device for the vehicle at fault is a signal" indicator variable
"other" – "traffic control device for the vehicle at fault is an «other regulatory sign or marking»" indicator variable
"sign" – "traffic control device for the vehicle at fault is an any sign" indicator variable

Table C.6 (Continued)

| # | Coefficient (t-ratio) | | | | | |
|---|---|---|---|---|---|---|
| | signal [$X_{31}$] | | $X_{33}$ | | $X_{34}$ | |
| | fatality | injury | fatality | injury | fatality | injury |
| 1 | | | | | | |
| ***2*** | | | | | | |
| 3 | | | 2.29 (6.96) | | .0127 (2.30) | |
| ***4a*** | | | | | | |
| ***4b*** | | | | | | |
| ***4c*** | | | | | | |
| ***5*** | | | | | | |
| 6 | | | | | | |
| 7 | | | | | | |
| ***8*** | | | | | | |
| 9 | | | | | | |
| 10 | | | | | | |
| ***11*** | | | | | | |
| 12 | | | | | | |
| 13 | | | 2.48 (2.00) | | .0103 (2.50) | .0103 (2.50) |
| ***14*** | | | | | | |
| 15 | | | 1.91 (2.87) | .793 (2.11) | .0153 (2.24) | |
| 16 | | | | | .0626 (3.10) | |
| ***17*** | | | | | | |
| 18 | | | | | | |
| 19 | | | | | | |
| ***20*** | | | | | | |
| 21 | | | 2.87 (2.56) | 2.87 (2.56) | -.0153 (-2.26) | -.0153 (-2.26) |
| 22a | | | 1.60 (4.89) | 1.60 (4.89) | | |
| 22b | | | 2.13 (5.16) | 2.13 (5.16) | | |
| 22c | | | 1.75 (2.41) | 1.75 (2.41) | | |
| 23 | | | | | | |
| 24 | | | 1.29 (2.97) | 1.29 (2.97) | | |
| 25 | | -.605 (-2.36) | | | | |
| ***26*** | | | | | | |
| 27 | | | 1.76 (2.24) | | | |
| 28 | | | | | | |
| ***29*** | | | 5.91 (3.29) | | .089 (2.80) | |
| 30 | | | 3.03 (2.43) | | | |

.... – positive coefficient
.... – negative coefficient

"signal" – "traffic control device for the vehicle at fault is an any signal" indicator variable
"$X_{33}$" – "at least one of the vehicles involved was on fire" indicator variable
"$X_{34}$" – "age (in years) of the driver at fault" quantitative variable

Table C.6 (Continued)

| # | Coefficient (t-ratio) | | | | | |
|---|---|---|---|---|---|---|
| | age1 [$X_{34}$] | | age2 [$X_{34}$] | | age5 [$X_{34}$] | |
| | fatality | injury | fatality | injury | fatality | injury |
| 1 | | | | | | |
| ***2*** | | | | | | |
| 3 | | | | | | |
| ***4a*** | | | -1.46 (-3.43) | -1.46 (-3.43) | | |
| ***4b*** | | | | | | |
| ***4c*** | -.789 (-2.09) | | | | | |
| ***5*** | | | | | | |
| 6 | | | | | | |
| 7 | | | | | | |
| ***8*** | | | | | | |
| 9 | | | | | | |
| 10 | | | | | | |
| ***11*** | | | | | | |
| 12 | | | | | | |
| 13 | | | | | | |
| ***14*** | | | | | | |
| 15 | | | | | | |
| 16 | | | | | | |
| ***17*** | | | | | | |
| 18 | | | | | | |
| 19 | | | | | | |
| ***20*** | | | | | | |
| 21 | | | | | | |
| 22a | | | | | | |
| 22b | | | | | | |
| 22c | | | | | | |
| 23 | | | | | | |
| 24 | | | | | .291 (3.46) | .291 (3.46) |
| 25 | | | | | | |
| ***26*** | | | | | | |
| 27 | | | | | | |
| 28 | | | | | | |
| ***29*** | | | | | | |
| 30 | | | | | | |

…. – positive coefficient
…. – negative coefficient

"age1" – "age of the driver at fault is ≥ 18 and < 24" indicator variable
"age2" – "age of the driver at fault is ≥ 24 and < 30" indicator variable
"age5" – "age of the driver at fault is ≥ 50 years" indicator variable

Table C.6 (Continued)

| # | Coefficient (t-ratio) | | | | | |
|---|---|---|---|---|---|---|
| | $X_{35}$ | | v3o [$X_{26}$] | | voldo [$X_{26}$] | |
| | fatality | injury | fatality | injury | fatality | injury |
| 1 | | | | | | |
| ***2*** | | | | | | |
| 3 | | .226 (4.92) | | | | |
| ***4a*** | | | | | | |
| ***4b*** | | | | | | |
| ***4c*** | | | | | | |
| ***5*** | | | | | | |
| 6 | .439 (2.22) | .439 (2.22) | | | | |
| 7 | .356 (2.02) | .356 (2.02) | | | | |
| ***8*** | | | | | | .735 (2.96) |
| 9 | | | | | | |
| 10 | | .450 (3.55) | | | | |
| ***11*** | | | | | | |
| 12 | | .444 (3.59) | | | | |
| 13 | | | | | | |
| ***14*** | | | | | | |
| 15 | | .330 (5.16) | | | | |
| 16 | | | | -.332 (-3.14) | | |
| ***17*** | .892 (3.16) | .892 (3.16) | | | | |
| 18 | | .483 (3.86) | | | | |
| 19 | | | | | | |
| ***20*** | | | | | | |
| 21 | | | | | | |
| 22a | | | | | | |
| 22b | | | | | | |
| 22c | | | | | | |
| 23 | | | | | | |
| 24 | | .278 (3.93) | | | | |
| 25 | .439 (2.32) | .439 (2.32) | | | | |
| ***26*** | | | | | .625 (2.65) | .625 (2.65) |
| 27 | .293 (2.91) | .293 (2.91) | | | | |
| 28 | -.537 (-3.57) | -.537 (-3.57) | | | | |
| ***29*** | | | | | | |
| 30 | | | | | | |

…. – positive coefficient
…. – negative coefficient

"$X_{35}$" – "gender of the driver at fault: 1 – female, 0 – male" indicator variable
"v3o" – "age of the oldest vehicle involved is > 1 and ≤ 3 years" indicator variable
"voldo" – "age of the oldest vehicle involved is > 7 years" indicator variable

Table C.6 (Continued)

| # | Coefficient (t-ratio) | | | | | |
|---|---|---|---|---|---|---|
| | maxpass [$X_{27}$] | | age2y [$X_{34}$] | | olddrv [$X_{34}$] | |
| | fatality | injury | fatality | injury | fatality | injury |
| 1 | .457 (2.58) | .182 (3.14) | | | | |
| ***2*** | | | | | | |
| 3 | | | | | | |
| ***4a*** | -.661 (-6.51) | | | | | |
| ***4b*** | | | | | | |
| ***4c*** | | | | | | |
| ***5*** | | | | | | |
| 6 | | | | | | |
| 7 | | | | | | |
| ***8*** | | | | | | |
| 9 | | | | | | |
| 10 | .140 (2.68) | .140 (2.68) | | | | |
| ***11*** | | | | | | |
| 12 | | | | | | |
| 13 | | | | | | |
| ***14*** | | | | | | |
| 15 | | | | | | |
| 16 | | .138 (4.77) | | | | |
| ***17*** | | | | | | |
| 18 | | | | | | |
| 19 | | .290 (5.67) | | | | |
| ***20*** | | | | | | |
| 21 | | | | | | |
| 22a | | .275 (9.91) | | | | |
| 22b | .143 (5.93) | .143 (5.93) | | | | |
| 22c | .128 (3.22) | .128 (3.22) | | | | |
| 23 | | | .512 (2.76) | .512 (2.76) | | |
| 24 | | | | | | |
| 25 | | | | | .133 (2.40) | .133 (2.40) |
| ***26*** | | | | | | |
| 27 | | | | | | |
| 28 | .133 (2.61) | | | | | |
| ***29*** | | | | | | |
| 30 | | | | | | |

…. – positive coefficient
…. – negative coefficient

---

"maxpass" – "the largest number of occupants in all vehicles involved" quantitative variable
"age2y" – "age of the youngest driver is ≥ 24 and < 30 years" indicator variable
"olddrv" – "the driver at fault is older than the other driver involved" indicator variable

Table C.6 (Continued)

| # | Coefficient (t-ratio) | | | | | |
|---|---|---|---|---|---|---|
| | age0o [$X_{34}$] | | age2o [$X_{34}$] | | age4o [$X_{34}$] | |
| | fatality | injury | fatality | injury | fatality | injury |
| 1 | | | | | | |
| ***2*** | | | | | | |
| 3 | | | | | | |
| ***4a*** | | | | | | |
| ***4b*** | | | | | | |
| ***4c*** | | | | | | |
| ***5*** | | | | | | |
| 6 | | | | | | |
| 7 | | | 1.75 (1.99) | | | |
| ***8*** | | | | | | |
| 9 | | | | | | |
| 10 | | | | | | |
| ***11*** | | | | | | |
| 12 | | | | | | |
| 13 | | | | | | |
| ***14*** | | | | | | |
| 15 | | | | | | |
| 16 | | | | | | .214 (2.93) |
| ***17*** | | | | | | |
| 18 | | | | | | |
| 19 | | | | | | |
| ***20*** | | | | | | |
| 21 | | | | | | |
| 22a | -1.25 (-4.40) | -1.25 (-4.40) | | | | |
| 22b | -2.20 (-3.03) | -2.20 (-3.03) | | | | |
| 22c | | | | | | |
| 23 | | | | | | |
| 24 | | | | | | |
| 25 | | | | | | |
| ***26*** | | | | | | |
| 27 | | | | | | |
| 28 | | | | | | |
| ***29*** | | | | | | |
| 30 | | | | | | |

…. – positive coefficient
…. – negative coefficient

---

"age0o" – “age of the oldest driver is < 18 years” indicator variable
"age2o" – “age of the oldest driver is ≥ 24 and < 30 years” indicator variable
"age4o" – “age of the oldest driver is ≥ 40 and < 50 years ” indicator variable

Table C.6 (Continued)

| # | Coefficient (t-ratio) | | | | | |
|---|---|---|---|---|---|---|
| | ff [$X_{35}$] | | mm [$X_{35}$] | | mf [$X_{35}$] | |
| | fatality | injury | fatality | injury | fatality | injury |
| 1 | | | | | | |
| ***2*** | | | | | | |
| 3 | | | | | | |
| ***4a*** | | | | | | |
| ***4b*** | | | | | | |
| ***4c*** | | | | | | .920 (2.84) |
| ***5*** | | | | | | |
| 6 | | | | | | |
| 7 | | | | | | |
| ***8*** | | | | | | |
| 9 | | | | | | |
| 10 | | | | | | |
| ***11*** | | | | | .534 (2.25) | .534 (2.25) |
| 12 | | | | | | |
| 13 | | | | | | |
| ***14*** | | | | | | |
| 15 | | | | | | |
| 16 | | | | -.240 (-3.39) | | |
| ***17*** | | | | | | |
| 18 | | | | | | |
| 19 | | | | | | |
| ***20*** | | | | | | |
| 21 | | | | | | |
| 22a | | | -.217 (-4.11) | -.217 (-4.11) | | |
| 22b | | | -.193 (-3.13) | -.193 (-3.13) | | |
| 22c | | | -.266 (-2.63) | -.266 (-2.63) | | |
| 23 | | | | -.451 (-2.64) | | |
| 24 | | | | | | |
| 25 | | | | | | |
| ***26*** | | | | | | |
| 27 | | | | | | |
| 28 | .725 (4.24) | .725 (4.24) | | | | |
| ***29*** | | | | | | |
| 30 | | | | | | |

…. – positive coefficient
…. – negative coefficient

---

"ff" – "two female drivers involved into a two-vehicle accident" indicator variable
"mm" – "two male drivers involved into a two-vehicle accident" indicator variable
"mf" – "male and female drivers involved into a two-vehicle accident" indicator variable

Table C.7 Tests of car-SUV separation in 2004 accident severity study[27]

| # | Model name | | | $M$ | $K$ | $LL(\boldsymbol{\beta}_{\mathbf{m}})$ | $\sum LL(\boldsymbol{\beta}_{\mathbf{m}})$ | df | p-value | conclusion* |
|---|---|---|---|---|---|---|---|---|---|---|
| 1 | County road | rural | (car/SUV)+(car/SUV) | 3 | 17 | -1636.3 | -1616.3 | 34 | 0.22 | Car = SUV |
| 2 | County road | rural | (car/SUV)+(truck) | 2 | 9 | -250.47 | -248.11 | 9 | 0.86 | Car = SUV |
| 4 | County road | urban | (car/SUV)+(car/SUV) | 3 | 10 | -683.95 | -678.39 | 20 | 0.94 | Car = SUV |
| 5 | County road | urban | (car/SUV)+(truck) | 2 | 5 | -104.80 | -100.71 | 5 | 0.15 | Car = SUV |
| 7 | Interstate | rural | (car/SUV)+(car/SUV) | 3 | 4 | -456.77 | -451.09 | 8 | 0.18 | Car = SUV |
| 8 | Interstate | rural | (car/SUV)+(truck) | 2 | 9 | -318.00 | -308.75 | 9 | 0.03 | Car ≠ SUV |
| 10 | Interstate | urban | (car/SUV)+(car/SUV) | 3 | 13 | -710.09 | -699.64 | 26 | 0.75 | Car = SUV |
| 11 | Interstate | urban | (car/SUV)+(truck) | 2 | 6 | -336.87 | -334.52 | 6 | 0.58 | Car = SUV |
| 13 | State route | rural | (car/SUV)+(car/SUV) | 3 | 12 | -1302.1 | -1292.1 | 24 | 0.69 | Car = SUV |
| 14 | State route | rural | (car/SUV)+(truck) | 2 | 11 | -362.86 | -359.27 | 11 | 0.78 | Car = SUV |
| 16 | State route | urban | (car/SUV)+(car/SUV) | 3 | 12 | -1649.1 | -1630.5 | 24 | 0.04 | Car ≠ SUV |
| 17 | State route | urban | (car/SUV)+(truck) | 2 | 5 | -173.40 | -172.27 | 5 | 0.81 | Car = SUV |
| 19 | City street | rural | (car/SUV)+(car/SUV) | 3 | 12 | -1358.7 | -1353.1 | 24 | 0.99 | Car = SUV |
| 20 | City street | rural | (car/SUV)+(truck) | 2 | 9 | -68.07 | -64.80 | 9 | 0.69 | Car = SUV |
| 22 | City street | urban | (car/SUV)+(car/SUV) | 3 | 23 | -14547 | -14524 | 46 | 0.53 | Car = SUV |
| 23 | City street | urban | (car/SUV)+(truck) | 2 | 10 | -885.98 | -877.16 | 10 | 0.06 | Car ≠ SUV |
| 25 | US route | rural | (car/SUV)+(car/SUV) | 3 | 11 | -978.44 | -969.45 | 22 | 0.71 | Car = SUV |
| 26 | US route | rural | (car/SUV)+(truck) | 2 | 12 | -270.09 | -262.03 | 12 | 0.19 | Car = SUV |
| 28 | US route | urban | (car/SUV)+(car/SUV) | 3 | 12 | -2345.1 | -2334.3 | 24 | 0.60 | Car = SUV |
| 29 | US route | urban | (car/SUV)+(truck) | 2 | 9 | -311.62 | -305.06 | 9 | 0.16 | Car = SUV |

* For models 8, 16 and 23 we find that "Car ≠ SUV", which means that for these models cars and SUVs must be considered separately in our accident severity study. For all other models we find that "Car = SUV", which means that cars and SUVs can be considered together.

[27] For explanation of these tests refer to footnote 24 on page 90.

Table C.8 Tests of car-SUV separation in 2006 accident severity study

| # | Model name | | | $M$ | $K$ | $LL(\boldsymbol{\beta}_{\mathbf{m}})$ | $\sum LL(\boldsymbol{\beta}_{\mathbf{m}})$ | df | p-value | conclusion* |
|---|---|---|---|---|---|---|---|---|---|---|
| 1 | County road | rural | (car/SUV)+(car/SUV) | 3 | 10 | -832.13 | -826.10 | 20 | 0.91 | Car = SUV |
| 2 | | | (car/SUV)+(truck) | 2 | 8 | -113.99 | -109.77 | 8 | 0.39 | Car = SUV |
| 4 | | urban | (car/SUV)+(car/SUV) | 3 | 2 | -782.85 | -777.77 | 4 | 0.04 | Car ≠ SUV |
| 5 | | | (car/SUV)+(truck) | 2 | 4 | -16.716 | -13.652 | 4 | 0.19 | Car = SUV |
| 7 | Interstate | rural | (car/SUV)+(car/SUV) | 3 | 9 | -459.31 | -447.45 | 18 | 0.16 | Car = SUV |
| 8 | | | (car/SUV)+(truck) | 2 | 6 | -271.78 | -269.97 | 6 | 0.73 | Car = SUV |
| 10 | | urban | (car/SUV)+(car/SUV) | 3 | 9 | -853.29 | -846.69 | 18 | 0.78 | Car = SUV |
| 11 | | | (car/SUV)+(truck) | 2 | 9 | -252.55 | -246.81 | 9 | 0.24 | Car = SUV |
| 13 | State route | rural | (car/SUV)+(car/SUV) | 3 | 11 | -560.40 | -550.05 | 22 | 0.54 | Car = SUV |
| 14 | | | (car/SUV)+(truck) | 2 | 5 | -72.403 | -71.805 | 5 | 0.95 | Car = SUV |
| 16 | | urban | (car/SUV)+(car/SUV) | 3 | 12 | -3225.9 | -3212.7 | 24 | 0.34 | Car = SUV |
| 17 | | | (car/SUV)+(truck) | 2 | 5 | -192.97 | -192.09 | 5 | 0.88 | Car = SUV |
| 19 | City street | rural | (car/SUV)+(car/SUV) | 3 | 8 | -1068.2 | -1061.9 | 16 | 0.67 | Car = SUV |
| 20 | | | (car/SUV)+(truck) | 2 | 2 | -146.45 | -145.98 | 2 | 0.63 | Car = SUV |
| 22 | | urban | (car/SUV)+(car/SUV) | 3 | 17 | -9505.6 | -9481.4 | 34 | 0.05 | Car ≠ SUV |
| 23 | | | (car/SUV)+(truck) | 2 | 12 | -520.63 | -517.29 | 12 | 0.88 | Car = SUV |
| 25 | US route | rural | (car/SUV)+(car/SUV) | 3 | 9 | -380.51 | -367.95 | 18 | 0.12 | Car = SUV |
| 26 | | | (car/SUV)+(truck) | 2 | 8 | -237.97 | -231.83 | 8 | 0.14 | Car = SUV |
| 28 | | urban | (car/SUV)+(car/SUV) | 3 | 7 | -967.22 | -961.02 | 14 | 0.57 | Car = SUV |
| 29 | | | (car/SUV)+(truck) | 2 | 7 | -185.87 | -182.84 | 7 | 0.53 | Car = SUV |

* For models 4 and 22 we find that “Car ≠ SUV”, which means that for these models cars and SUVs must be considered separately in our accident severity study. For all other models we find that “Car = SUV”, which means that cars and SUVs can be considered together.